\documentclass[a4paper,twocolumn,accepted=2024-09-03,longbibliography]{quantumarticle}
\pdfoutput=1
\usepackage[utf8]{inputenc}
\usepackage[english]{babel}
\usepackage[numbers,sort&compress]{natbib}

\usepackage{hyperref}

\usepackage{graphicx,times}
\usepackage{color} 
\usepackage{amssymb}

\usepackage{subeqnarray}
\usepackage{diagbox}
\usepackage{makecell}
\usepackage{amsmath}
\usepackage{mathtools}
\usepackage{empheq}

\usepackage[dvipsnames]{xcolor}

\definecolor{aprcolor}{rgb}{0.1,0.2,0.9}

\newcommand{\msf}[1]{\mathsf{#1}}
\newcommand{\mcl}[1]{\mathcal{#1}}
\newcommand{\bsb}[1]{\boldsymbol{#1}}
\newcommand{\mbf}[1]{\mathbf{#1}}

\DeclareSymbolFont{greekletters}{OML}{FiraSans}{m}{n}

\DeclareMathOperator{\bV}{\bsb{V}}

\DeclareMathOperator{\bphi}{\bsb{\phi}}
\DeclareMathOperator{\bPhi}{\bsb{\Phi}}
\DeclareMathOperator{\bq}{\bsb{q}}
\DeclareMathOperator{\bQ}{\bsb{Q}}

\DeclareMathOperator{\bI}{\bsb{I}}

\DeclareMathOperator{\bz}{\bsb{z}}

\DeclareMathOperator{\bzeta}{\bsb{\zeta}}

\DeclareMathOperator{\msC}{\msf{C}}
\DeclareMathOperator{\msD}{\msf{D}}
\DeclareMathOperator{\msK}{\msf{K}}

\DeclareMathOperator{\msN}{\msf{N}}

\DeclareMathOperator{\msY}{\msf{Y}}

\DeclareMathOperator{\msF}{\msf{F}}
\DeclareMathOperator{\msL}{\msf{L}}
\DeclareMathOperator{\msG}{\msf{G}}

\DeclareMathOperator{\msP}{\msf{P}}

\DeclareMathOperator{\msE}{\msf{E}}

\newcommand{\mone}{\mathbbm{1}}

\newcommand{\dd}{{\mathrm{d}}}

\begin{document}
\title{Geometrical description and Faddeev--Jackiw quantization of electrical networks}
\author{A. Parra-Rodriguez}
\email{adrian.parra.rodriguez@gmail.com}
\affiliation{Institut Quantique and Département de Physique, Université de Sherbrooke, Sherbrooke, Qu\'ebec J1K 2R1, Canada}
\affiliation{Instituto de Física Fundamental IFF-CSIC, Calle Serrano 113b, 28006 Madrid, Spain}
\author{I. L. Egusquiza}
\email{inigo.egusquiza@ehu.es}
\affiliation{Department of Physics, University of the Basque Country UPV/EHU, Apartado 644, 48080 Bilbao, Spain}
\affiliation{EHU Quantum Centre, University of the Basque Country UPV/EHU, Apartado 644, 48080 Bilbao, Spain}

\begin{abstract}
  In lumped-element electrical circuit theory, the problem of solving Maxwell's equations in the presence of media is reduced to two sets of equations, the constitutive equations encapsulating local geometry and dynamics of a confined energy density, and  the Kirchhoff  equations enforcing  conservation of charge and energy in a larger, topological, scale. We develop a new geometric and systematic description of the dynamics of general lumped-element electrical circuits as first
  order differential equations, derivable from a Lagrangian and a Rayleigh dissipation function. Through the Faddeev-Jackiw method we identify and classify the
  singularities that arise in the search for Hamiltonian descriptions of general networks. The core of our solution relies on the correct identification of the reduced manifold in which the circuit state is expressible, e.g., a mix of
  flux and charge degrees of freedom, including the presence of compact ones. We apply our fully programmable method to obtain (canonically quantizable)  Hamiltonian descriptions of nonlinear and
  nonreciprocal circuits which would be cumbersome/singular if pure
  node-flux or loop-charge variables were used as a starting configuration space. We also propose a specific assignment of topology for the branch variables of energetic elements, that when used as input to the procedure gives results consistent with classical descriptions as well as with spectra of more involved quantum circuits. This work unifies diverse existent geometrical pictures of electrical network theory, and will prove useful, for instance, to automatize the computation of exact Hamiltonian descriptions of superconducting quantum chips.
\end{abstract}

\keywords{}
\maketitle

\section{Introduction}
The lumped element model of Maxwell's equations, also known as the
quasi-static approximation, has been extremely successful in
describing the low-energy dynamics of electrical
circuits~\cite{Jackson:1999,Feynman:2010}, including the rapidly
growing field  of superconducting quantum
circuits~\cite{CaldeiraLeggett:1981,YurkeDenker:1984,Blais:2021}. In
doing so, the distributed partial differential field equations
(infinite-dimensional state space) are reduced to ordinary
differential equations (finite-dimensional state space), with an
intrinsic high-frequency cutoff embedded in this
mapping~\cite{ParraRodriguezPhD:2021}.

The analogy between lumped circuits and mechanical dynamical systems
was known from the origins of the field, and consequently
lumped element circuit theory is periodically concerned, especially in the problem of   synthesis~\cite{Foster:1924,Cauer:1926,Cauer:1929,Brune:1931,Tellegen:1948a,Belevitch:1950,Cauer:1954,Newcomb:1966,Anderson:1975}, with the applicability of methods of
analytical mechanics. Thus, already by 1938 Wells
\cite{Wells:1938,Wells:1945} mentioned as well known that Lagrangians
could be constructed for some classes of circuits \cite{MacFarlane:1969,Chua:1974,Kwatny:1982,Shragowitz:1988}. Hamiltonian
approaches have also been regularly undertaken
\cite{Bernstein:1989,Maschke:1995,Weiss:1997}, in particular in the
port-Hamiltonian perspective
\cite{VanDerSchaft:2006,VanDerSchaft:2014}.

In more recent times the technological explosion in superconducting
circuits in the quantum regime has propelled the investigation of
their Lagrangian and Hamiltonian formulations
\cite{Devoret:1997,Burkard:2004,Burkard:2005,Ulrich:2016,ParraRodriguez:2022,Egusquiza:2022}. Descriptions
in terms of (loop) charge variables were initially dominant
\cite{YurkeDenker:1984,Chakravarty:1986,Yurke:1987}, but later the
(node) flux presentation took over
\cite{Devoret:1997,Paladino:2003,Blais:2004,Houck:2008,Bourassa:2009,Clerk:2010,Koch:2010,Filipp:2011,Bourassa:2012,Bergenfeldt:2012,Peropadre:2013,Nigg:2012,Devoret:2013,Solgun:2014,Sundaresan:2015,Solgun:2015,Malekakhlagh:2016,Mortensen:2016,Roy:2016,Vool:2017,Malekakhlagh:2017,Roy:2018,ParraRodriguez:2018}. Mixing as variables both
effective loop charges and effective node fluxes of the
network~\cite{Jeltsema:2009,Ulrich:2016,Mariantoni:2020,ParraRodriguez:2022} is being
developed currently. This is proven to be necessary for a general
treatment~\cite{ParraRodriguez:2019,Rymarz:2021,ParraRodriguez:2022,ParraRodriguez:2022b,Egusquiza:2022} of nonreciprocal circuits~\cite{Kerckhoff:2015,Sliwa:2015,Lecocq:2017,Chapman:2017,Mahoney:2017,Barzanjeh:2017,Rosario:2018,Navarathna:2023}. 

Analytical mechanics is a highly geometrical theory, and thus geometrical considerations considerations have entered these descriptions since the work of Brayton and Moser \cite{Brayton:1964a,Brayton:1964b} and Smale \cite{Smale:1972}. This has led, among other lines of inquiry, to the deep concept of Dirac
structures \cite{Weinstein:1983,Courant:1990} applied to electrical
circuits \cite{Yoshimura:2006}. The main reason is that the dynamics
of the circuit is not simply due to the energetics of the individual
elements (capacitors and inductors). Rather, it is \emph{constrained}
by charge and energy conservation in the connections of the
elements~\cite{Kirchhoff:1847}.  Now, as constrained dynamics is the central object of study in the
geometrical approach to mechanics, we posit that a geometrical
approach is essential to further develop this aspect of circuit theory
and to achieve our goal of a quantum mechanical description of
superconducting circuits. Furthermore, in that context some variables
can be compact, such as fluxes for superconducting islands, and
topological considerations come into play.

In this article, we address the central question in circuit theory: Can we construct Lagrangian and Hamiltonian descriptions of lumped-element circuits composed of nonlinear inductors, capacitors, voltage and current sources,  transformers~\cite{Belevitch:1950} and gyrators~\cite{Tellegen:1948a}, connected arbitrarily? Our constructive answer is in the  affirmative. We use a geometric language that ensures that compact variables are inherently considered. In doing so, we drastically extend the range of circuits
for which these descriptions exist, with direct applications in canonical circuit quantization~\cite{YurkeDenker:1984,Devoret:1997,Burkard:2004,Burkard:2005,Solgun:2015,ParraRodriguez:2019,Egusquiza:2022}, for instance in the extension of the blackbox quantization approach to multiport nonreciprocal linear systems~\cite{Nigg:2012,Solgun:2014,Solgun:2015,ParraRodriguez:2019,ParraRodriguez:2022b}. Furthermore, within the framework of classical physics, we incorporate linear multiport resistive elements into the method by introducing a Rayleigh dissipation function.

Our systematic approach is algorithmizable,  in computer algebra or numerical software, offering significant improvements over existing methods~\cite{Gely:2020,Minev:2021,Chitta:2022,Rajabzadeh:2023}. We employ a first-order Lagrangian and use the Faddeev--Jackiw method (FJ) for symplectic reduction~\cite{Faddeev:1969,Faddeev:1988,Jackiw:1993,Jackiw:1994}, which prepares for quantization to be tackled, both canonically and with a path integral formalism. In cases where obstacles to a Hamiltonian description arise, our method  detects and characterizes these hindrances~\cite{Rymarz:2023,Miano:2023}. 

The geometric approach is valid, as a method, independently of the actual compact or extended character of the branch variables. Even so, we conjecture a concrete assignment of topology (extended or compact) for the branch variables of energetic elements, and analyze its consequences. This proposal is compatible with all the experiments of which we are aware. In particular, for well-posed circuits with classical elements the final effective physics will be described in our proposal with extended variables. So will the fluxonium~\cite{Manucharyan:2009}, for instance, in contraposition to compact nodal phase for the Cooper pair box~\cite{Buettiker:1987}.

This paper is organized as follows. In the next section we introduce the geometrical description of the class of circuits under consideration, introducing differential geometric concepts as required. We postulate the assignment of degenerate two-forms to capacitive, inductive and source branches.  We then  in Section \ref{sec:textproof} prove that the dynamical system that follows from applying constraints to the energy function and the two form is precisely the one that corresponds to the circuit. Once the geometric perspective has been shown to apply, we can identify singular dynamical systems, and we propose the application of the Faddeev-Jackiw method in Section \ref{sec:FJtextJ}. 


Focused on the application of the method to quantize lumped-element superconducting circuits, we put forward in Sec.~\ref{sec:quantization_superc_circuits} a proposal for topology assignment to energetic elements, and analyze is consequences, such as (classical) decompactification of phase for the fluxonium or the dualmon~\cite{ThanhLe:2019} circuits. In Section \ref{sec:circuit_examples} we present more involved examples for which the geometric approach provides systematic descriptions and for which no other systematic procedure had been presented to date. We defer to appendices a more detailed look at the proof, as well as general introductions to other geometrical concepts and to the Faddeev-Jackiw method. We also present in appendices the method as it pertains to sources, in particular time dependent magnetic fluxes and the corresponding emfs. Additional circuit examples are provided in App. \ref{sec_app:examples}, and classically singular circuits are addressed in the last App.~\ref{sec:nonl-sing-circ}. A lightning review of differential geometry, symplectic geometry, and canonical quantization can be found in App. \ref{sec:mathapp}.

\section{Geometrical description of classical electrical circuits}
\label{sec:geomtext}
In traditional circuit theory \cite{Guillemin:1953} the
electrical state is determined by the values of voltage drop $v^b$ and current intensity $i^b$ in each port/branch $b$. These variables are separately  linear, and this variable space for each branch is of the form $\mathbb{R}\times\mathbb{R}$. We investigate flux-charge elements, for which the state is given by branch fluxes $\phi^b$ and charges $q^b$, whose time derivatives are $v^b$ and $i^b$ respectively. We do not consider other elements of the Chua classification \cite{Chua:1980}, involving higher integrals or derivatives, as these would systematically lead to Ostrogradsky singularities, i.e., energetic instability \cite{Ostrogradsky:1850,Woodard:2015}. The manifold of states for each branch has as tangent $\mathbb{R}\times\mathbb{R}$, with distinct voltage and intensity directions. Thus the only possibilities for fluxes and charges as variables are $\mathbb{R}$ and $S^1$ or intervals thereof \cite{nakahara2003geometry}. In traditional circuit theory compact variables were not considered, but the introduction of superconducting circuits has led to compact fluxes (in presence of superconducting islands) or compact charges (for quantum phase slips). Hence, the model for a circuit will have a state manifold of the form $\mathcal{M}_{2B}=\mathbb{R}^{2B-k}\times(S^1)^k$, with $k$ the number of compact variables and $B$ the number of branches in the circuit graph. In the following, we will use the notation $\mathbbm{T}^k$ (the k-torus) for $(S^1)^k$.

\begin{figure}
	\includegraphics[width=1\linewidth]{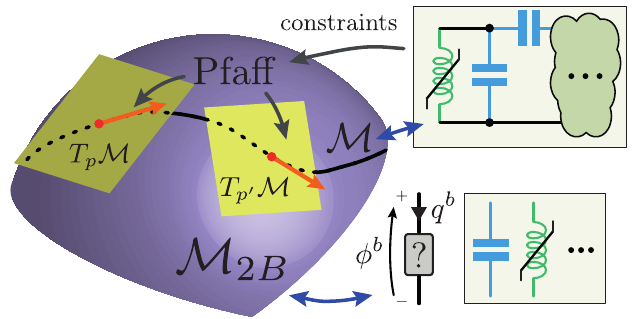}
	\caption{Sketch of the geometrical picture. There are two variables per branch, and they span $\mathcal{M}_{2B}$. Connecting the branches in a circuit gives rise to constraints, collected in a Pfaff system that fixes the possible directions at each point on $\mathcal{M}_{2B}$. The integration of those directions results in the integral manifold $\mathcal{M}$ that describes the constrained physics.}
	\label{fig:Geometric_sketch}
\end{figure}

The concept of conjugate variables is present in several branches of physics, with different understandings of what ``conjugation" means. However, these different meanings are not wholly unrelated. In the context of electrical circuits charge current and voltage are conjugate both in that one looks at the response in one of them induced by a change in the other (immitances), and in an energy argument. Our concern, because of our quantization goal, is mechanical conjugation. As we use the integrated version of these electrical variables, and because of Eqs. \eqref{eq:conscl} below, we expect conjugation between the flux and the charge of each branch, so $\mathcal{M}_{2B}$ is to be styled as a phase space
or precursor thereof. However, there are constraints. Most importantly, the voltage and current Kirchoff constraints are a
Pfaff system of equations for one forms \cite{Choquet-Bruhat:1968} in
$T^*\mathcal{M}_{2B}$,
    \begin{align}
        \sum_{b\in \mathcal{N}}\mathrm{d}q^b =0\,\mathrm{(KCL)}\,,\qquad
        \sum_{b\in  \mathcal{P}}\mathrm{d}\phi^b =0\,\mathrm{(KVL)}\,,
    \end{align}
with $b\in \mathcal{N}$ denoting all branches incident on node $\mathcal{N}$ and $b\in \mathcal{P}$ all branches in loop $\mathcal{P}$. A short review of differential geometry is included as App. \ref{sec:mathapp}. Here and in what follows, and in line with frequent abuse of notation, we use  $\mathrm{d}z$ to denote the $z$ coordinate one-form also if the corresponding manifold is $S^1$, in spite of it not being an exact form. 

This set of equations is of rank $B$, i.e. the number of independent equations is $B$. This follows from standard graph theory, first applied to circuits by Weyl  \cite{Weyl:1923}. An additional property of these equations is that all linear combinations are closed. It follows that the Frobenius compatibility \cite{Choquet-Bruhat:1968} conditions are automatically satisfied, and therefore the system is integrable. This means that each point of $\mathcal{M}_{2B}$ is included in one and only one manifold of dimension $B$ that is a solution of the system. For practical purposes, this entails finding a parametrization with $B$ parameters of solutions of the equations. When the constraints are just the Kirchhoff ones the flux and the charge variables are solved in terms of separate flux and charge parameters and the standard node-flux analysis provides us with mechanisms to do so. Furthermore, if there are no compact directions in $\mathcal{M}_{2B}$ the Kirchhoff Pfaff equations devolve into a set of algebraic linear equations, in line with usual presentations. We shall below show the impact of nontrivial topologies in this construction by means of examples.

Additionally, to describe canonical representations for linear time-independent (LTI) passive multiport devices, one needs to take into account ideal NR elements like gyrators, circulators, and transformers.  In the terminology of superconducting technology~\cite{Blais:2021}, such multiport descriptions are often referred to as ``blackboxes"~\cite{Nigg:2012,Solgun:2014,Solgun:2015}, as they do not reveal internal microscopic dynamics. These additional elements  impose further linear restrictions on branch voltage and intensity configurations~\cite{Foster:1924,Cauer:1926,Tellegen:1948,Belevitch:1950}. These elements, therefore, correspond to additional Pfaff equations, which involve both the branch variables of their ports and of other elements.

In summary, the constraints, both Kirchhoff and constraining element ones, are applied to branch voltages and currents and represented by a matrix $\msF$, constant over the manifold, that annihilates the differential state vector $\dd{\bzeta}^T=(\dd{\bphi}^T, \dd{\bq}^T)$, that is $\msF\dd{\bzeta}=0$. If the Kirchhoff constraints are only augmented with ideal transformers the constraint equations separate in a flux subset and a charge subset, and the solutions are expressed separately with flux parameters and charge parameters respectively, albeit no longer directly given by node-flux analysis. However, the presence of gyrators\footnote{We use the gyrator as a representative of ideal non-reciprocal (NR) elements throughout the manuscript.} mixes fluxes and charges. At any rate, because the matrix $\mathsf{F}$ is constant we can conclude that the system of constraints is again integrable.

Even restricting ourselves to the case of only Kirchhoff constraints, importantly, the number of charge type and flux type variables need not be equal, thus
precluding the integral manifolds from being understood as phase
spaces $T^*\mathcal{X}$ for some underlying configuration space
$\mathcal{X}$.  Hence the issues in all approaches to the construction of Hamiltonians for circuits of lumped elements, in one way or another.
However, we know that large classes of circuits do indeed admit Hamiltonian descriptions. On reflection, we observe that the  constructions for those descriptions are associated with the specific distribution of energetic elements of different types on  specific graph structures, in intimate interplay with the topology of the circuit~\cite{Burkard:2004,Burkard:2005}.

Therefore, we consider circuits for which each branch belongs to one
of the following categories, all of them ideal: linear and nonlinear capacitors ($\mathcal{C}$) and
inductors  ($\mathcal{L}$),  voltage  ($\mathcal{V}$) and current
($\mathcal{I}$) sources,  linear resistors  ($\mathcal{R}$), and 
transformer ($\mathcal{T}$) and gyrator  ($\mathcal{G}$)
branches. Only the reactive     ($\mathcal{C}$ and $\mathcal{L}$) and
source 
($\mathcal{I}$  and $\mathcal{V}$) branches do actually present conjugate pairs
of variables  from the dynamical perspective, while the dissipative
$\mathcal{R}$  set and the $\mathcal{T}$ and $\mathcal{G}$ sets will
be  constraints. The constitutive relations for capacitive $c\in\mathcal{C}$ and
inductive $l\in\mathcal{L}$ branches are \begin{align}
  \label{eq:conscl}
  \dot{\phi}^c= v^c= \frac{\partial h_c}{\partial q^c}\,,\qquad
  \dot{q}^l = i^l = \frac{\partial h_l}{\partial \phi^l}\,,
\end{align}
with a total energy function 
\begin{align}
    H=
\sum_{l\in\mathcal{L}}h_l(\phi^l)+\sum_{c\in\mathcal{C}}h_c(q^c).\label{eq:Henergy_C_L}
\end{align}
Please observe that this energy function depends on the point in the manifold $\mathcal{M}_{2B}$, but that we always express it in terms of a coordinate system.

As to the external voltage  ($v_v(t)$) and current ($i_i(t)$) sources, they will be described  by the (time-dependent) energy terms 
\begin{align}
    H_d(t)=\sum_{v\in \mcl{V}} q^v v_v(t)+\sum_{i\in \mcl{I}}\phi^i i_{i}(t)\label{eq:Henergy_drive_V_I}.
\end{align}
In the following, we will denote the total energy contributions by $H_T=H+H_d(t)$, combining the passive and active components.

Both sets of relations \eqref{eq:conscl} seem to be half of a  Hamiltonian pair, with opposing signs for capacitive and inductive elements.
Thus, we associate a degenerate two-form on
$\mathcal{M}_{2B}$ to the actual content of the circuit as
\begin{align}
  \label{eq:twoform2b}
  \begin{aligned}
      \omega_{2B}&= \frac{1}{2}
\left[\sum_{l\in\mathcal{L}}\mathrm{d}\phi^{{l}}\wedge\mathrm{d}q^l+
  \sum_{c\in\mathcal{C}}
  \mathrm{d}q^c\wedge\mathrm{d}\phi^{{c}}\right]\\
&  +  \frac{1}{2} \left[\sum_{i\in\mathcal{I}}\mathrm{d}\phi^{{i}}\wedge\mathrm{d}q^i +
  \sum_{v\in\mathcal{V}}
  \mathrm{d}q^v\wedge\mathrm{d}\phi^{{v}}\right].
  \end{aligned}
\end{align}
We restate again that, if necessary, we incur in the standard abuse of notation $\mathrm{d}\alpha$ for
the winding one form in
an $S^1$ variable, by writing it in terms of the angle $\alpha$ in one
coordinate patch. We also associate 
a degenerate  symmetric 2-covariant tensor field 
\begin{align}
  \label{eq:g2b}
  g_{2B}=\frac{1}{2}\sum_{r\in\mathcal{R}} \mathrm{d}q^r \mathrm{d}\phi^{r}\,,
\end{align}
that encodes dissipation,
and a source one form $\sigma$, formally
\begin{align}
    \sigma &= H_d(t) \mathrm{d}t\,,
\end{align}
in the extended manifold $\mathcal{M}_{2B}\times \mathbb{R}$, to account for the external time dependence through sources.

\begin{figure}[!ht]
	\includegraphics[width=1\linewidth]{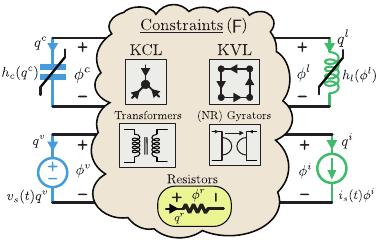}
	\caption{The lumped ideal elements under consideration. In the periphery the dynamical elements, determined by energy functions. Anticlockwise from the top right, inductor ($\mathcal{L}$), capacitor ($\mathcal{C}$), voltage source ($\mathcal{V}$) and current source ($\mathcal{I}$). In the cloud, various constraints govern the connection of dynamical elements, including Kirchhoff's current (KCL) and voltage (KVL) laws, transformers ($\mathcal{T}$), gyrators ($\mathcal{G}$), and linear resistances ($\mathcal{R}$). The constitutive equations of the (multi-terminal) constraint elements can be expressed in the scattering matrix formalism $\mathsf{S}$, as systems of linear differential equations for charges and fluxes, $R\left[\mathsf{P}_{\alpha b}+\mathsf{S}_{\alpha b}\right]\mathrm{d}q^b=\left[\mathsf{P}_{\alpha b'}-\mathsf{S}_{\alpha b'}\right]\mathrm{d}\phi^{b'}$, with $\mathsf{P}$ a projector onto the intervening branches.}
	\label{fig:IntroFig}
\end{figure}

The constraints $\mathcal{R}$,  $\mathcal{T}$ and
$\mathcal{G}$ that arise from connecting the elements,  together with the Kirchoff constraints,   constitute an
external Pfaff system that, again, is integrable. The restriction
(pullback under the immersion map $\imath:\mathcal{M}\to\mathcal{M}_{2B}$) of
$\omega_{2B}$ to integral manifolds $\mathcal{M}$ of the Pfaff system is a (possibly
degenerate) homogeneous closed two-form, 
\begin{align}
    \omega=\imath^* \omega_{2B}=\frac{1}{2}\omega_{\alpha\beta} \dd z^\alpha\wedge \dd z^\beta,\label{eq:twoform_restricted}
\end{align}
with $z^\alpha$ coordinates in the restricted submanifold $\mcl{M}$.For clarity, for any local system of coordinates such that we express the embedding as $\zeta^\mu(z^\alpha)$ the pullback of 
\begin{align}
    \omega_{2B}=\frac{1}{2}\tilde{\omega}_{\mu\nu}\dd\zeta^\mu\wedge\dd\zeta^\nu
\end{align} is
\begin{align}
    \omega=\frac{1}{2}\tilde{\omega}_{\mu\nu}\frac{\partial\zeta^\mu}{\partial z^\alpha}\frac{\partial\zeta^\nu}{\partial z^\beta}\dd z^\alpha\wedge \dd z^\beta\,.
\end{align}
As we have a linear set of homogeneous constraints, computing the restricted manifold entails finding the kernel of $\msF$, e.g., by Gaussian elimination, as we shall see presently.
The dimension of the final
manifold $\mathcal{M}$, $\mathrm{dim}(\mathcal{M})=B-r_P\leq B$
depends on $r_P$, the relative rank of the resistor, transformer and gyrator
constraint system with respect to the Kirchhoff constraints. The first central result of
this work is that,  in the absence of
dissipation, the equations of motion that would follow from standard circuit theory, i.e. those that follow from the constitutive equations 
when the  constraints are imposed, are the same as the Euler--Lagrange equations of motion for the Lagrangian (using Einstein's convention)
\begin{align}
  \label{eq:explag}
  L= L_{\omega}-H
  -S_\alpha(t){z}^\alpha\,,
\end{align}
where $L_{\omega}=\tfrac{1}{2}\omega_{\alpha\beta}z^\alpha\dot{z}^\beta$, with $z^\alpha$ coordinates in the restricted submanifold $\mathcal{M}$, and $\omega_{\alpha\beta}$ the matrix elements of the restricted two-form in that coordinate system. Analogously, $S_\alpha(t)$ has been constructed by restricting the external driving energy  $H_d(t)$ to $\mathcal{M}$. Generically, the $z^\alpha$ coordinates need not be strictly node fluxes or loop
charges. In fact, in the presence of non-reciprocal elements, with characteristic impedance parameter, they will be a mixture of those. The choice of integral manifold solution to the constraints is practically understood as a choice of integration constants. For our objectives we have to remember at this stage that we also have an energy function, and the restricted energy function will depend on the choice of integral manifold generically. Thus, we would generically have families of models at this stage. However the dynamics does not depend on a homogeneous shift of the energy function, so many of those will be equivalent. Additionally, a crucial criterion to be used is that the critical points of the initial Hamiltonian be included in the integral manifold to be chosen. In more practical terms, we can say that the integration constants that determine an integral manifold will only appear through the energy function $H$, so they can generically be chosen such that the origin of coordinates is an energy minimum.

If there is dissipation, the equations are those derived from this Lagrangian together with the Rayleigh dissipation function determined by the restriction of $g_{2B}$.

\section{Summary of the proof}
\label{sec:textproof}
The proof of the previous result hinges on the Tellegen properties of the constraints. The original Tellegen theorem is that for any network that describes an electrical
circuit, the Kirchhoff constraints ensure that, 
independently of the actual dynamics, $\sum_b i^b v^b=0$. This follows from the graph theory perspective as a homology statement (``a boundary has no boundary"), and algebraically from the relation of the incidence matrix to the constraints. Thus it is independent of constitutive relations, and holds for all dynamics. For our purposes, this theorem is  equivalent to the statement that the restriction of the symmetric 2-covariant tensor 
 $\tau_{2B}=\sum_{b=1}^B \mathrm{d}\phi^b\mathrm{d}q^b$ to an
integral manifold of a Pfaff external system that includes the
Kirchhoff constraints is identically zero, $\imath^*\tau_{2B}=0$. If we
have a restriction that satisfies this last equation we say that it
presents the global Tellegen property. Thus, if the restriction flows from the Kirchhoff constraints and additional ones then it will present the global Tellegen property. It should be borne in mind that any of the Kirchhoff sets together with global Tellegen implies the other Kirchoff set. However, global Tellegen by itself does not imply the Kirchhoff laws. We shall only consider systems of constraints that include the Kirchhoff laws and their solutions will necessarily satisfy the global Tellegen property. In contrast, the local Tellegen
property concerns 2-covariant tensors that involve sums over  a restricted set of branches,
$\tau_{\mathcal{S}}=\sum_{s\in\mathcal{S}}\mathrm{d}\phi^s\mathrm{d}q^s$. The
restriction from $\mathcal{M}_{2B}$ to $\mathcal{M}$ is said to have
the local Tellegen property with respect to the set $\mathcal{S}$ if
$\imath^*\tau_{\mathcal{S}}=0$. Crucially, the
integral manifold for the Pfaff exterior system determined by the
Kirchhoff, linear resistors,  and transformer and gyrator constraints
satisfies both  the global Tellegen property and the local Tellegen
properties with respect to $\mathcal{T}$ and $\mathcal{G}$. Once
integrability and Tellegen have
been established, the proof is algebraic, by identification of the
circuit equations of motion with those derived from the Lagrangian
(and Rayleigh function for dissipation).

Explicitly, with collective branch coordinates $\boldsymbol{\zeta}^T=(\bq^T,\bphi^T)
$, current-voltage source vector $\boldsymbol{s}^T(t)=(\bI^T,\bV^T)(t)$ (with nonzero values only in the positions corresponding to the source branches), and projectors $\mathsf{P}^{\mathcal{S}}$ selecting the branches in a set $\mathcal{S}$, the constitutive equations for capacitors, inductors and sources can be written compactly as 
\begin{align}
  \label{eq:constmatrix}
  \begin{pmatrix}
    0&\mathsf{P}^{\mathcal{C}}+\mathsf{P}^{\mathcal{V}}\\ \mathsf{P}^{\mathcal{L}}+\mathsf{P}^{\mathcal{I}}&0
  \end{pmatrix}\dot{\boldsymbol{\zeta}}=\nabla_{\bzeta} H +
  \begin{pmatrix}
    0& \mathsf{P}^{\mathcal{V}}\\ \mathsf{P}^{\mathcal{I}}&0
  \end{pmatrix}\boldsymbol{s}\,.
\end{align}
On the other hand, the solution of the constraint equations is given by a matrix $\mathsf{K}$, that connects the derivatives of the final variables $\bz$ with those of $\boldsymbol{\zeta}$ in the pair
\begin{align}\label{eq:diffzetas}
\begin{aligned}
\dot{\boldsymbol{\zeta}}&=\mathsf{K}\dot{\bz},\\
\nabla_{\bz} f&=\mathsf{K}^T \nabla_{\bzeta} f.    
\end{aligned}
\end{align}
That is, $\mathsf{K}$ is the matrix that implements the pullback
$\imath^*$ and $\mathsf{K}^T$  the one implementing the pushforward
$\imath_*$. As previously advanced, we compute it by codifying all the linear constraints as
$\mathsf{F}\mathrm{d}\boldsymbol{\zeta}=0$, and computing the kernel
of $\mathsf{F}$. Here, 
\begin{align}
\msF^T=\begin{pmatrix}
\msF_{\text{Kir}}^T&\msF_{T}^T&\msF_{R}^T&\msF_{G}^T
\end{pmatrix},    
\end{align}
where Kirchhoff's constraints have the well-known structure (see for instance \cite{Burkard:2004})
\begin{align}
\msF_\text{Kir}=\begin{pmatrix}
\msF_{\text{cut}}&0\\
0&\msF_{\text{loop}}
\end{pmatrix},    
\end{align}
with $\msF_{\text{cut}}$ ($\msF_{\text{loop}}$) the fundamental cut-set (loop) matrix that annihilates currents (voltages). $\msF_{T}$, $\msF_{R}$ and $\msF_{G}$ describe the linear relations of generic Belevitch transformers, (multi)port networks of resistors and general nonreciprocal elements, respectively. A basis of this kernel is then collected as columns
of the matrix $\mathsf{K}$, i.e. $\mathsf{FK}=0$. See more details for the explicit construction in Appendix~\ref{sec:const_geom_alg}.

Thus, system
(\ref{eq:constmatrix}) after the imposition of constraints becomes
\begin{align}\begin{aligned}
\mathsf{K}^T
  \begin{pmatrix}
    0&\mathsf{P}
   \\ -\mathsf{P}&0
  \end{pmatrix}\mathsf{K}\dot{\bz}=&\,\nabla_{\bz} H + \mathsf{K}^T  \begin{pmatrix}
       0&\mathsf{P}^{\mathcal{V}}\\ \mathsf{P}^{\mathcal{I}}&0
     \end{pmatrix}\boldsymbol{s}\\
     &+\frac{1}{2}\mathsf{K}^T
 \begin{pmatrix}
   0&
   \mathsf{P}^{\mathcal{R}}\\  \mathsf{P}^{\mathcal{R}}&0
 \end{pmatrix}\mathsf{K}\dot{\bz} \,,
 \end{aligned}\label{eq:finaleoms}
\end{align}
where
 $ \mathsf{P}=  \mathsf{P}^{\mathcal{C}}+\mathsf{P}^{\mathcal{V}}+\left(\mathsf{P}^{\mathcal{R}}+\mathsf{P}^{\mathcal{G}}\right)/2$
and the  global (Kirchhoff) and local ($\mathcal{T}$ and
$\mathcal{G}$) Tellegen properties have been used.
These are the equations of motion from Lagrangian (\ref{eq:explag})
with Rayleigh dissipation function 
\begin{align}\label{eq:rayleigh}
\begin{aligned}
    \mathcal{F}&=\frac{1}{2}\sum_{r\in\mathcal{R}}\dot{q}^r(z)\dot{\phi}^r(z)=\frac{1}{4}\mathcal{F}_{\alpha\beta}\dot{z}^\alpha\dot{z}^\beta\\
&=\frac{1}{4}\dot{\bz}^T \mathsf{K}^T
  \begin{pmatrix}
    0&\mathsf{P}^{\mathcal{R}}\\ \mathsf{P}^{\mathcal{R}}&0
  \end{pmatrix}\mathsf{K}\dot{\bz}\,.
\end{aligned}
\end{align}
That is, Eq.~\eqref{eq:finaleoms} is exactly the system of equations
\begin{align}\label{eq:finaleomscoords}
    \omega_{\alpha\beta}\dot{z}^\beta= \frac{\partial H}{\partial z^\alpha}+ S_\alpha +\frac{1}{2}\mathcal{F}_{\alpha\beta}\dot{z}^\beta\,.
\end{align}

For completeness, we signal that the source term in the Lagrangian is 
\begin{align}
  \label{eq:sourceterm}
  S_\alpha z^\alpha = \bz^T\mathsf{K}^T\begin{pmatrix}
       0&\mathsf{P}^{\mathcal{V}}\\ \mathsf{P}^{\mathcal{I}}&0
     \end{pmatrix}\bsb{s}\,,
\end{align}
whence we can read the functions $S_\alpha(t)$. For more details, refer to App.~\ref{sec:algebraic-proof}.

\section{Degenerate forms, the Faddeev-Jackiw method  and quantization} 
\label{sec:FJtextJ}
We focus on the case without dissipation or sources, see App.~\ref{sec:meth-fadd-jack} for further details. Some basic concepts from differential and symplectic geometry are collected in App.~\ref{sec:mathapp}, for reference. The concept of rank and nondegeneracy for differential two-forms is stated in App.~\ref{sec:sympgeomapp} . 

The two-form that appears in Lagrangian (\ref{eq:explag})  is homogeneous by construction, and thus of homogeneous rank. If it is full rank, i.e., nondegenerate, then
it can be inverted and a local Darboux basis can be found by symplectic
Gram--Schmidt \cite{de2006symplectic}. If it is global, and it corresponds to a cotangent bundle,  one can proceed with standard canonical quantization. The generic case, however, is that the two-form is not full rank. For this reason we chose a first-order lagrangian formalism, Eq. (\ref{eq:explag}),  so as to apply the method of Faddeev and Jackiw \cite{Faddeev:1969,Faddeev:1988,Jackiw:1993}.

Thus, once we have obtained  the 
the two-form $\omega$ on a solution of the constraint Pfaff equations, Eq. \eqref{eq:twoform_restricted}, it is our task to compute its kernel, i.e., a linearly independent set of vectors
$W=\left\{\boldsymbol{W}_I\right\}_{I=1}^{|W|}$, with
cardinality $|W|= B-r_P-\mathrm{rank}(\omega)$. In components, each of these vectors satisfies $W_I^\alpha\omega_{\alpha\beta}=0$. Therefore, contracting each of these vectors with both sides of  Eq. \eqref{eq:finaleomscoords}, in the case without dissipation ($\mathcal{F}_{\alpha\beta}=0$) and sources ($S_\alpha=0$), we conclude that if the equations are satisfied then we must have
\begin{align}\label{eq:Wconstrain}
 W_I(z)=   W^\alpha_I \frac{\partial H}{\partial z^\alpha}=0\,.
\end{align}
Turning this result on its head, we read the $|W|$ equations \eqref{eq:Wconstrain} as required for consistent evolution, and thus as dynamical constraints.
\begin{figure}\centering
\includegraphics[width=.85\linewidth]{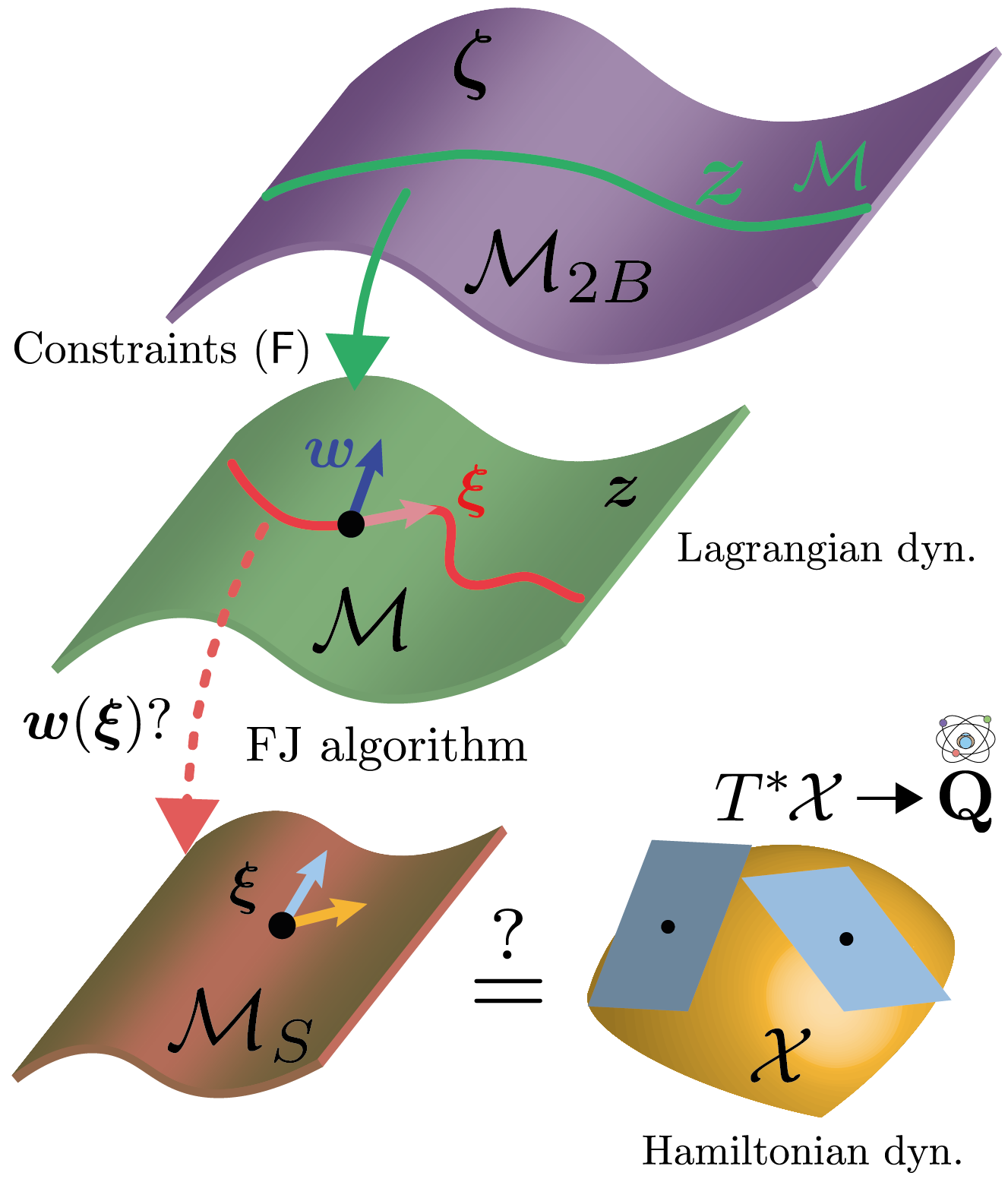}
\caption{Sketch of the geometrical reduction by the Faddeev-Jackiw method. The initial manifold $\mcl{M}_{2B}$ of branch (charge and flux) variables is first reduced to a submanifold $\mcl{M}$ by solving the geometrical linear equations of the constraints encoded in matrix $\msF$ (KCL, etc.). In a second step, we may need to apply the Faddeev-Jackiw algorithm to remove the zero-modes of the Lagrangian (\ref{eq:explag}) (both from the two form and the energy terms), to obtain classical Hamiltonian dynamics in a well-defined symplectic manifold. Finally, to perform canonical quantization, we further require that the symplectic manifold be diffeomorphic to a cotangent bundle.}	\label{fig:Geometrical_reduction}
\end{figure}
 
Some of these can be \emph{gauge} constraints (i.e., identically satisfied) revealing redundancies in the description. If only Kirchhoff and transformer relations have been used to reach $\omega$, the gauge vectors are purely in charge or purely in flux directions, this not being the case for a dimensionful constraint. A charge (flux) gauge constraint signals an effective loop (node cutset\footnote{A cutset in a graph is a set of edges whose removal increases the number of connected components. A node cutset is the set of all edges connected to a node, whose removal increases the number of connected components in exactly one.}) in which only inductive (capacitive) elements take part. 

In this first round of the method as applied to electrical circuits the zero-mode vectors $\boldsymbol{W}_I$ commute, as $\omega$ is homogeneous and therefore so are $\boldsymbol{W}_I$. We say that this set of vectors is integrable because there exist submanifolds invariant under the action of these vectors. In other words, we can use coordinatizations of the manifold structured as $z\to\left\{w^I\right\}\cup\left\{\xi^\mu\right\}$, with $w^I$ those associated with the zero-mode vectors.   

If the functions $\left\{W_I(z)\right\}_{I=1}^{|W}$ are infinitely differentiable and the Jacobian matrix $W_{IJ}(z)=\partial^2_{w^I w^J}H$  is everywhere invertible there exists a submanifold of $\mcl{M}$ satisfying the constraints. We now pull back the two-form to this reduced submanifold and repeat the analysis. If the new two-form has homogeneous full rank, then we have completed this process. Lastly, if homogeneous but not full rank, we compute its zero modes and repeat. The new vectors need no longer be homogeneous, and integrability is not a given. If the new two form is not homogeneous in rank we have a different type of obstruction.  
 The process is repeated until either an obstruction is found or homogeneous full rank has been reached. In the following sections, we refer as $\mcl{M}_S\subseteq \mcl{M}$ to the final symplectic submanifold with a homogeneous full rank two form. Finally, to proceed with canonical quantization, we require that the symplectic submanifold is diffeomorphic to a cotangent bundle for some base manifold $\mcl{X}\subseteq \mcl{M}_S$, i.e.,  $\mcl{M}_S=T^*\mcl{X}$, see Fig.~\ref{fig:Geometrical_reduction}. It is to be noted that this final requirement directly discards (even dimensional) symplectic manifolds of the type $\mcl{M}_S=\mathbbm{R}^n\times \mathbbm{T}^m$ with $n<m$. E.g., the 2-dimensional torus ($\mathbbm{T}^2$) is not a cotangent bundle and does not allow standard canonical quantization\footnote{An alternative approach to canonical quantization for symplectic manifolds that are not cotangent bundles, 
known as \emph{geometric quantization}~\cite{Kostant:1970,Souriau:1997}, has been proposed. In this article we do not deal with this approach.}. Particular methods can exist for specific obstructions~\cite{Miano:2023}.

Summarizing, there are three types of obstruction, with different impact on quantization. First, lack of global Darboux coordinates or nontrivial topology of phase space, as in the $0-\pi$ qubit~\cite{Egusquiza0pi:2022}, giving rise to inequivalent quantizations~\cite{galindo:2012,Reed:1975}. Second, nonhomogeneity in rank of $\omega$. In examples, as previously noted in~\cite{Rymarz:2023} with the dual Dirac--Bergmann analysis, this is due to a bifurcation in the underlying classical system, that admits a local Hamiltonian description, such that quantization might be possible in some cases. Third, nonintegrability of dynamical constraints, such that there is no classical Hamiltonian description.

\section{Quantization of superconducting circuits: a pre-canonical manifold for the branch variables}\label{sec:quantization_superc_circuits}
In the previous section, we have described the general method for performing canonical quantization of the classical dynamics of electric circuits and advanced some of the issues that can be encountered in this process, i.e., obstructions to the construction of consistent Hamiltonian dynamics in a reduced submanifold. 
Let us emphasize again that our general procedure works for any given topology of the pre-canonical manifold $\mathcal{M}_{2B}$, i.e., the method is agnostic to that choice (as long as it is in the family of $\mathbb{R}^{2B-k}\times \mathbbm{T}^{k}$). However, in its practical application, the user must make such a choice, and consequently, different classical symplectic submanifolds (possibly cotangent bundles) and canonical quantizations (when applicable) thereof may follow.

As previously advanced, the main motivation for obtaining canonically quantized descriptions is the field of circuit quantum electrodynamics~\cite{Blais:2021}, and thus, we will now focus on lumped-element descriptions of superconducting circuits, where the fundamental unit of charge (flux) is $2e$, i.e., that of two electrons,  ($\Phi_Q=h/2e$, i.e., the superconducting flux quantum). In particular, we  propose a specific choice for the topology of the branch variables for the passive energetic elements that matches the typical phenomenological assumptions on the topology of the node fluxes and loop charges, e.g., that a quantum operator representing the phase variable of a superconducting island has a compact continuous spectrum $\sigma(\hat{\varphi}_{\text{island}})=S^1$, whereas its conjugate operator (the excess of Cooper pairs in the island) has an integer spectrum $\sigma(\hat{n}_{\text{island}})=\mathbbm{Z}$. Interestingly, the topology of the final phase-space  has been the core of a very long debate on the quantization of circuits with Josephson or phase-slip junctions with or without superconducting islands and closed loops~\cite{Apenko:1989,Zaikin:1990,Schoen:1990,Likharev:1985,Nakamura:1999,Koch:2007,Koch:2009,Devoret:2021,Sonin:2022,Mooij:2006,ThanhLe:2019,ThanhLe:2020,Koliofoti:2023}. In fact, it has been suggested that the \emph{compactification} of the phase variable is nothing but an effective viewpoint, given the apparent indistinguishability of the a.c. response~\cite{Koch:2009,Devoret:2021} from the alternative assumption where both opeartors have a real spectrum $\sigma(\hat{\varphi}_{\text{island}})=\sigma(\hat{n}_{\text{island}})=\mathbbm{R}$\footnote{To solve the apparent ambiguity, an extremely challenging experiment has been  proposed~\cite{ThanhLe:2020} to distinguish between the two perspectives. In essence, this requires disrupting a superconducting wire to form two superconducting islands.}. We will now show how one should proceed if variables representing phase differences (loop charges) between superconducting islands (superconducting loops) are to be considered compact, which is mathematically more involved than if all of the initial variables are considered to be in $\mathbbm{R}^{2B}$. This is because if one assumes that $\mcl{M}_{2B}=\mathbbm{R}^{2B}$, the first reduction of the linear constraints provided by the matrix $\msF$ is, trivially, a submanifold $\mathbbm{R}^{2B-n}$ with $n\geq B$. 

\begin{figure}\centering
\includegraphics[width=.75\linewidth]{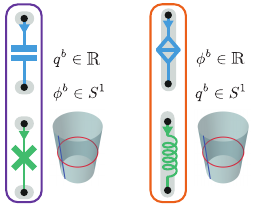}
\caption{The conjectured topology of the manifold of branch variables for the passive elements in Fig.~\ref{fig:IntroFig}.  Capacitors and pure Josephson junction nonlinearities (two-island elements) would have charge (flux) variables in the real line (circle). Reversely, phase slips and linear inductors (one-island elements) would have flux (charge) variables in the real line (circle). This is the only possible assignment such that the reduced manifold $\mcl{M}$ (node fluxes and loop charges for circuits with only KCL and KVL constraints) has the typically assumed topology~\cite{Ulrich:2016,Devoret:2021}.}	\label{fig:Conjectured_M2B}
\end{figure}

Let us first consider circuits with only four kinds of elements, i.e., linear inductors and capacitors,  nonlinear Josephson junctions~\cite{Josephson:1962} and phase-slip junctions~\cite{Mooij:2006,Astafiev:2012}, see Fig.~\ref{fig:Conjectured_M2B}. 
The four elements have been previously classified (colours) depending on how they enter in the pre-canonical two-form (\ref{eq:twoform2b}) and in the local energy terms (\ref{eq:Henergy_C_L}). In particular, the inductor, the Josephson junction, the capacitor, and the phase-slip junction are described by the energy functions
\begin{align}
\begin{aligned}
h_L(\phi_L)&=\frac{(\phi_L)^2}{2L},\, h_J(\phi_J)=-E_J\cos\left(\frac{2\pi\phi_J}{\Phi_Q}\right),\\
h_C(q_C)&=\frac{(q_C)^2}{2C},\, h_P(q_P)=-E_P\cos\left(\frac{\pi q_P}{e}\right).
\end{aligned}
\end{align}

Now, we axiomatically enrich the characterization of generic nonlinear inductors and capacitors using a fixed initial topology for the branch variables of the four mentioned elements. In particular, we set capacitors and Josephson junctions to have branch variables living in $q^b=\{q_C, q_J\}\in \mathbbm{R}$ and $\phi^b=\{\phi_C, \phi_J\}\in S^1$, and dually for inductors and phase-slip junctions, i.e., $q^b=\{q_L, q_P\}\in S^1$ and $\phi^b=\{\phi_L, \phi_P\}\in \mathbbm{R}$, see Fig.~\ref{fig:Conjectured_M2B}. This choice of the topology of the pre-canonical manifold is fundamentally inspired by arguments presented in Refs.~\cite{Apenko:1989,Ulrich:2016} for quantized descriptions of the capacitively- and inductively-shunted Josephson junction. Now, we will explore how, in combination with our geometrical reduction method, it enables us to regain the typical assumptions on the topology of the reduced manifold $\mathcal{M}$ (node fluxes and loop charges), and, when applicable and necessary, of $\mathcal{M}_S$.

\begin{figure*}\centering
\includegraphics[width=.75\linewidth]{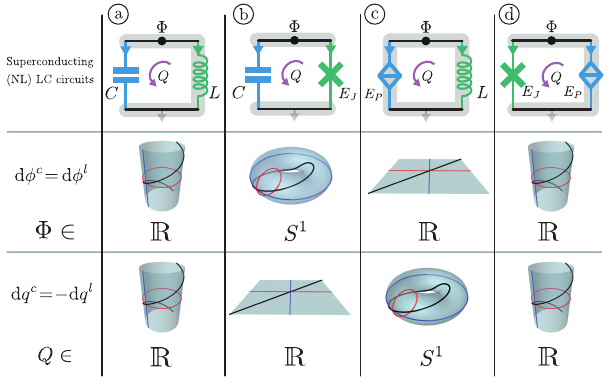}
\caption{Given the prescription provided in previous Fig.~\ref{fig:Conjectured_M2B} for the manifold of branch variables $\mcl{M}_{2B}$, the topology of the reduced submanifold $\mcl{M}(\equiv\mcl{M}_S)=\{\Phi,Q\}$ is a consequence of the geometrical equations for the constraints. For the  linear LC oscillator (a), and purely nonlinear Josephson/phase-slip circuit (d) both variables are \emph{decompactified} and we have submanifolds $\mcl{M}\in\mathbbm{R}^2$. For the capacitively-shunted Josephson junction (b) and the inductively-shunted phase-slip (c) one variable remains compact, i.e., the reduced submanifolds are $\mcl{M}\in S^1\times\mathbbm{R}$, giving later rise to quantum operators of charge $\hat{Q}$ (b) and flux $\hat{\Phi}$ (c) with discrete spectrum.}	\label{fig:NL_LC_oscillators}
\end{figure*}

Consider the linear LC oscillator in Fig.~\ref{fig:NL_LC_oscillators}(a). The starting point is a 4 dimensional space $\mcl{M}_{2B}=\mathbbm{R}^{2}\times \mathbbm{T}^2$, with $\{\phi_C,q_L\}\in S^1$ and $\{q_C,\phi_L\}\in\mathbbm{R}$. Following our proposed method, the pre-canonical two-form (\ref{eq:twoform2b}) reads
\begin{align}
    \omega_{2B}=\frac{1}{2}\left(\mathrm{d}q_C\wedge\mathrm{d}\phi_C+\mathrm{d}\phi_L\wedge\mathrm{d}q_L\right)
\end{align}
The circuit has only Kirchhoff constraints, written in terms of the matrix $\msF_{\text{Kir}}$, i.e., $\msF_{\text{Kir}}\dd{\bzeta}=0$, whose kernel $\msK$ provides us with a basis of (loop) currents and (node) voltages $\dd {\bz}^T=(\dd{Q},\dd \Phi)$ of the circuit. Explicitly, \begin{subequations}\label{eq:Pfaff_LC}
\begin{align}
\dd{\phi_C}-\dd{\phi_L}=&0,\label{eq:Pfaff_NL_LC_phi}\\
    \dd{q_C}+\dd{q_L}=&0.\label{eq:Pfaff_NL_LC_q}
\end{align}
\end{subequations}
The integration of these Pfaff system of equations provides us with a basis of (loop) charges and (node) fluxes
\begin{subequations}\label{eq:Pfaff_LC_solution}
\begin{align}
    q_C&=Q,\qquad q_L=Q\mod{2e},\\
    \phi_L&=\Phi,\qquad \phi_C=\Phi\mod{\Phi_Q},
\end{align}
\end{subequations}
where both variables are now extended $\{Q,\Phi\}\in\mathbbm{R}$, and integration constants have been set to zero, see Fig.~\ref{fig:NL_LC_oscillators}(a). Interestingly, the topology of  $\mcl{M}$ would have also been $\mathbbm{R}^2$ if one had imposed $\mcl{M}_{2B}=\mathbbm{R}^3\times S^1$, or $\mcl{M}_{2B}=\mathbbm{R}^4$.

Performing the immersion map, we obtain a canonical two-form 
\begin{align}
    \omega=\imath^*\omega_{2B}=\frac{1}{2}\left(\mathrm{d}Q\wedge\mathrm{d}\Phi-\mathrm{d}\Phi\wedge\mathrm{d}Q\right)\label{eq:omega_LC}
\end{align}
and a final Lagrangian expression (\ref{eq:explag}) that consists of the canonical first-order term
  \begin{align}
    L_{\omega}&=\frac{1}{2}\left(Q \dot{\Phi}-\Phi \dot{Q}\right),\label{eq:Lomega_LC}
\end{align}  
and the energy function (\ref{eq:Henergy_C_L}), now the Hamiltonian, 
\begin{align}
     \begin{aligned}
        H&=h_C(q_C(Q))+h_L(\phi_L(\Phi))=\frac{Q^2}{2C}+\frac{\Phi^2}{2L}.
    \end{aligned}
\end{align}
The conjugate variables have the canonical Poisson bracket $\{\Phi,Q\}=1$, read from Eq.~(\ref{eq:Lomega_LC}), and quantization follows in the standard case by promoting classical variables to operators and the Poisson bracket to the commutator $[\hat{\Phi},\hat{Q}]=i\hbar$.

Having understood this case, it is immediate to obtain the results for the other examples in Fig.~\ref{fig:NL_LC_oscillators}. For instance, the geometrical equations for the second circuit (b) with a pure Josephson junction and a capacitor, read 
\begin{subequations}\label{eq:Pfaff_CJJ}
\begin{align}
\dd{\phi_C}-\dd{\phi_J}=&0,\label{eq:Pfaff_CJJ_phi}\\
    \dd{q_C}+\dd{q_J}=&0.\label{eq:Pfaff_CJJ_q}
\end{align}
\end{subequations}
The initial pre-canonical manifold is again $\mcl{M}_{2B}=\mathbbm{R}^{2}\times \mathbbm{T}^2$ but now $\{\phi_C,\phi_J\}\in S^1$ and $\{q_C,q_J\}\in\mathbbm{R}$. Integration of the equations provides a submanifold parameterized by a flux and a charge variables, such that
\begin{subequations}\label{eq:Pfaff_CJJ_solution}
\begin{align}
    q_C&=q_J=Q\in\mathbbm{R},\\
    \phi_C&=\phi_J=\Phi\in S^1,
\end{align}
\end{subequations}
where now $Q\in\mathbbm{R}$ and $\Phi\in S^1$, i.e., the classical phase space is an infinite cylinder. Notice that again, integration constants have been, for now, set to zero. The rest of the procedure is completely analogous, and one obtains the classical CPB Hamiltonian
\begin{align}\label{eq:H_JJ_classical}
\begin{aligned}
H&=h_C(q_C(Q))+h_J(\phi_J(\Phi))\\
    &=\frac{Q^2}{2C}-E_J\cos(2\pi \Phi/\Phi_Q).
\end{aligned}
\end{align}
As it is well known, the canonical quantization of the classical dynamics of a massive particle in the circle leads to inequivalent Hamiltonians~\cite{Egusquiza0pi:2022} (with inequivalent spectra). In this case, a c-number shift $Q_g$ of the charge (a different integration constant, i.e., $q_C=Q-Q_g$) leads to equivalent classical dynamics but can be understood as a marker of inequivalent quantum dynamics~\cite{Koch:2007}. Adimensionalizing the charge and flux variables as $n=Q/(2e)$ and $\varphi=2\pi\Phi/\Phi_Q$, and promoting the conjugate classical variables to quantum operators with charge number and phase commutation relations, i.e., $[\hat{n},e^{i\hat{\varphi}}]=ie^{i\hat{\varphi}}$, we obtain the final family of quantum  Hamiltonians
\begin{align}
    H=4E_C(\hat{n}-n_g)^2-E_J\cos(\hat{\varphi}),\label{eq:H_JJ_quantum}
\end{align}
with $E_C=e^2/2C$, parameterized by the shift $n_g=Q_g/2e$.

Applying the same procedure to the circuit in Figs.~\ref{fig:NL_LC_oscillators}(c), we obtain the reduced phase space $\mcl{M}=\{\Phi,Q\}=\mathbbm{R}\times S^1$. This case  (c) is the total dual to (b), i.e., the  quantized Hamiltonian 
\begin{align}
    H=4E_L(\hat{\varphi}-\varphi_g)^2-E_P\cos(2\pi\hat{n}),\label{eq:H_PS_quantum}
\end{align}
with the canonical commutator $[\hat{\varphi},e^{i\hat{n}}]=e^{i\hat{n}}$, where the operator with integer spectrum is now $\hat{\varphi}$, counting the number of flux quanta in the superconducting loop. 

On the other hand, the geometric reduction for the circuit in Fig.~\ref{fig:NL_LC_oscillators}(d), also known as the \emph{dualmon} \cite{ThanhLe:2019}, results again in a \emph{decompactification} of both charge and flux variables ($\mcl{M}=\mathbbm{R}^2$), like in the LC oscillator case. Indeed, this matches the underlying microscopic physics that there is no superconducting loop, nor two superconducting islands. Reproducing the same steps of the method, we obtain the quantized Hamiltonian
\begin{align}
    H=-E_P\cos(2\pi\hat{n})-E_J\cos(\hat{\varphi}),\label{eq:H_dualmon_quantum}
\end{align}
where now both $\hat{n}$ and $\hat{\varphi}$ are self-adjoint operators with real spectrum, and commutator $[\hat{n},\hat{\varphi}]=i$.
\begin{figure}
\centering
\includegraphics[width=.85\linewidth]{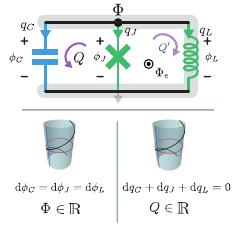}
\caption{Following the prescription of Fig.~\ref{fig:Conjectured_M2B}, the solution of the geometrical equations for the inductively shunted Josephson junction, provide a final reduced (cotangent space) submanifold $\mcl{M}_S=\{Q,\Phi\}\in \mathbbm{R}^2$. That is, the inductor \emph{decompactifies} the node flux, see the difference with the two-island circuit in Fig.~\ref{fig:NL_LC_oscillators}(b), the Cooper-Pair Box~\cite{Buettiker:1987}. Notice that in a first reduction $\mcl{M}=\{\Phi,Q,Q'\}$ has dimension three, where $Q'$ is a gauge loop charge that can be arbitrarily fixed.}	\label{fig:Fluxonium}
\end{figure}

Let us now consider the three-branch example in Fig.~\ref{fig:Fluxonium}, the archetypal circuit for building fluxonium~\cite{Manucharyan:2009} and blochnium qubits~\cite{Pechenezhskiy:2020}. The Kirchhoff constraints now read
\begin{subequations}
\begin{align}
    \dd{q_C}&=-\dd{q_J}-\dd{q_L},\\
    \dd{\phi_C}&=\dd{\phi_J}=\dd{\phi_L},
\end{align}
\end{subequations}
thus, their integration provides us with 
\begin{subequations}\label{eq:Pfaff_Fluxonium_solution}
\begin{align}
    q_C&=Q,\qquad q_J=-(Q+Q'),\qquad q_L=Q',\\
    \phi_L&=\Phi+\Phi_e,\qquad \phi_C=\phi_J=\Phi\mod{\Phi_Q},
\end{align}
\end{subequations}
where $Q\in \mathbbm{R}$, $Q'\in \mathbbm{R}$,
and $\Phi\in \mathbbm{R}$, and we have made explicit an integration constant $\Phi_e$ (the constant external flux threading the superconducting loop), which in this case plays a role in both the classical and quantum dynamics. Following the method, the pullback of the two form reads once more
\begin{align}
    \omega=\dd{Q}\wedge \dd{\Phi},
\end{align}
i.e., $Q$ and $\Phi$ are canonical conjugate variables once more, and the final Hamiltonian has the form
\begin{align}
    H=\frac{Q^2}{2C} +\frac{1}{L}(\Phi+\Phi_e)^2-E_J \cos(2\pi \Phi/\Phi_Q).\label{eq:H_fluxonium_classical}
\end{align}

Observe that here $Q'$ is a gauge charge that can be arbitrarily fixed to complete the reduction. The canonical quantization of the dynamics of (\ref{eq:H_fluxonium_classical}) is analogous to that of the LC oscillator, i.e., the operators $\hat{\Phi}$ and $\hat{Q}$ have real spectrum. It is worth mentioning that if the external flux in the superconducting quasi-loop is time-dependent, the method requires breaking the loop and adding an external voltage source to account for all possible gauges~\cite{You:2019}. See below a general theory for time dependent magnetic fields in App.~\ref{sec:time-depend-magn}, and examples in App.~\ref{sec:time-depend-extern} and \ref{sec:miano}.

\begin{figure}
\centering
\includegraphics[width=.67\linewidth]{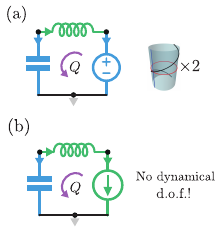}
\caption{Geometrical reduction with voltage (a) and current (b) sources. Voltage and current sources either do not play a role in the geometrical reduction, as in (a), or they trivialize the dynamics, as in (b). Naturally, dual circuits (exchanging series by parallel connections and dual elements) would have dual properties.
.}	\label{fig:Geometrical_reduction_V_I}
\end{figure}

Now that we have discussed the main four passive elements used in the modeling of superconducting circuits, let us briefly comment on the geometrical reduction in the presence of voltage and current sources, analyzing the simple examples in Figs.~\ref{fig:Geometrical_reduction_V_I}(c) and (d), concerning the role played by external classical forces in the geometrical reduction. In doing so, we assume that the voltage (current) source branches have the same topology as the capacitors (inductors)\footnote{In circuit theory, it is standard practice to consider a voltage source as an infinite capacitor, whose voltage remains independent of the current through it, i.e.,  
$
\lim_{{q^v, C_v \to \infty}} \frac{q^v}{C_v} = V_s(t)$. Similarly, this applies to current sources and inductors.
}. In the case with the voltage source (c), the series connection summarizes in the equations
\begin{subequations}\label{eq:CLV_example_Kirchhoffs}
\begin{align}
    \dd{\phi}_C&=\dd{\phi}_L+\dd{\phi}_V,\\
    \dd{q}_C&=-\dd{q}_L=-\dd{q}_V.
\end{align}
\end{subequations}
The geometrical solutions are
\begin{subequations}
\begin{align}
    q_C=Q,\quad\, q_L=-Q\,\,\mathrm{ mod }\, (2e), \quad\, q_V=-Q,\\
    \phi_C=\Phi',\quad \phi_L=\Phi,\quad \phi_V=(\Phi'-\Phi)\,\,\mathrm{ mod }\, (\Phi_Q),
\end{align}
\end{subequations}
with $\{Q,\Phi\}\in \mathbbm{R}$, and $\Phi'\in S^1$, such that the two form is written in terms of the extended variables, i.e., $\omega=\dd{Q}\wedge\dd{\Phi}$, and the total time-dependent Hamiltonian reads
\begin{align}
    H_T=\frac{Q^2}{2C}+\frac{\Phi^2}{2L}-Q V_s(t).
\end{align}

Now, let us replace the voltage source by a current source, as in Fig.~\ref{fig:Geometrical_reduction_V_I}(d), and derive the known result that this system does not have dynamical degrees of freedom. The constraint equations are the same as in Eqs.~(\ref{eq:CLV_example_Kirchhoffs}), replacing the subindex $V\rightarrow I$, but now,  their solutions, given the assumed topology (the current source branch variables sharing the same as for the induector),

\begin{subequations}
\begin{align}
    q_C=Q,\quad\, q_L=q_I=-Q\,\,\mathrm{ mod }\, (2e),\\
    \phi_C=\Phi+\Phi',\quad \phi_L=\Phi,\quad \phi_I=\Phi',
\end{align}
\end{subequations}
with $\{Q,\Phi,\Phi'\}\in\mathbbm{R}$. The two form, written in terms of this parameterization of the submanifold is  $\omega=\dd{Q}\wedge(\dd{\Phi}+\dd{\Phi'})$, which is not in canonical form. The energy term reads 
\begin{align}
    H_T=\frac{Q^2}{2C}+\frac{\Phi^2}{2L}+\Phi' I_s(t).
\end{align}
Making a coordinate transformation $\bar{\Phi}=\Phi+\Phi'$, we obtain the canonical two form $\dd{Q}\wedge\dd{\bar{\Phi}}$, whereas the energy changes to
\begin{align}
    H_T=\frac{Q^2}{2C}+\frac{\Phi^2}{2L}+(\bar{\Phi}-\Phi) I_s(t).\label{eq:Henergy_CLIs}
\end{align}
Observe however, that the flux $\Phi$ is a zero mode of the two form, and thus, we for consistency with the equations of motion, the equation ($\partial_\Phi H=\Phi/L-I_s(t)=0$). Reintroducing its solution in Eq.~(\ref{eq:Henergy_CLIs}) we finally obtain the time-dependent Hamiltonian 
\begin{align}
    H_T=\frac{Q^2}{2C}+\bar{\Phi} I_s(t),
\end{align}
which has trivial classical dynamics.

With this, we conclude our discussion on the conjectured intrinsic topology for the branch variables of the main lumped elements used in building superconducting networks. We note that further work will be required to extend the topological analysis in the reduction method for circuits more generally, including the constraints of ideal transformers, nonreciprocal elements, and resistors, under the assumption that compact variables exist, and with the proposed assignments in Fig.~\ref{fig:Conjectured_M2B}.

\section{Other circuit examples}
\label{sec:circuit_examples}
Let us further illustrate our systematic method to obtain
Hamiltonians by applying it to two circuit examples
that would be singular, or defective, within the pure node-flux~\cite{Devoret:1997} or loop-charge methods~\cite{Ulrich:2016}, see Fig.~\ref{fig:Circuit_examples}. Further archetypal examples with voltage sources (modelling time-dependent external magnetic fluxes~\cite{You:2019}) and resistances, as well as nonlinear singular circuits~\cite{Rymarz:2023,Miano:2023} are treated in App.~\ref{sec_app:examples}.

\subsection{Nonlinear star circuit}
\begin{figure}[h]\centering
	\includegraphics[width=1.05\linewidth]{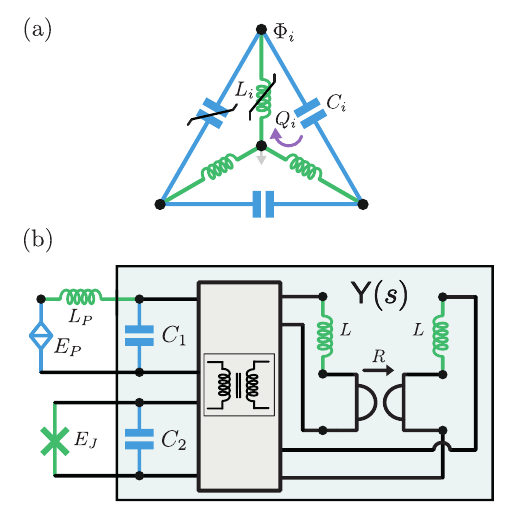}
\caption{(a) The star circuit: a prototypical singular one. Conservation of current (inner node) and voltage (outer loop) may lead to reduction of variables. Elements are named clockwise with indices $i\in\{1,2,3\}$ starting from the ones with label. (b) Josephson and phase-slip junctions coupled to a 2-port nonreciprocal linear system described with an admittance matrix $\msY(s)$, which is canonically-decomposed in a network of inductors, capacitors, gyrator and transformers.}	\label{fig:Circuit_examples}
\end{figure}

Consider the circuit in Fig.~\ref{fig:Circuit_examples}(a) with 3 nodes in a circle connected by capacitors forming a closed loop, and each of them connected to a common inner fourth node through inductors. We first apply our method to the linearized version of this circuit. Using as configuration space coordinates only node-flux variables $\Phi_i$ (loop-charge variables $Q_i$), the kinetic matrix $\tilde{\msC}$ ($\tilde{\msL}$) is singular, the Lagrangian is defective~\cite{Devoret:1997,Ulrich:2016}, and no Hamiltonian can be obtained by Legendre transformation. 

Following our method, we find a complete basis for voltages and currents (one-forms $\dd{\Phi_i}$ and $\dd{Q_i}$) solving the kernel of $\msF_{\text{Kir}}$. Under the assumed topology for the branch variables for inductors and capacitors introduced in the previous section (see Fig.~\ref{fig:Conjectured_M2B}), we can integrate the geometrical equations in $\mathbbm{R}^6\times \mathbbm{T}^6$ to obtain the submanifold $\mathbbm{R}^6$ coordinated by the basis of node-fluxes $\Phi_i$ and loop-charges $Q_i$. Independent of the topology of the submanifold, the pullback of the pre-canonical two form (\ref{eq:twoform2b}), $\omega=\imath^* \omega_{2B}$, becomes
\begin{align}\label{eq:twoform_starcircuit}
    \omega =&\,\dd Q_1\wedge\left(\dd\Phi_1-\dd\Phi_2\right)+\dd Q_2\wedge\left(\dd \Phi_2-\dd \Phi_3\right)\nonumber\\
    &+\dd Q_3\wedge\left(\dd \Phi_3-\dd \Phi_1\right).
\end{align}
that has zero vectors,
\begin{subequations}
\label{eq:Wl_Wc_star_circuit}
    \begin{align}
        \mbf{W}_l&= \frac{\partial}{\partial \Phi_1}+\frac{\partial}{\partial \Phi_2}+\frac{\partial}{\partial \Phi_3}\,,\\
        \mbf{W}_c&= \frac{\partial}{\partial Q_1}+\frac{\partial}{\partial Q_2}+\frac{\partial}{\partial Q_3},
    \end{align}
\end{subequations}
associated with  coordinates $w_c$ and $w_l$. Under the pertinent change of variables (see App.~\ref{sec:star_circuit}) we obtain $\omega= \dd Q_a\wedge\dd \Phi_a+\dd Q_b\wedge\dd \Phi_b$, i.e., $\{\Phi_\alpha, Q_\beta\}=\delta_{\alpha\beta}$. If all the elements are linear, the energy function in this Darboux basis $\{\xi^\mu\}\cup\{w^I\}\equiv(Q_a,Q_b,\Phi_a,\Phi_b,w_c,w_l)$ is written as

\begin{align}
\begin{aligned}
    H=&\frac{{\left(Q_a + w_c	\right)}^{2}}{2 C_{1}} + \frac{w_c^{2}}{2 C_{2}} + \frac{{\left(Q_b + w_c\right)}^{2}}{2 C_{3}}\\
	&+ \frac{{\left(w_l+ \Phi_{b}\right)}^{2}}{2 L_{1}} + \frac{{\left(w_l + \Phi_a + \Phi_b\right)}^{2}}{2 L_{2}} + \frac{w_l^{2}}{2 L_{3}}.
\end{aligned}
\end{align}
Under elimination of the constraints (solving the linear equations $\mbf{W}_l(H)=\mbf{W}_c(H)=0$) we derive the final Hamiltonian
\begin{align}
	H&=\frac{1}{2}\left(\bQ^T\msC^{-1}\bQ+\bPhi^T\msL^{-1}\bPhi\right),
\end{align}
with  $\msC^{-1}$ and $\msL^{-1}$ now full-rank $2\times2$ matrices, see App.~\ref{sec:star_circuit} for all details of the computation. Naturally, this linear circuit is amenable to node-flux (loop-charge) quantization after making use of the star-mesh equivalence~\cite{Kennelly:1899}, followed by a subsequent reduction of one of the variables. It is worth emphasizing that this and all other linear reduction equivalences are inherently integrated into the method.

Next, let the third capacitor and the first inductor be nonlinear, with energy functions $h_c(q_{c3})$ and $h_l(\phi_{l1})$ respectively, with $q_{cj}$ the $j$-th capacitor's branch charge and $\phi_{lj}$ the $j$-th inductor's branch flux. Let us describe the dynamics, for now, in terms of the same rotated basis of fluxes and charges $\{\xi^\mu\}\cup\{w^I\}\equiv(Q_a,Q_b,\Phi_a,\Phi_b,w_c,w_l)$ that allow us to write the canonical two-form, without assigning a particular topology to those variables. In contrast with the previous linear case, there are now two nonlinear zero mode equations $\mbf{W}_c(H)=\frac{Q_a+w_c}{C_1}+\frac{w_c}{C_2} +h_c'(Q_b+w_c)=0$, and $\mbf{W}_l(H)=\frac{w_l+\Phi_a+\Phi_b}{L_2}+\frac{w_l}{L_3}-h_l'(-(w_l+\Phi_b))=0$, that implement charge and energy conservation for consistent reduction of the dynamics to two degrees of freedom. Under the conditions $h_c''(x)+\frac{1}{C_1}+\frac{1}{C_2}>0$ and $ h_l''(-y)+\frac{1}{L_2}+\frac{1}{L_3}>0$,  explicit solutions exist for the variables $w_c$ and $w_l$ in terms of these equations, resulting in a simplified Hamiltonian and canonical two-form, for two degrees of freedom, and canonical quantization follows. If not satisfied, this belongs to a more general class of singularities which are under active research~\cite{Rymarz:2023,Miano:2023}, see App.~\ref{sec:nonl-sing-circ}. The analysis of this nonlinear case has been carried out assuming the same topology for the integrated submanifold. Indeed, it is easy to observe that if the nonlinear capacitor and inductor are precisely a phase slip and a Josephson junctions, respectively, the integration of the Pfaff equations with the assumed topology of the branch variables in Fig.~\ref{fig:Conjectured_M2B} results, again, in a sub-manifold $\mathbbm{R}^6$ (coordinated again by the node fluxes and loop charges $\{\Phi_i, Q_i\}$).
\subsection{Nonreciprocal blackbox-admittance quantization}
As a second example, our method enables a natural extension of the highly effective black-box quantization approach~\cite{Nigg:2012, Solgun:2014, Solgun:2015} to nonreciprocal environments with multiple ports and poles. We remark that here we are not considering any  topological assignment for the initial manifold $\mcl{M}_{2B}$, and we are just considering local solutions. Consider the circuit in
Fig.~\ref{fig:Circuit_examples}(b), a canonical nonreciprocal 2-port linear system  with  admittance matrix $\msY(s)=\msD_\infty s + (\msD_1 s + \msE_1)/(s^2+\Omega_1^2)$~\cite{Newcomb:1966} coupled to a Josephson junction and a phase-slip junction at the ports. Following our systematic method, we find that the two-form ($\omega=\imath^* \omega_{2B}$) is degenerate, with a single gauge zero mode. After its systematic elimination, we reach Hamiltonian
\begin{align}
\begin{aligned}
    H=&\frac{1}{2}\left(\bQ^T\msC^{-1}\bQ+\bPhi^T\msL^{-1}\bPhi+\bQ^T\msG\bPhi\right)\\&+h_P(Q_{1}) + h_J(\Phi_2),
\end{aligned}
\end{align}
with three pairs of conjugate variables $\{\bPhi, \bQ^T\}=\mone$, ready for canonical quantization. The computation is set out explicitly in App.~\ref{sec:Y-bb-circuit}. Note that this Hamiltonian can be obtained with more effort by following the algebraic approach proposed in~\cite{Egusquiza:2022} with an adhoc method to eliminate
transformer constraints followed by a Williamson's reduction~\cite{Williamson:1936} of the nonreciprocal zero mode.

\section{Conclusions \& Outlook} In this article, we have developed a new and systematic method, fully programmable, to write the dynamics of general lumped-element electrical circuits, i.e., Kirchhoff's and constitutive equations, as derived from a Lagrangian in first order and a Rayleigh dissipation function, following a consistent fully geometrical description. Using the Faddeev-Jackiw method, we identify and classify singularities. We provide a Hamiltonian description for all non-singular and for important classes of singular circuits. Our solution is based on the correct identification of the circuit-state degrees of freedom (mix of flux and charge variables), including both compact and extended ones. We have illustrated the method with pedagogical  circuits with
nonreciprocal and nonlinear elements, emphasizing those not
susceptible to treatment or singular in  a pure node-flux/loop-charge
configuration space approach. Furthermore, we have put forward a proposal for the assignment of topological structure to the branch manifolds of the basic energetic elements. This proposal is inspired by phenomenological and microscopic arguments and is consistent with all experiments we are aware of.

Further work will be required to add nonlinear dissipation to the classical dynamics of circuits within this scheme~\cite{Chua:1974}. Additionally, it will be necessary to extend the concrete set of assumptions provided in Sec.~\ref{sec:quantization_superc_circuits} for the geometrical reduction of multiport superconducting quantum circuits that are partially described by ideal transformer and nonreciprocal constraints. Finally, an extension to quasi-lumped element
circuits~\cite{Janhsen:1992,Minev:2021}, 
as well as to construct fully quantum mechanical models of dissipation for quantized electrical networks, e.g.,
following the Caldeira-Leggett
formalism~\cite{CaldeiraLeggett:1981,Devoret:1997,Burkard:2004,Hassler:2019}, should be natural by taking appropiate continuous limits of the models here obtained~\cite{ParraRodriguez:2018,ParraRodriguez:2022}.

{\it Note: while finishing this manuscript we became aware of Ref. \cite{Osborne:2023} (later published in \cite{Osborne_v2:2024}), which arrives at some of the results presented here for a more restricted class of circuits, i.e., with only non-dissipative (reciprocal) two-terminal elements.}

\begin{acknowledgments}
A. P.-R. thanks the Canada First Research Excellence Fund and Juan de la Cierva fellowship FJC2021-047227-I. I. L. E. acknowledges support by the Basque Government through Grant No. IT1470-22, and project PCI2022-132984 financed by MICIN/AEI/10.13039/501100011033 and the European Union NextGenerationEU/PRTR.
\end{acknowledgments}

\appendix

\section{Mathematical elements}
\label{sec:mathapp}
This work requires only some very basic concepts of differential and symplectic geometry, because of the specific nature of the systems under study. Nonetheless, we include a lightning review of those. Good introductory texts to differential geometry for physicists can be \cite{nakahara2003geometry} or \cite{Choquet-Bruhat:1982}. For symplectic geometry in the context of classical mechanics, we suggest \cite{Jose:1998} or \cite{Abraham:2008}. 
\subsection{Differential geometry}
The object of study of differential geometry are (smooth) differentiable manifolds $\mathcal{M}$, which are locally homeomorphic to $\mathbb{R}^n$ for some $n$, the dimension, and such that the local descriptions are compatible with (infinitely) differentiable transition functions. At each point of the manifold (tangent) vectors are defined as equivalence classes of curves tangent at that point. However, there are alternative equivalent definitions. For our purposes, the most relevant is that they are derivations on (the algebra of germs of) differentiable functions at the point. Given a local coordinate system, $x^i$, a vector $\boldsymbol{V}_p$ at the point $p$ acts of functions as 
\begin{align}
    \boldsymbol{V}_p(f)=V^i\frac{\partial f}{\partial x^i}\,,
\end{align}
where $V^i$ are the components of the tangent vector. This extends to vector \emph{fields}, by stitching together smoothly tangent vectors at different points. This is properly understood as a section of the tangent bundle $T\mathcal{M}$, which is the space of pairs $(p,\boldsymbol{V}_p)$. We write locally a vector field as
\begin{align}
    \boldsymbol{V}= V^i\frac{\partial}{\partial x^i}\,.
\end{align}
We say that a vector field vanishes nowhere (has no critical points) if in each chart the components do not all simultaneously vanish. In such a case, it provides us with a direction at each point of the manifold, and thus a local coordinate $v$ such that $\boldsymbol{V}(f)=\partial f/\partial v$.

It is immediate to construct the commutator of two vector fields, $\left(\left[\boldsymbol{V},\boldsymbol{W}\right]\right)(f)= \boldsymbol{V}\left(\boldsymbol{W}\left(f\right)\right)-\boldsymbol{W}\left(\boldsymbol{V}\left(f\right)\right)$. The case of commuting vector fields is a particular case of involution, i.e. of the commutator belonging at each point to the span of the two tangent vectors and the commutator being a vector field. If we have vector fields in involution,  we have local coordinates for the underlying manifold that divide into those corresponding to the vector fields in involution and those invariant under the action of those vector fields. Simply put, locally we have coordinates of the form $\left\{v,w\right\}\cup\left\{\xi^\alpha\right\}$ such that the vector fields in involution only involve the first two. I.e., we have locally adapted coordinates.

At each point of the manifold we can consider the dual of the vector space of tangent vectors. These are the dual vectors, cotangent vectors (covectors) or, more suitably for the present context, one-forms. Let $\theta$ be a one form at point $p$, $\boldsymbol{V}_i=\partial/\partial x^i$ the coordinate tangent vectors at that point in some coordinate system. Then $\left\langle\theta, \boldsymbol{V}_i\right\rangle=\theta_i$ are the components of the one form in that system, and locally we can write $\theta=\theta_i\mathrm{d}x^i$, where $\mathrm{d}x^i$ form the dual basis to $\boldsymbol{V}_i$. This is smoothly extended to a differential one-form or covector field $\theta$ by analogy to vector fields, and the process extends to tensor fields. An $r$-covariant, $q$-contravariant tensor fields has components in local coordinates with $r$ lower and $q$ upper indices, which transform respectively as those of one-forms and tangent vectors.

However, there are only two types of tensor fields that we shall be considering besides vectors and covectors. Namely $r$-covariant fully antisymmetric ones, or differential forms, and $2$-covariant symmetric tensor fields.  Let $V_p$ be the vector space of tangent vectors at that point. A linear $r$-form at point $p$ maps linearly the $r$-tensor product $V_p\otimes V_p\otimes\cdots\otimes V_p$  (with $r$ factors) to numbers. An exterior $r$-form is a completely antisymmetric linear $r$ form. We denote the vector space of exterior $r$-forms at $p$ by $\Omega_p^r$. The exterior or wedge product of exterior forms maps $\Omega_p^r\times\Omega_p^q$ to $\Omega_p^{(r+q)}$ in the obvious manner. Thus, the wedge product of the two 1-forms $\theta_1$ and $\theta_2$ acts on the tensor product $\boldsymbol{V}\otimes\boldsymbol{W}$ as 
\begin{align}
\left(\theta_1\wedge\theta_2\right)\left(\boldsymbol{V}\otimes\boldsymbol{W}\right)=\theta_1(\boldsymbol{V})\theta_2(\boldsymbol{W})-\theta_1(\boldsymbol{W})\theta_2(\boldsymbol{V})\,.
\end{align}
The corresponding tensor fields are the differential forms, with $T^*\mathcal{M}$, the cotangent bundle of the manifold $\mathcal{M}$, being the space of pairs $(p,\theta)_p$ with $\theta_p\in\Omega_p^1$, smoothly connected, and other bundles being similarly defined. We can use the notation $\Omega^r(\mathcal{M})$ to denote the space of smooth differential $r$-forms on $\mathcal{M}$, and thus $\Omega^1(\mathcal{M})=T^*\mathcal{M}$. If the dimension of the manifold is $n$, the dimension of the vector space $\Omega_p^r$ is ${n}\choose{r}$. Notice that for $r>n$ $\Omega_p^r$ is empty and so is $\Omega^r(\mathcal{M})$.

There is an important additional concept to mention, that of the exterior derivative. This is a map $\mathrm{d}:\Omega^r(\mathcal{M})\to \Omega^{r+1}(\mathcal{M})$, that acting on the form with local expression
\begin{align}
    \omega= \frac{1}{r!}\omega_{{i_1}{i_2}\cdots{i_r}}\,\mathrm{d}x^{i_1}\wedge\cdots\wedge \mathrm{d}x^{i_r}
\end{align}
(implicit summation) 
produces the $r+1$ differential form
\begin{align}
    \mathrm{d}\omega= \frac{1}{r!}\frac{\partial\omega_{{i_1}{i_2}\cdots{i_r}}}{\partial x^j}\,\mathrm{d}x^j\wedge\mathrm{d}x^{i_1}\wedge\cdots\wedge \mathrm{d}x^{i_r}\,.
\end{align}
Crucially, $\mathrm{d}^2=0$. We say that a differential form $\theta$ is \emph{closed} if its exterior derivative is identically zero, i.e. if $\mathrm{d}\theta=0$. A differential form $\theta$ is \emph{exact} if it equals the exterior derivative of another form, i.e. if $\beta$ exists such that $\theta=\mathrm{d}\beta$. These definitions are geometric, i.e. independent of coordinate choice. Clearly, all exact differential forms are closed, but not all closed differential forms need be exact. The existence or otherwise of closed but not exact differential forms is the topic of de Rahm cohomology, and depends on the topology of the underlying manifold.  

The most important example for our purposes is that of the circle $S^1$. We cannot cover the whole circle with a single coordinate patch, as we would have discontinuities. Nonetheless,  at each point of the circle we have the one form $\theta$ that is dual to the local generator of translations, in coordinates $\partial_\varphi$ . This one form is locally $\mathrm{d}\varphi$ for an angular coordinate, but it is not exact. It is however necessarily closed, as $\Omega^2(S^1)=\emptyset$. 

The introduction of differential forms and the use of the notation $\mathrm{d}$ for the exterior derivative are intimately related with the concept of integration over manifolds. For our purposes, the most relevant aspect is that integration of an $n$-form  $\omega$ over an $n$-dimensional orientable manifold $\mathcal{M}$ is well-defined by matching Lebesgue or Riemann integration of the (single) coefficient function with $\int_\mathcal{M}\omega$.  In particular we have to pay attention to the integration of one-forms on smooth curves, amongst which closed loops are of special relevance. Thus, consider the integration of $\mathrm{d}\varphi$ (as a local expression of $\theta$) on $S^1$: the result is $2\pi$. If the integral of a one form over all possible closed loops on a manifold is zero then it is an exact form. This is the underlying reason for the Stokes and Gau\ss~ theorems. In order to characterize the topology of a manifold (in the cases that enter this work) we have to focus on the existence or otherwise of nontrivial cycles, i.e. those loops that are not smoothly contractible to a point. We also use the name nontrivial cycle for each corresponding class of homotopically equivalent loops. For instance, in the torus we have two independent nontrivial cycles. Although we will not emphasize this perspective, the identification of nontrivial cycles in the manifold that results from applying the constraints is fundamental to achieve a complete understanding of the final dynamics.

A delicate topic is that of embedding and immersion. Let $\mathcal{M}$ be a manifold of dimension $m$ and $\mathcal{N}$  a manifold of dimension $n<m$. Not all smooth maps $f$ from $\mathcal{N}$ to $\mathcal{M}$ are such that   the differential structure that $f(\mathcal{N})$ inherits from $\mathcal{M}$ matches the one that would follow from $\mathcal{N}$. For instance, imagine that $S^1$ is mapped to the plane, in such a way that the image of $S^1$ has self-intersections, even if it is a closed curve and otherwise smooth. This means that as we follow a trajectory on $S^1$ we also follow smoothly a trajectory on the image of $S^1$, even through the self-intersection. As vectors are equivalence classes of trajectories through a point, tangent vectors on $S^1$ are smoothly mapped to tangent vectors of $\mathbb{R}^2$ at each point of $S^1$. We say that we have an immersion of $S^1$ in the plane. However, because there are at least two points of $S^1$ with the same image in the plane, we would have an ambiguity in going back to $S^1$ and its tangent bundle. When these ambiguities do not exist, i.e. when, for the example, there is no self-intersection in the image, then we say that $\mathcal{N}$ has been embedded in $\mathcal{M}$, and that $f(\mathcal{N})$ is a submanifold of $\mathcal{M}$. Whenever we speak of submanifolds of a manifold $\mathcal{M}$, we mean that they can be understood as the embedding  $f(\mathcal{N})$ of another manifold $\mathcal{N}$, or, equivalently, that the submanifold is diffeomorphic to $\mathcal{N}$. The intuitive image to have in mind is that a submanifold does not have self-intersections.
\subsection{Symplectic geometry}
\label{sec:sympgeomapp}
In the development of analytical mechanics, the tangent bundle and the contangent bundle were recognized as the spaces on which the Lagrangian and the Hamiltonian act. In connection to this, the need appeared for a map of vector fields to differential forms. These maps are given by 2-covariant tensor fields, as they maps pairs of vector fields to functions on the underlying manifold. The fully symmetric 2-covariant tensors, a metric if positive and non-degenerate, are not as relevant in this context, and the object of interest are differential two-forms. In particular, closed non-degenerate differential two-forms, for which the name \emph{symplectic form} is reserved. These structures are natural for cotangent spaces, although not restricted to those.  

Not all differential two-forms are symplectic. Both conditions, closedness and nondegeneracy, have to be fulfilled. In practical terms, the analysis of nondegeneracy is carried out in coordinates. Namely,
given a differential two-form and a chart we obtain a matrix function of the coordinates. Frequently the same symbol is used for the two-form and its associated matrix functions, by assuming implicitly some particular coordinates, as the matrix itself will change under a change of variables. On the other hand, the matrix invariants under similarity transformations will be preserved, and thus are inherent properties of the two form. In particular, rank. Thus, the question of degeneracy of the differential two-form at a point is independent of the coordinate system, and the determinant of the matrix function at a point in any set of coordinates informs us to that effect. We use the terms full-rank and nondegenerate interchangeably.

The concept of full rank in the case of two-forms is in direct correspondence to that of full rank for linear operators, and signifies that at all points of the underlying manifold the two-form has no null vectors. To clarify, at each point of the manifold the two-form takes two vectors (two elements of the tangent space at that point) and maps the pair to a number. The vector $\boldsymbol{W}_p$ at a point $p$ is a null vector of the two-form if for all $\boldsymbol{V}_p$ vectors at that point   we have $\omega(\boldsymbol{W}_p,\boldsymbol{V}_p)=0$. The rank of a two-form at a point $p$ is the dimension of the manifold minus the dimension of the kernel of the two-form at that point. Notice that rank is a local quantity; inhomogeneous rank is a possibility in general.

Let us now portray the natural symplectic structure of cotangent bundles $T^*\mathcal{M}$. These are  manifolds in their own right, and local coordinates $\left\{q^i,p_j\right\}$ can always be chosen, where $q^i$ are coordinates of the \emph{configuration space} $\mathcal{M}$ and $p_i$ are the components of a one-form at that point of $\mathcal{M}$ in the associated basis of covectors. Then the canonical one form $\theta$ with local expression $p_i \mathrm{d}q^i$ has the exterior derivative $\omega=\mathrm{d}\theta=\mathrm{d}p_i\wedge\mathrm{d}q^i$, which is closed (as it is exact) and nondegenerate.

An even dimensional manifold $\mathcal{M}$ on which a symplectic form $\omega$ exists is called, together with that particular form, symplectic. Not all symplectic manifolds are diffeomorphic to a cotangent bundle. For example, the torus $S^1\times S^1$ is symplectic with  symplectic form $\theta^1\wedge\theta^2$, where $\theta^i$ is the one-form for the corresponding compact direction. In fact, all orientable two-dimensional manifolds can be provided with a symplectic form. However, compact manifolds cannot be cotangent bundles. On the other hand,  local coordinates exist for all  symplectic manifolds such that locally the symplectic form can be expressed as $\mathrm{d}p_i\wedge\mathrm{d}q^i$. We refer to these as Darboux coordinates. 

The symplectic form is central to Hamiltonian dynamics because it assigns to each smooth function $H$ on $\mathcal{M}$ a vector field $\boldsymbol{X}_H$ whose integral lines are the solutions of the dynamical system $H$ and $\omega$ determine. This vector field is defined by the condition that \begin{align}
    \left\langle\mathrm{d}H,\boldsymbol{V}\right\rangle=\omega(\boldsymbol{X}_H,\boldsymbol{V})
\end{align}
for all vector fields $\boldsymbol{V}$. The condition of non-degeneracy is required for the function $H$ to fully determine the vector field. The equations of motion in Darboux coordinates are the Hamiltonian ones,
\begin{align}
    \dot{q}^i&=\frac{\partial H}{\partial p_i}\,,\nonumber\\
    \dot{p}_i&=-\frac{\partial H}{\partial q^i}\,.
\end{align}

This map from functions to vector fields also allows the construction of a special algebraic structure on the space of smooth functions on $\mathcal{M}$, defined by the Poisson bracket
\begin{align}
    \left\{f,g\right\}=\omega(\boldsymbol{X}_f,\boldsymbol{X}_g)\,.
\end{align}
Skew-symmetry of $\omega$ guarantees $\left\{f,g\right\}=-\left\{g,f\right\}$. The fact that $\mathrm{d}(gh)= g\mathrm{d}h+h\mathrm{d}g$, together with $\left\langle g\theta,\boldsymbol{V}\right\rangle= g\left\langle \theta,\boldsymbol{V}\right\rangle$ for $g$ a function, $\theta$ a differential one-form and $\boldsymbol{V}$ a vector field, ensures that the Poisson bracket acts as a derivative, 
\begin{align}
    \left\{f,gh\right\}=g\left\{f,h\right\}+h\left\{f,g\right\}\,.
\end{align}
A final property of the Poisson bracket, the Jacobi identity, follows in this definition from the symplectic form being closed.
\subsection{On canonical quantization}
For the purposes of this paper, we consider canonical quantization to be  a family of constructions of  quantum mechanical models which takes as data a Hamiltonian function of a cotangent bundle $T^*\mathcal{M}$. Even in this restricted case many subtleties can arise \cite{Carosso:2022}, so in fact we will further limit our considerations to configuration spaces diffeomorphic to $\mathbb{R}^n\times\left(S^1\right)^k$. In that case, and setting aside ordering issues and the Groenewald-van Hove theorem, it is enough for our purposes to look at $T^*S^1$. Our perspective is that the quantum model that is constructed should reflect crucial properties of the classical model we start from. The first one is that the classical Hamiltonian is a function on the cylinder, with different functional forms depending on the coordinates used to describe the cylinder. Secondly, there is a  vector field that generates  rotations around the axis. The function on the cylinder that generates that vector field using the natural symplectic structure  is in fact the coordinate function along the axis. Classically, a full rotation around the axis must leave the system unchanged. Quantum mechanically we must allow for the possibility of a phase arising out of such a rotation, as the probabilities that the theory must be able to predict are computed from interfering amplitudes. Therefore, we must consider that there is an observable that carries this r\^ole of generating rotations, and such that it can be associated with a monodromy. There is a natural candidate family: that of self-adjoint extensions of $-i\partial_\varphi$ acting on functions on an  interval $(\varphi_0,\varphi_0+2\pi)$. This family is parameterised by the group $U(1)$, precisely the monodromy that we normally express as the generalized periodicity condition $\psi(\varphi_0+2\pi)=\exp(-i\gamma)\psi(\varphi_0)$. Notice that these observables have a discrete spectrum, $\sigma(\hat{n}_\gamma)=\mathbb{Z}- \gamma/2\pi$, for a definite value of $\exp(i\gamma)$. In the quantum circuit context it is frequent to term $n_\gamma=\gamma/2\pi\in[0,1)$ the gate charge. Quantization on the circle means using this operator and expanding functions in the corresponding basis, or, by means of a unitary transformation, using $\hat{n}_0-\gamma/2\pi$, where  $\hat{n}_0$ is the self-adjoint extension with strictly periodic boundary conditions, and expanding in the periodic basis.  Notice that a function on the cylinder, when using the corresponding coordinatization, must also be expressed in an adequate periodic basis. 

This example shows a number of things. First of all, that the emphasis cannot be on the Hilbert space, which cannot distinguish an open interval from the circle, but on the observables. Secondly, that kinematic symmetries are the guiding principle to the identification of the central observables of the quantum model. Thirdly, that inequivalent quantizations will definitely appear even for the simplest observables if the classical model presents compact directions.

One of the central objectives of this work is to propose a systematic process to identify a classical description of the dynamics of a circuit as being Hamiltonian on a tangent bundle space, if it exists, and otherwise to identify the obstructions to the existence of such a description. In this work we do not look into alternative quantization prescriptions, nor into the possibility of inequivalent quantizations nor into the subtleties of canonical prescriptions.

\section{Constraints, geometry and algebra}
\label{sec:const_geom_alg}

Let us be explicit here about the construction of the embedding matrix
$\mathsf{K}$, and illustrate it by way of a simple example. Moreover, this example will also show additional points of the process we want to stress later on.

First consider the case with purely Kirchhoff constraints. Since the work of Kirchhoff himself  \cite{Kirchhoff:1847}, the algebraic construction has been well established in terms of a pair of related matrices, the fundamental loop matrix $\mathsf{F}_L$ (please note that we use ``loop'' as is customary in the circuit literature to name what is termed ``cycle'' in the mathematical literature on graphs) and the fundamental cutset matrix $\mathsf{F}_C$, that furthermore satisfy the Tellegen property that $\mathsf{F}_L\mathsf{F}_C^T=0$. To obtain them, one selects a numbering of nodes, index $n$, and branches, index $b$, and an orientation of the branches in the graph that represents the lumped element circuit (the incorporation of graph theory to the analysis dates to Weyl \cite{Weyl:1923}). The incidence matrix of the graph, $\mathsf{A}_{nb}$, will have rank $\mathrm{min}(N,B)-N_C=N-N_C$, where $N$ is the number of nodes, $B$ the number of branches, and $N_C$ the number of connected components of the graph. If the graph corresponds to a nontrivial circuit then there exists at least one loop, which implies $B\geq N$. The fundamental cutset matrix $\mathsf{F}_C$ is the reduced row echelon form by Gaussian elimination of this incidence matrix, after discarding the $N_C$ identically zero rows. For readers unfamiliar with these algebraic constructions, observe that they are equivalent to the identification of a minimal set of node fluxes that determine all branch voltage drops. The advantage of using this canonical form is clearly seen in the next step: this means that the kernel of $\mathsf{F}_C$ is readily constructed. Take a basis of this kernel space as vectors in $\mathbb{R}^B$. Writing the basis vectors in row form provides us with the fundamental loop matrix $\mathsf{F}_L$, and the Tellegen property $\mathbf{i}\cdot\mathbf{v}=0$ follows.

The Kirchoff constraints, written in a format that connects with the notations of this paper, are the statement
\begin{align}
  \label{eq:kirchoff}
 \mathsf{F}_{\mathrm{Kir}}\dd \boldsymbol{\zeta}= \begin{pmatrix}
    \mathsf{F}_C&0\\0& \mathsf{F}_L
  \end{pmatrix}\dd \boldsymbol{\zeta}=0 \,,
\end{align}
which is a Pfaff exterior system. Solving these constraints means that we obtain a basis of the kernel space of $\mathsf{F}_{\mathrm{Kir}}$, and arrange its vectors as columns of a matrix $\mathsf{K}$, i.e. $\mathsf{F}_{\mathrm{Kir}}\mathsf{K}=0$. As we have chosen a canonical form for the Kirchhof constraints we can write $\mathsf{K}$ explicitly if those are the only constraints,
namely
\begin{align}
  \label{eq:Kkirchof}
  \mathsf{K}_{\mathrm{Kir}}=
  \begin{pmatrix}
    \mathsf{F}_L^T&0\\ 0& \mathsf{F}_C^T
  \end{pmatrix}\,.
\end{align}
Observe however that alternative algorithms exist and can be
implemented systematically, were we not to start from a canonical
form. As the Kirchhoff constraints separate in two sets, the current
conservation constraint or Kirchhoff current law (KCL) on the one hand
and the Kirchhoff voltage law on the other, the constrained manifold
under just $\mathsf{F}_{\mathrm{Kir}}$ has clearly separated current and voltage
directions.
\begin{figure}[!ht]
  \centering
  \includegraphics[width=0.5\linewidth]{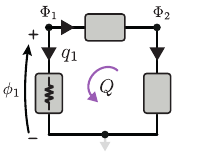}
  \caption{Series connection of three two-terminal lumped elements.}
    \label{fig:Circuit_series_loop}
\end{figure}
Consider for instance the graph associated with the circuit in
Fig. \ref{fig:Circuit_series_loop}. It is easy to construct the
fundamental matrices in a canonical form,
\begin{align}
  \mathsf{F}_C &=
                 \begin{pmatrix}
                   1&0&1\\
                   0&1&-1
                 \end{pmatrix}\,,\\
  \mathsf{F}_L &=
                 \begin{pmatrix}
                   1&-1&-1
                 \end{pmatrix}\,.
\end{align}
The minimal set of variables that is required to describe \emph{all} possible circuits constructed with the topology of Fig. \ref{fig:Circuit_series_loop} and one port element is therefore composed of one loop charge and two node fluxes, and we have an explicit expression for $\mathsf{K}_{\mathrm{Kir}}$,
\begin{align}
\label{eq:explicitkkir}
  \dd \boldsymbol{\zeta} =
	  \begin{pmatrix}
	    \dd \boldsymbol{q}\\ \dd \boldsymbol{\phi}
	  \end{pmatrix}=
  \begin{pmatrix}
    1&0&0\\-1&0&0\\-1&0&0\\0&1&0\\0&0&1\\0&1&-1
  \end{pmatrix}
  \begin{pmatrix}
    \dd Q\\
    \dd \Phi_1\\ \dd \Phi_2 
  \end{pmatrix}\,.
\end{align}

This separated structure is also mantained if just transformer
constraints are added to the mix, since a transformer fixes currents
in terms of currents and voltages in terms of voltages, namely there
exists a pair of matrices $\mathsf{T}$ and $\mathsf{T}'$ such that
$\mathsf{T}'\mathsf{T}^T=0$ (local Tellegen property). Connecting with
the projector notation that has been introduced in the main text (MT),
observe that
\begin{align}
  \mathsf{T}\mathsf{P}^{\mathcal{T}}= \mathsf{T}
\end{align}
is required, and similarly for $\mathsf{T}'$. The Pfaff
system they provide is 
\begin{align}
  \label{eq:transformerpfaff}
  \mathsf{F}_T \dd \boldsymbol{\zeta}=\begin{pmatrix}
    \mathsf{T}&0\\ 0& \mathsf{T}'
  \end{pmatrix}\dd \boldsymbol{\zeta}=0\,.
\end{align}

Therefore in a system in which we only have Kirchhoff and transformer
constraints, the Pfaff system again partitions into a current set and
a voltage set, and the computation of the kernel is facilitated by
observing that in such a case
\begin{align}
  \label{eq:kernelsum}
  \mathrm{ker}\left(\mathsf{F}_{\mathrm{Kir}}\right)\cap\mathrm{ker}\left(\mathsf{F}_{\mathrm{T}}\right)=  \mathrm{ker}\left(\mathsf{F}_{\mathrm{cur}}\right)\oplus \mathrm{ker}\left(\mathsf{F}_{\mathrm{vol}}\right)\,,
\end{align}
where
\begin{subequations}\label{eq:fcurandvol}
  \begin{align}
  \mathsf{F}_{\mathrm{cur}}&=
                             \begin{pmatrix}
                               \mathsf{F}_C\\
                               \mathsf{T}
                             \end{pmatrix}\qquad
    \mathrm{and}\label{eq:fcur}\\
    \mathsf{F}_{\mathrm{vol}}&=
                             \begin{pmatrix}
                               \mathsf{F}_L\\
                               \mathsf{T}'
                             \end{pmatrix}\label{eq:fvol}
  \end{align}
\end{subequations}
act on $\mathbb{R}^B$, unlike $\mathsf{F}_{\mathrm{Kir}}$ and
$\mathsf{F}_T$, that act on $\mathbb{R}^{2B}$. Thus, under these
conditions we are led to a description of the system in terms of
\emph{effective} loop charges and flux nodes (see to this point \cite{Solgun:2015}). Concentrating on the
effective loop charges, we determine those by again applying Gaussian
reduction to $\mathsf{F}_{\mathrm{cur}}$ so that it is written in
reduced row echelon form (discarding possible null rows),
\begin{align}
  \label{eq:fisttoeffectiveloop}
  \mathsf{F}_{\mathrm{cur}}\sim
  \begin{pmatrix}
    \mone_{B-\tilde{L}} & \tilde{\mathsf{A}}
  \end{pmatrix}\,.
\end{align}
Here $\tilde{\mathsf{A}}$ is a $\left(B-\tilde{L}\right)\times\tilde{L}$ matrix,
and all currents are independently determined by $\tilde{L}$ effective
loop charges $\left\{Q_i\right\}_{i=1}^{\tilde{L}}$,
\begin{align}
  \label{eq:effectiveloop}
  \mathbf{i}=
  \begin{pmatrix}
    -\tilde{\mathsf{A}}\\ \mone_{\tilde{L}}
  \end{pmatrix}
  \begin{pmatrix}
    \dot{Q}_1\\ \vdots \\ \dot{Q}_{\tilde{L}}
  \end{pmatrix}\,.
\end{align}
The interpretation is that the number of independent cycles of the
graph is reduced because of the transformer, and we are left with a
set of effective independent cycles. An analogous construction holds
for the effective node fluxes.

The  separability of current and flux constraints for
Kirchhoff and transformer constraints is reflected in the fact that
these two cases are determined by pure numbers, incidences in the case
of Kirchhoff, and turn ratios for transformers. On the other hand
both resistor and gyrator constraints must include parameters with
dimensions of impedance, as they do mix currents
and voltages.

As a trivial example, imagine that the first branch of the example whose
Kirchhoff constraints are realized in Eq. (\ref{eq:explicitkkir}) is a
resistence with constitutive equation $R \dd q_1= \dd \phi_1=\dd \Phi_1$, see Fig.\ref{fig:Circuit_series_loop}.  Then
Eq. (\ref{eq:explicitkkir}) is substituted by
\begin{align}\label{eq:explicitk}
  \dd \boldsymbol{\zeta} =
  \begin{pmatrix}
    \dd \boldsymbol{q}\\ \dd \boldsymbol{\phi}
  \end{pmatrix}=
  \begin{pmatrix}
    1&0\\-1&0\\-1&0\\ R&0\\ 0&1\\R&-1
  \end{pmatrix}
\begin{pmatrix}
    \dd Q\\
    \dd \Phi_2
  \end{pmatrix}\,.
\end{align}
More generally, if there are resistive elements there will be matrix
elements with dimensions of either impedance or admittance in
$\mathsf{K}$. However, if there are no non-reciprocal elements the
final coordinates can be chosen as purely loop-charge or flux-node, as
we see in the example.

In fact, the constraint elements (resistive, transformers and
gyrators) can be described in a unified manner with the scattering matrix formalism, which we now briefly summarize. As stated in the
MT, we partition the branches of the circuit graph in different
sets with label
$\mathcal{S}\in\left\{\mathcal{C},\mathcal{L},\mathcal{V},\mathcal{I},\mathcal{R},\mathcal{T},\mathcal{G}\right\}$,
which denote, in sequence, capacitor, inductor, voltage source,
current source, resistence, transformer, and gyrator. The 
set of constraint elements is
$\mathcal{R}\cup\mathcal{T}\cup\mathcal{G}$. Then, acting on
$T\mathcal{M}_{2B}$ homogenously, we have that for the subset
$\mathcal{S}$ of the constraint element set the Pfaff system is
\begin{align}
  \label{eq:generalSmatrix}
  \mathsf{F}_S\dd \boldsymbol{\zeta}=
  \begin{pmatrix}
    R\left(\mathsf{P}^{\mathcal{S}}+\mathsf{S}^{\mathcal{S}}\right)  &|&  \left(\mathsf{S}^{\mathcal{S}}-\mathsf{P}^{\mathcal{S}}\right)
  \end{pmatrix}\dd \boldsymbol{\zeta}=0\,.
\end{align}
$R$ is a characteristic impedance, while the $\mathsf{S}^{\mathcal{S}}$ scattering
matrices are adimensional. $\mathsf{P}^{\mathcal{S}}$ is the
projector onto the $\mathcal{S}$ set. The  scattering matrices
$\mathsf{S}^{\mathcal{S}}$ satisfy
\begin{align}
    \label{eq:smatrixprojector}
    \mathsf{P}^{\mathcal{S}}\mathsf{S}^{\mathcal{S}}=
  \mathsf{S}^{\mathcal{S}}\mathsf{P}^{\mathcal{S}} =  \mathsf{S}^{\mathcal{S}}\,.
\end{align}
The constraint is lossless if the scattering matrix is unitary, and
reciprocal if it is symmetric. Ideal transformers are lossless and
reciprocal, and therefore the eigenvalues of the corresponding
scattering matrix (on its support space $\mathcal{S}$) are either $1$
or $-1$, which implies that they are correctly described by the
$\mathsf{T}$ and $\mathsf{T}'$ matrices mentioned earlier, satisfying
the local Tellegen property. The separation between the $+1$ and $-1$
eigenspaces means that the impedance parameter drops out of the
transformer constraint. An ideal  gyrator, on the other hand, corresponds to the
purely nonreciprocal part of a lossless scattering matrix, and
therefore mixes the voltage and current variables, and the impedance
parameter is required. Nonetheless,
because it is lossless it satisfies the local Tellegen property. The
simplest gyrator constraint, with two branches, is given by
\begin{align}
  \label{eq:elementalgyrator}
  \begin{pmatrix}
    i_1\\ i_2
  \end{pmatrix} = \frac{1}{R}
  \begin{pmatrix}
    0&1\\-1&0
  \end{pmatrix}
             \begin{pmatrix}
               v_1\\ v_2
             \end{pmatrix}\,,
\end{align}
or, more compactly,
\begin{align}
  \label{eq:elemgyrvect}
  \mathbf{i}_G=\mathsf{Y}_G\mathbf{v}_G=\frac{1}{R}\, i\sigma^y\,\mathbf{v}_G\,,
\end{align}
where $\sigma^y$ is the imaginary Pauli matrix. More generally, the
canonical form of a gyrator constraint matrix is
\begin{align}
  \label{eq:generalgyrator}
  \mathsf{F}_G =
  \begin{pmatrix}
    \mathsf{P}^{\mathcal{G}}&|&-\mathsf{Y}_G
  \end{pmatrix}\,,
\end{align}
where the admittance matrix $\mathsf{Y}_G$ is real, satisfies
$\mathsf{P}^{\mathcal{G}}\mathsf{Y}_G=\mathsf{Y}_G\mathsf{P}^{\mathcal{G}}=\mathsf{Y}_G$,
and is antisymmetric on its support.

Thanks to the scattering matrix formalism we could unify all these
ideal linear constraints into a single scattering matrix and a single
impedance parameter, and that approach might prove useful in the
automatization of our prescription. However, from a conceptual
perspective and with a view to proving general statements it is
convenient to present as separate transformer, gyrator, and linear
resistive constraints.

Thus, the full set of Pfaff exterior equations that we
consider will have the structure
\begin{align}\label{eq:fullPfaff}
  \mathsf{F}\dd \boldsymbol{\zeta}=
  \begin{pmatrix}
    \mathsf{F}_{\mathrm{Kir}}\\
    \mathsf{F}_T\\
    \mathsf{F}_R\\
    \mathsf{F}_G
  \end{pmatrix}\dd \boldsymbol{\zeta}=0\,.
\end{align}
As before, the task is to compute the kernel of $\mathsf{F}$ and collect
a basis as column vectors in a matrix $\mathsf{K}$. Then
$\mathsf{FK}=0$ will hold, and we express the embedding of the
resulting integral manifold $\mathcal{M}$, with coordinates $z^\alpha$,
by means of
\begin{align}
  \dd \boldsymbol{\zeta}= \mathsf{K}\dd {\bz}\,,
\end{align}
and correspondingly for the vectors
\begin{align}
  \nabla_{\bz} = \mathsf{K}^T \nabla_{\bzeta}\,.
\end{align}
These coordinates will generally be neither loop-charge- nor
node-flux-like. Rather they will mix them adequately.

Having achieved the description of all the branch variables, $\zeta$, in term of
a set of prima facie independent ones, $z^\alpha$, we can now carry
out the task of constructing $\omega$, by using the $\mathsf{K}$
matrix. As we shall now see, this simply means that
\begin{align}
  \omega_{\alpha\beta}= \left[\mathsf{K}\right]_{b\alpha} \left[\omega_{2B}\right]_{bb'}\left[\mathsf{K}\right]_{b'\beta}\,,
\end{align}
using coordinates.

The global and local Tellegen properties are the statements that
\begin{align}
  \label{eq:tellegenmatrix}
  \begin{aligned}
    \mathsf{K}^T
  \begin{pmatrix}
    0&\mone\\ 0&0
  \end{pmatrix}\mathsf{K}&=\mathsf{K}^T
  \begin{pmatrix}
    0& \mathsf{P}^{\mathcal{T}}\\  0&0
  \end{pmatrix}\mathsf{K}\\
  &=\mathsf{K}^T
  \begin{pmatrix}
    0& \mathsf{P}^{\mathcal{G}}\\ 0&0
  \end{pmatrix}\mathsf{K}=0,
  \end{aligned}
\end{align}
together with their transposes. The first one is simply the statement
that $\mathbf{i}\cdot\mathbf{v}$, for all possible $\mathbf{i}$ and
$\mathbf{v}$ that satisfy the Kirchhoff constraint. As to the second
one, observe that the matrix $\mathsf{T}$ we introduced above can be
written in canonical right Belevitch \cite{Belevitch:1950,Belevitch:1968} form $\mathsf{T}\to
\begin{pmatrix}
  \mathsf{N}&\mone
\end{pmatrix}$ acting only on the branches in $\mathcal{T}$, and then
similarly $\mathsf{T}'\to
\begin{pmatrix}
  \mone &-\mathsf{N}^T
\end{pmatrix}$ on just the branches in $\mathcal{T}$. Here
$\mathsf{N}$ is the turns ratio matrix. Then the currents and voltages
in the branches of $\mathcal{T}$ must be of the form
\begin{align}
  \mathsf{P}^{\mathcal{T}}\mathbf{i}&\to
  \begin{pmatrix}
    \mone\\ -\mathsf{N}
  \end{pmatrix}\tilde{\mathbf{I}}\qquad \mathrm{and}\\
  \mathsf{P}^{\mathcal{T}}\mathbf{v}&\to
  \begin{pmatrix}
    \mathsf{N}^T\\ \mone
  \end{pmatrix}\tilde{\mathbf{V}}\,,
\end{align}
for some $\tilde{\mathbf{I}}$ and  $\tilde{\mathbf{V}}$, not
necessarily generic, since other constraints will have to be
fulfilled. 
From this structure it follows that
$\left(\mathsf{P}^{\mathcal{T}}\mathbf{i}\right)\cdot\left(\mathsf{P}^{\mathcal{T}}\mathbf{v}\right)=0$. The
local Tellegen property for $\mathcal{G}$ follows analogously from the
definitions introduced by Tellegen \cite{Tellegen:1948}, that
required no dissipation for the ideal element. Alternatively, it flows
from the unitarity of the scattering matrix $\mathsf{S}^{\mathcal{G}}$.

\section{Algebraic proof}
\label{sec:algebraic-proof}
The two-form $\omega_{2B}$ in Eq.~(\ref{eq:twoform2b}) 
can be
written more compactly as
\begin{align}
  \label{eq:matricesandforms}
  \omega_{2B}&= \frac{1}{2}
               \dd \boldsymbol{q}^T\wedge\left(\mathsf{P}^{\mathcal{C}}+\mathsf{P}^{\mathcal{V}}-\mathsf{P}^{\mathcal{L}}-\mathsf{P}^{\mathcal{I}}\right)\dd \boldsymbol{\phi}\nonumber\\
  &=\frac{1}{2}\dd \boldsymbol{\zeta}^T\wedge
    \mathsf{E}_{2B}\dd \boldsymbol{\zeta}\,,
\end{align}
with
\begin{align}
  \label{eq:2E2B}
&\mathsf{E}_{2B}=\\
&\frac{1}{2} \begin{pmatrix}
      0&
      \mathsf{P}^{\mathcal{C}}+\mathsf{P}^{\mathcal{V}}-\mathsf{P}^{\mathcal{L}}-\mathsf{P}^{\mathcal{I}}\\ -\mathsf{P}^{\mathcal{C}}-\mathsf{P}^{\mathcal{V}}+\mathsf{P}^{\mathcal{L}}+\mathsf{P}^{\mathcal{I}}&0
    \end{pmatrix},\nonumber
\end{align}
which is the antisymmetric part of
\begin{align}
  \label{eq:etilde}
  \tilde{\mathsf{E}}_{2B}= \begin{pmatrix}
      0&
      \mathsf{P}^{\mathcal{C}}+\mathsf{P}^{\mathcal{V}}\\ \mathsf{P}^{\mathcal{L}}+\mathsf{P}^{\mathcal{I}}&0
    \end{pmatrix}\,,
  \end{align}
precisely the matrix that appears in the constitutive equations,
Eq.~(\ref{eq:constmatrix})
.  The transformation $\mathsf{A}\to
\mathsf{K}^T\mathsf{A}\mathsf{K}$ for an antisymmetric matrix is
 the  expression in components of the pullback of a two-form by the
immersion map $\imath:\mathcal{M}\to \mathcal{M}_{2B}$. Thus we see
that the matrix acting on $\dot{\bz}$ in the LHS of
Eq.~(\ref{eq:finaleoms}) 
is 
\begin{align}
  \label{eq:matrixid}
  \mathsf{K}^T
{\mathsf{E}}_{2B}\mathsf{K}=\mathsf{K}^T
  \begin{pmatrix}
    0&
    \mathsf{P}\\
      -\mathsf{P}
        &0
  \end{pmatrix}\mathsf{K}\,,
\end{align}
where
 $ \mathsf{P}=  \mathsf{P}^{\mathcal{C}}+\mathsf{P}^{\mathcal{V}}+\left(\mathsf{P}^{\mathcal{R}}+\mathsf{P}^{\mathcal{G}}\right)/2$
thanks to the global (Kirchhoff) and local ($\mathcal{T}$ and $\mathcal{G}$) Tellegen properties, Eq. (\ref{eq:tellegenmatrix}) and
its transposes.

For additional reference, we remind the reader that a Lagrangian
$L(\left\{q^i,\dot{q}^i\right\})$ and a Rayleigh dissipation function
$\mathcal{F}(\left\{\dot{q}^i\right\})$ provide us with the equations
of motion \cite{Goldstein:2002}
\begin{align}
  \label{eq:eomswithrayleigh}
  \frac{\dd }{\dd t}\left(\frac{\partial
    L}{\partial\dot{q}^i}\right) = \frac{\partial L}{\partial q^i}-
\frac{\partial \mathcal{F}}{\partial\dot{q}^i}\,.
\end{align}

\section{The method of Faddeev and Jackiw}
\label{sec:meth-fadd-jack}

Both the original paper by Faddeev and Jackiw \cite{Faddeev:1988}  and the later very
pedagogical presentation by Jackiw \cite{Jackiw:1994}  are good sources to understand the method. Nonetheless here we shall provide a brief introduction in support of application of the method to the analysis of electrical circuits. This presentation starts as local, in that we are using coordinates explicitly, and is then also rephrased in differential geometric terms.

The starting point is a first order lagrangian, that we write as
\begin{align}
  \label{eq:lFJgeneric}
  L= A_\alpha(z) \dot{z}^\alpha- H(z)\,.
\end{align}
The Euler--Lagrange equations of motion derived from this Lagrangian are
\begin{align}
  \label{eq:eomFJ}
  \omega_{\alpha\beta}(z)\dot{z}^\beta=\frac{\partial H}{\partial z^\alpha}\,,
\end{align}
with
\begin{align}
  \label{eq:omFJ}
  \omega_{\alpha\beta}(z) = \frac{\partial A_\beta}{\partial z^\alpha}-\frac{\partial A_\alpha}{\partial z^\beta}\,.
\end{align}
It is frequently convenient to think in terms of differential forms,
in which case we have
\begin{subequations}
  \label{eq:diffformFJ}
  \begin{align}
    A &= A_\alpha(z) \dd z^\alpha\,,\label{eq:A1form}\\
    \omega &= \frac{1}{2}\omega_{\alpha\beta}
             \dd z^\alpha\wedge\dd z^\beta\,,\label{eq:om2form}\\
    \omega &= \dd A\,.\label{eq:omexteriorA}
  \end{align}
\end{subequations}

Looking again at Eq. \eqref{eq:eomFJ}, observe that if $\mathrm{det}(\omega)\neq0$ everywhere we are in presence of the
Hamiltonian equations of motion derived from Hamiltonian $H$ and
Poisson bracket given by the inverse of the matrix $\omega$,
\begin{align}
  \label{eq:Poisson}
  \left\{z^\alpha, z^\beta\right\}= \omega^{\alpha\beta}\,,
\end{align}
with the convention that
\begin{align}
  \omega^{\alpha\beta}\omega_{\beta\gamma}=\delta^\alpha_\gamma\,.
\end{align}
However, as we shall see in circuital examples, it is frequently the
case that $\mathrm{det}(\omega)=0$ in some locus. In the examples of
interest to us we shall have $\omega$ homogeneous in the first step of
the method, while inhomogeneous rank will possibly appear in later
steps of the process.

At any rate, if $\mathrm{det}(\omega)=0$ somewhere (or everywhere) the
motion is not guaranteed to be Hamiltonian. Let us consider the zero modes of $\omega$
at a singularity $p$ (which is implicit in the following, where required), $\left\{\mathbf{W}_I\right\}$. These are vector
fields at $p$, such that for \emph{all} vector fields $\mathbf{V}$ at the same
point we have
\begin{align}
  \omega_p\left(\mathbf{W}_I,\mathbf{V}\right)=0\,.
\end{align}
In coordinates, we have
\begin{align}\label{eq:Wonomega}
 \left( W_I\right)^\alpha \omega_{\alpha\beta}= \omega_{\alpha\beta}\left(W_I\right)^\beta=0\,.
\end{align}
Thus a zero mode of the two-form corresponds to a constraint, since
using Eq. (\ref{eq:Wonomega}) with Eq. (\ref{eq:eomFJ}) entails the constraint
\begin{align}\label{eq:constraintexplicit}
  W_I=\left(W_I\right)^\alpha\frac{\partial H}{\partial z^\alpha}=0
\end{align}
at that point, where we define the value $W_I$ and, if
$\mathrm{det}(\omega)=0$ holds in a region with homogeneous rank, $W_I$ is understood as a
function on that region.

At this stage the question arises whether one can reduce the
dimensionality of the problem by imposing the constraints. Let us
assume that $\omega$ is of homogeneous rank. Then, locally, one can
find coordinates $z\to \left\{w^I\right\}\cup\left\{\xi^\mu\right\}$
such that the constraints are
\begin{align}
  \label{eq:rewriteconstraint}
  \frac{\partial H(\xi,w)}{\partial w^I}=0\,,
\end{align}
while the one form reads
\begin{align}
  \label{eq:oneformnew}
  A= \frac{1}{2}f_{\mu\nu}\xi^\mu\dot{\xi}^\nu\,,
\end{align}
and $f_{\mu\nu}$ is full rank.
 If $H$ is a nonlinear
function of $w^I$ one can solve Eq. (\ref{eq:rewriteconstraint}) for
$w^I$. Thus, at this point, it is convenient to solve for as many as
possible and reduce dimensions. Observe that the existence of these
coordinates is guaranteed \emph{locally}, and obstructions to a global
reduction can exist for nontrivial topologies of
$\mathcal{M}$. Furthermore, if the Hamiltonian were to  dependend
linearly on $w^I$ it is
necessary to carry out additional steps.

There is an alternative approach in which instead of reducing the
dimension of the dynamical system as above we treat all the
constraints in the same manner, and introduce a new coordinate for
each constraint. The reduction in dimensionality will be achieved
obliquely, by identifying conserved quantities in the final
Hamiltonian motion. There exist applications in which both reduction and increase in the number of dimensions are used for the study of  the same model. An important instance of this procedure is in the realm of loop quantum cosmology and loop quantum gravity \cite{Thiemann:2008}. Frequently there are assumptions such as homogeneity or isotropy that are translated into symmetries of the model, and the dynamical system is correspondingly reduced in dimension. On the other hand diffeomorphism invariance is a crucial element of general relativity, and this entails that what would ordinarily be understood as the Hamiltonian is actually a constraint, determining a locus. This means that there is no actual evolution. On the other hand, it should be a predictive theory, for instance, associated with the thermal history of the universe, so a presentation with a time evolution is required. This is achieved by increasing the dimension with a coordinate dual to the Hamiltonian constraint, the lapse variable.

Both increasing and reducing procedures fail if, at some point in the iterations,
$\omega$ is not of homogeneous rank. If the system under investigation
is a  physical model this situation will arise when there is a
breakdown in the validity of the model.

Let us now examine the increase in dimensions associated with
constraints. This is carried out by associating a new dimension to
each constraint, and one form $\dd \lambda^I$ dual to its
tangent. Then we modify the one-fom $A$ as
\begin{align}
  \label{eq:addingdims}
  A\to A+ W_I\dd \lambda^I\,,
\end{align}
leading to
\begin{align}
  \label{eq:addinginomega}
  \omega\to \omega+\dd W_I\wedge\dd \lambda^I\,.
\end{align}
Were this enlarged two-form nondegenerate, we would have achieved
the goal of a Hamiltonian description for the dynamics. We have not pursued the increase in dimension route in this work, but do not exclude it being useful in particular circuital models in the future. We should point out that in contexts such as quantum cosmology the system will not be coupled to other ones, whereas the circuits we consider can be probed in a multitude of manners. This leads us to consider the minimum dynamical description as preferable, and thus to adopt reduction procedures.

\section{Time dependent external magnetic fluxes and sources}
\label{sec:time-depend-magn}

There has been some discussion in the literature of superconducting
circuits concerning the inclusion of external fluxes in their
study (see, for instance, \cite{You:2019}). The main issue seems to be that there does not appear to be a
unique prescription, and the doubts that this creates as to the
uniqueness of their description. Here we study this question from our
perspective. The results are, first, that indeed there is freedom, as
in many other steps of the process, yet, because of the underlying
geometric structure, the physics in all choices is equivalent, and,
two, that \emph{gauge} variables might appear, and can be eliminated
as part of our procedure.

The central point to be made is that an external flux is univocally
associated with a loop in the circuit. Thus a
prescription, illustrated in Fig. \ref{fig:ExtFluxToVoltage}, for describing an external flux in terms of a voltage
source goes as follows:
\begin{figure}[!ht]
  \centering
  \includegraphics[scale=.9]{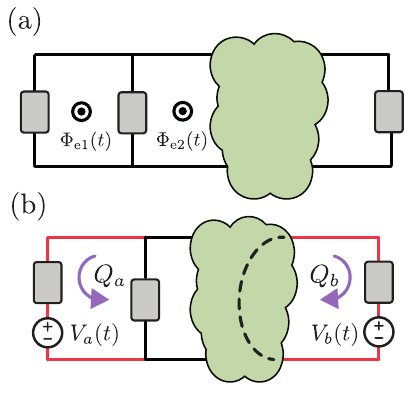}
  \caption{a) Fluxes $\Phi^e_1$ and $\Phi^e_2$ thread a circuit, and
    are represented b) by voltage sources $V_a$ and $V_b$
    corresponding to the loop charges, with the external fluxes being
    partitioned into sources according to a cycle decomposition. }
  \label{fig:ExtFluxToVoltage}
\end{figure}
\begin{enumerate}
\item Choose a tree and a corresponding chord set to carry out the
  Kirchhoff constraint analysis, before imposing transformer and
  gyrator constraints. Each chord corresponds to a unique loop, and
  any loop can be written as a linear superposition of the chord
  loops.
\item Specifically, the choice of a tree means that the fundamental cutset matrix has been identified and has been written in
  Gau\ss--Jordan form as $\begin{pmatrix}
    \mathbbm{1}&\mathsf{A}_{\mathrm{chord}}
  \end{pmatrix}$. In particular, this also entails that the branches
  have been ordered in such a way that the chords are the last
  branches to be enumerated. The loop basis is given by the column
  matrices of
  \begin{equation}
    \label{eq:loopbasis}
    \mathsf{B}_{\mathrm{loop}}=
    \begin{pmatrix}
      -\mathsf{A}_{\mathrm{chord}}\\ \mathbbm{1}
    \end{pmatrix}\,.
  \end{equation} Let us denote those column vectors by $\mathbf{c}_i$,
  where $i$ ranges from 1 to the number of chords.
  \item Given an external flux, the associated loop is given by a
    sequence of edges. These edges are oriented according to the sign
    of the flux, and the flux loop is expressed with a column vector
    with entries $\pm1$, if the corresponding edge is in the loop, or
    0 otherwise. Let $\mathbf{l}$ be that column vector. If there is
    more than one external flux let $e$ be an index for the fluxes,
    $\Phi^e(t)$, and denote the corresponding loop vectors as $\mathbf{l}_e$.  
  \item Now compute the coefficients of $\mathbf{l}_e$ in the basis
    $\left\{\mathbf{c}_i\right\}$,
    \begin{equation}
      \label{eq:loopbasisexpansion}
      \mathbf{l}_e = L_e^{\phantom{e}i}\mathbf{c}_i\,.
    \end{equation}
    We now see that we can expand the \emph{flux vector}
    $\Phi^e\mathbf{l}_e$ in the loop basis,
    \begin{equation}
      \label{eq:flux vector}
      \Phi^e\mathbf{l}_e= \tilde{\Phi}^i \mathbf{c}_i = \Phi^e L_e^{\phantom{e}i}\mathbf{c}_i\,,
    \end{equation}
    using this to define its components in that basis,
    $\tilde{\Phi}^i=\Phi^e L_e^{\phantom{e}i}$.
  \item Starting from the previous directed graph $G$ that describes
    the circuit $C$, we now obtain a new graph $G'$ (and a new circuit
    $C'$) by breaking each chord
    into two edges (i.e. inserting a new node per chord), with one of
    the edges retaining its previous character (capacitive, inductive,
    transformer branch...) and the other becoming a voltage source,
    with voltage function $V_i(t)=\dot{\tilde{\Phi}}^i$. We are working
    in the maximal case in which none of the $\tilde{\Phi}^i$ are
    zero. If there are zeroes, then we do not break that chord, and
    the subsequent counting has to be properly modified, without
    changing the result in any essential way. We continue the
    description only with the maximal case.
    Observe that alternatives exist. For instance, instead of breaking
    a chord sometimes it is more convenient (and equivalent) to break
    a node into two and having the new connecting branch be the source
    branch. We shall actually use this alternative in a later example,
    Fig. \ref{fig:time-dep-ext-flux-SQUID} in section \ref{sec:time-depend-extern}.
  \item By following this procedure we have created a new circuit with
    a new number of branches, $B'$, and nodes $N'$, so we have
    increased the dimension of the initial manifold to $2B'$. The
    number of components, $P$, does not change. Let the
    number of chords of the initial circuit be denoted by $C_h$, and
    correspondingly the number of chords of the resulting circuit by
    $C_h'$. Then we have the following set of relations that issue
    from the procedure:
    \begin{align}
      \label{eq:bprimenprime}
      C_h&=B-N+P\,,\nonumber\\
      B' &= B+ C_h= 2B-N+P \,,\\
      N' &= N + C_h= B+P\,,\nonumber\\
      C_h'&=B'-N'+P = C_h\nonumber\,.
    \end{align}
  \item The rank of the initial two form $\omega_{2B}$ is also
    modified, in that
    \[\mathrm{rank}\left(\omega_{2B'}\right)=\mathrm{rank}\left(\omega_{2B}\right)+
      2(B-N+P)\,,\]
    as we have introduced one conjugate pair per (broken) chord.
  \item However, it is clear that there will be redundancies: in a
    language adequate for Kirchhoff constraints, we have not increased
    the number of loop charges, which is still $C_h$, but we have increased the number of
    nodal fluxes to $N'-P$. Therefore, on imposing the linear
    constraints, now on $\mathcal{M}^{2B'}$, the resulting
    presymplectic form might have additional zero modes. Even more,
    some of those zero modes will be gauge modes. By this we mean that
    the constraint function built as
    $\Omega_v=\mathbf{v}\cdot\nabla_zH$ for the zero mode $\mathbf{v}$
    is identically zero on the constraint. 
  \item It is important to observe that these gauge modes will also  appear
    for other choices of tree and chords in the initial circuit. The
    result of the procedure is independent, from the physics point of
    view, of the choice of tree and chords for imposing Kirchhoff
    constraints. In fact, our description of the procedure
     is completely independent of the character of the chord being
    split up. Admittedly, there are
    choices in which the number of gauge modes is smaller: if the
    chord is inductive then there will be a gauge mode, but not so for
    the capacitive case. Nonetheless, the procedure is systematic and
    allows the identification of these modes.
  \end{enumerate}

  As a final comment, observe that alternative prescriptions exist, in
  which the Kirchhoff constraints are rewritten to be no longer
  linear, but \emph{affine}. This perspective could be brought to use
  in a contact geometric analysis, as well. However, as we show in the
  next section, the specificity of the dynamical systems associated to
  the class of circuits under consideration means that contact
  geometry is not required, and thus we provide this recipe that
  mantains the Kirchhoff constraints as linear.

\section{Dynamical constraints  in the presence of sources and dissipation}
\label{sec:fj-method-presence}

In this work we are studying the equations of motion
for a lumped element circuit
whose reactive elements are capacitors
and inductors, resistances are linear, and possibly  ideal
transformers and gyrators are present, and we have shown that they can be written
as derived from a first order lagrangian and a quadratic Rayleigh
dissipation function,
\begin{subequations}
  \label{eq:restatement}
  \begin{align}
    L& = \frac{1}{2}\omega_{\alpha\beta}z^\alpha\dot{z}^\beta-H
       -S_\alpha{z}^\alpha\,,\label{eq:restatementlag}\\
    \mathcal{F}&= \frac{1}{2}\sum_{r\in\mathcal{R}}\dot{q}^r\dot{\phi}^r=\frac{1}{2}\mathcal{F}_{\alpha\beta}\dot{z}^a\dot{z}^\beta\,,\label{eq:restatementrayleigh}
  \end{align}
\end{subequations}
where the coefficients of the Rayleigh function are
\begin{align}
  \label{eq:rayleighmatrixexplicit}
  \mathcal{F}_{\alpha\beta}=\frac{1}{2}\left[\mathsf{K}^T
  \begin{pmatrix}
    0&\mathsf{P}^{\mathcal{R}}\\ \msP^{\mathcal{R}}&0
  \end{pmatrix}\mathsf{K}\right]_{\alpha\beta}\,.
\end{align}
As mentioned above and shown in examples, generically $\omega$ is
degenerate, and this gives rise to constraints that can be analyzed
with the Faddeev--Jackiw method in the absence of sources and
dissipation. Here we shall extend the analysis of dynamical constraints when thery are present.

To do so, observe that the equations of motion can be written in the
form
\begin{align}
  \label{eq:reordereom}
  \left(\omega_{\alpha\beta}-\mathcal{F}_{\alpha\beta}\right)\dot{z}^\beta
  = \frac{\partial}{\partial z^\alpha}\left[H+ S_\beta z^\beta\right]\,.
\end{align}
Therefore, if (with obvious notation) the matrix $\omega-\mathcal{F}$
is invertible we have a system of ordinary differential equations,
non-autonomous if $S_\beta$ are time-dependent, and all variables
$z^\alpha$ are dynamical. Observe that if $\omega$ is degenerate and $\omega-\mathcal{F}$
is invertible the system is only amenable to a Hamiltonian plus
dissipation description by enlarging the phase space.

On the other hand, if $\omega-\mathcal{F}$ is not invertible we have
(time-dependent) constraints, that must be imposed
consistently. Namely, let $W^\alpha$ be the components of a left
zero-vector of $\omega - \mathcal{F}$,
\begin{align}
  \label{eq:leftzerovector}
  W^\alpha\left(\omega_{\alpha\beta}-\mathcal{F}_{\alpha\beta}\right)=0\,.
\end{align}
Then the following condition is a dynamical constraint on the system
of equations in Eq. \eqref{eq:reordereom},
\begin{align}
  \label{eq:dynconst}
  W^\alpha \frac{\partial}{\partial z^\alpha}\left[H+ S_\beta z^\beta\right]=0\,,
\end{align}
in lieu of the FJ constraint \eqref{eq:constraintexplicit}. A local
coordinatization will again
exist in terms of the degenerate ($w$) and non-degenerate ($\xi$)
directions, and the dynamical constraint is the requirement of
independence of the total energy function $H+S_\beta z^\beta$ with
respect to the degenerate directions. If the constraints are solvable
we will obtain a reduced dynamical system. Generally this dynamical
system will not be Hamiltonian, and non-Hamiltonian techniques will be
required for further analysis. At any rate, our description allows for
an identification of the actual dynamical content of the circuit description.

Let us now concentrate on the non-dissipative case with sources. In
this case an extension of the Faddeev--Jackiw is possible, with
subtleties. The issue of time dependent constraints has been examined
in the literature, both from the Dirac--Bergmann \cite{Mukunda:1980,Evans:1993,Gitman:2012} and the
Faddeev--Jackiw \cite{Belhadi:2016} perspectives. We give a general
description first, then we compute a couple of simple examples in our
context, and we finish this section with a general discussion of FJ
reduction for our nondissipative circuits. The starting point
is the (pre-)contact structure, defined on $\mathcal{M}\times\mathbb{R}$,
where the real line is the time parameter,
\begin{align}
  \label{eq:omegac}
  \Omega_c= \omega+\dd \left(H+S_\alpha(t) z^\alpha\right)\wedge\dd t\,.
\end{align}
For the systems under study, with equations of motion
\begin{align}
  \label{eq:eomforcontact}
  \omega_{\alpha\beta}\dot{z}^\beta= \frac{\partial H}{\partial z^\alpha}+S_\alpha(t)\,,
\end{align}
the two-form $\Omega_c$ is an integral invariant. 
If we have a set of time independent constraints that determine a
submanifold $\mathcal{M}'\subset \mathcal{M}$ at all times, then there
is a pullback of the contact form that has the same structure, with the
pullback of $\omega$ and the time dependent energy function
separately. 
However, if there exist time dependent constraints there will be an
additional term in the pullback of $\Omega_c$, of the form $B\wedge
\dd t$, and unless $B$ is a closed form modulo $\dd t$ we
cannot understand the
evolution as Hamiltonian. More concretely, let the embedding be, in
coordinates, $z^\alpha(\xi^\mu,t)$. Then the one form $B$ is
\begin{align}
  \label{eq:bonefomr}
  B = -\omega_{\alpha\beta}\frac{\partial z^\alpha}{\partial
  t}\frac{\partial z^\beta}{\partial \xi^\mu}\dd \xi^\mu\,.
\end{align}
Thus, explicitly, $\dd B$ is
\begin{align}
  \label{eq:dboneform}
  \dd B&= \frac{\partial}{\partial
  \xi^\nu}\left[\omega_{\alpha\beta}\frac{\partial z^\alpha}{\partial
  t}\frac{\partial
               z^\beta}{\partial\xi^\mu}\right]\dd \xi^\mu\wedge\dd \xi^\nu\nonumber\\
  &\,+\frac{\partial}{\partial
  t}\left[\omega_{\alpha\beta}\frac{\partial z^\alpha}{\partial
  t}\frac{\partial
               z^\beta}{\partial\xi^\mu}\right]\dd \xi^\mu\wedge\dd t\,,
\end{align}
which means that $B$ is locally  closed module $\dd t$ if the
first term is zero.  If $B$ is locally closed modulo $\dd t$ and there are no global
obstructions (which could be  the case in presence  of compact
variables), then we can determine a function $K$ such that
$B=\dd K$ modulo $\dd t$. Let us additionally assume that
the pullback of $\omega$ is nondegenerate. Then the evolution under the
constraints is Hamiltonian, with complete Hamiltonian
$H+S_\mu\xi^\mu+K$.

Let us show a simple example in which there are no issues, that of a
voltage driven LC oscillator.
\begin{figure}[!ht]
  \centering
  \includegraphics[scale=0.5]{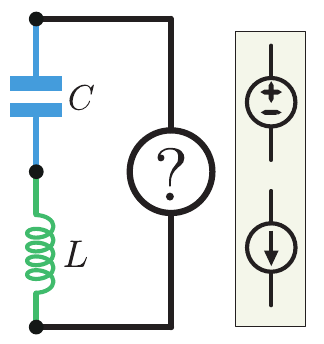}
  \caption{Series LC oscillator driven by a voltage or a current source. These systems can be described with a time-dependent Hamiltonian derived from the method, by imposing a constraint.}
  \label{fig:Dynamical_constraints}
\end{figure}
The circuit graph is a ring with three
branches and three nodes, as represented in Fig. \ref{fig:Dynamical_constraints}. Applying our systematic procedure we have
three variables, $\bz =(Q,\Phi,w)^T$, with two-form
\begin{align}
  \label{eq:simpletwoform}
  \omega = \dd Q\wedge\dd \Phi
\end{align}
and total energy function $h_1(Q)+h_2(\Phi)+Q V(t)$. The zero-mode
coordinate $w$ does not appear anywhere, the correspoding constraint
$\partial_w(H+QV)=0$ is automatically satisfied and we have a
Hamiltonian system, with canonical coordinates.

The situation is a bit more involved if it is a current source instead
of a voltage source. In that case the null direction for the two form
is again a coordinate $w$ and  \eqref{eq:simpletwoform} again holds. The total energy function is however
$h_1(Q)+h_2(w-\Phi)-w I(t)$. Thus there is a nontrivial constraint,
namely $I(t)=h_2'(w-\Phi)$, which is to be solved for $w$. Then the
system is Hamiltonian under the substitution of $w$ by its constrained
value.

\begin{figure}[!ht]
  \centering
  \includegraphics[scale=0.18]{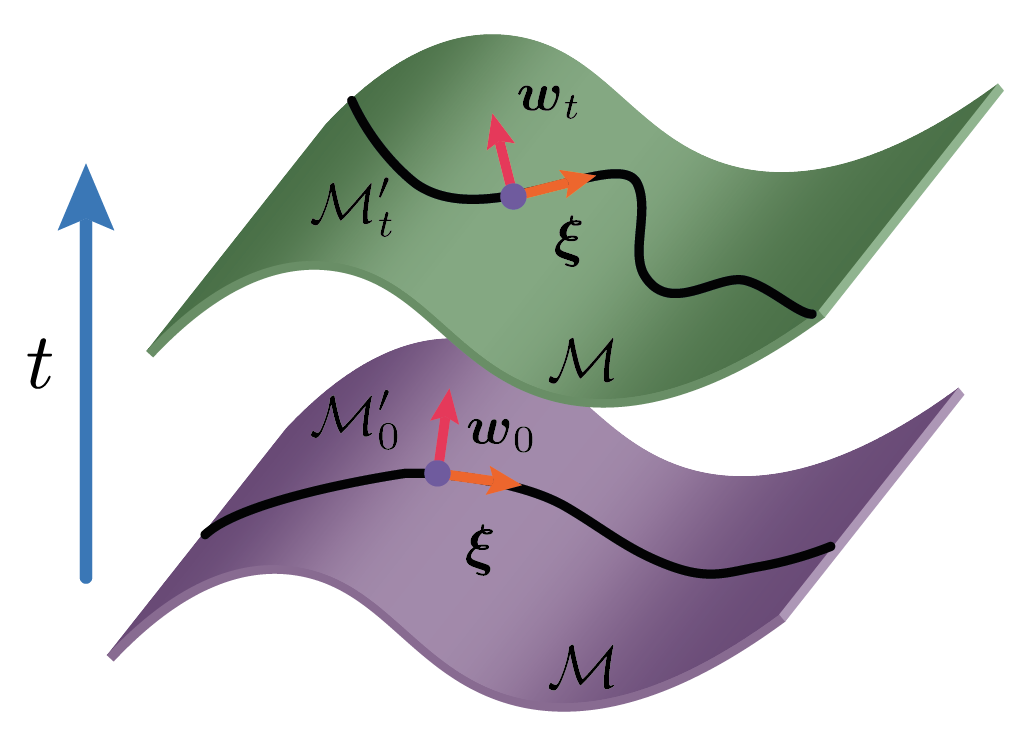}
  \caption{The manifold $\mathcal{M}$ is restricted by a time
    dependent constraint to a submanifold $\mathcal{M}'_t$ at time
    $t$. The constraint is made explicit by solving $w^I$ as functions
    of $\xi^\mu$ and $t$. Because $\omega$ is homogeneous and time
    independent, the coordinates $\xi^\mu$ can always describe
    $\mathcal{M}'_t$ for all times $t$.}
  \label{fig:Geometrical_time_evolution}
\end{figure}
Let us now show that the FJ reduction we explained in section
\ref{sec:meth-fadd-jack} is directly applicable in our context. For an
illustration of the geometric setup, see Fig. \ref{fig:Geometrical_time_evolution}. In a
nutshell, in the context of circuits $B$ is identically zero, and thus
trivially closed. In the systems of interest, as we have indicated
above, $\omega$ constructed from our algorithm is
homogeneous. If it is singular, its zero directions have associated
coordinates $w^I$, and $\omega$ is nondegenerate in directions
associated with coordinates $\xi^\mu$. In other words,
\begin{align}
  \label{eq:omegaxi}
  \omega=\frac{1}{2}f_{\mu\nu}\dd \xi^\mu\wedge\dd \xi^\nu\,,
\end{align}
with homogeneous $f_{\mu\nu}$.
The constraints are
\begin{align}
  \label{eq:rewriteconst}
  \frac{\partial}{\partial w^I}\left(H(\xi,w)+ S_\mu \xi^\mu+ S_I w^I\right)=0\,,
\end{align}
and we determine $w^I(\xi,t)$ hence, if possible. The coordinates
$\xi^\mu$ are free, and in fact are the coordinates for
$\mathcal{M}'_t$ (at each instant of time). Therefore $B=0$. It
follows that, if we can solve $w^I$ from the constraints
\eqref{eq:rewriteconst}, we can effect Hamiltonian reduction in this
case. 

\section{Examples}
\label{sec_app:examples}

In this section, we utilize our method to construct systematically  (extended) Lagrangian and Hamiltonian dynamics of circuits containing all the lumped elements introduced in the main text. In doing so, we recover standard results from flux-node~\cite{Devoret:1997,Burkard:2004} or loop-charge analysis~\cite{Ulrich:2016}, as well as more general (nonreciprocal) black-box analysis~\cite{Nigg:2012,Solgun:2015,ParraRodriguez:2019,ParraRodriguez:2022b}.
\subsection{RLC circuit}
\begin{figure}[!ht]
  \centering
  \includegraphics[scale=0.6]{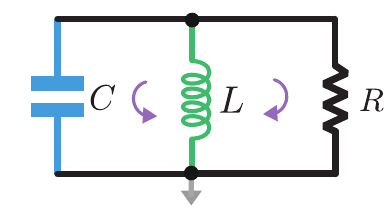}
  \caption{Parallel RLC circuit. In the classical depiction there are two loops and one active node involved in general. The resistor is considered a constraint, allowing the formulation of dynamical equations using just one set of paired conjugate variables.}
  \label{fig:RLC_Circuit}
\end{figure}
The formalism we have presented  includes the systematic  treatment of linear resistences, making use of the
Rayleigh dissipation function formalism~\cite{Mariantoni:2020}. To illustrate this point and to give a very simple example we now
consider the parallel RLC circuit, depicted in Fig. \ref{fig:RLC_Circuit}. We choose to order the three parallel branches in the sequence capacitor/inductor/resistor. The cutset matrix and resistence constraint matrix are
\begin{align}
	\msF_C&=\begin{pmatrix}
		1&1&1
	\end{pmatrix},\\
	\msF_R&=\begin{pmatrix}
		0&0&R&0&0&-1
	\end{pmatrix}
\end{align}
to construct $\msF$, with kernel 
\begin{align}
	\msK= \begin{pmatrix}1 & 0 \\ 0 & 1 \\ -1 & -1 \\ -R & -R \\ -R & -R \\ -R & -R \end{pmatrix}
\end{align}
such that we have two independent charges $\bz^T=
\begin{pmatrix}
  Q_1& Q_2
\end{pmatrix}
^T$, and the precanonical two-form is simply
\begin{align}
	\msE_{2B}= \begin{pmatrix} 0 & 0 & 0 & 1 & 0 & 0 \\ 0 & 0 & 0 & 0 & 0 & 0 \\ 0 & 0 & 0 & 0 & 0 & \frac{1}{2} \\ -1 & 0 & 0 & 0 & 0 & 0 \\ 0 & 0 & 0 & 0 & 0 & 0 \\ 0 & 0 & -\frac{1}{2} & 0 & 0 & 0 \end{pmatrix}.
\end{align}
Mapping to the reduced manifold, 
\begin{align}
	\msE=\msK^T\msE_{2B}\msK=\begin{pmatrix}
		0&-R\\R&0
	\end{pmatrix}\,
\end{align}
or, equivalently,  $\{Q_1,Q_2\}=R$.
The Hamiltonian reads 
\begin{align}
	H= \frac{Q_1^{2}}{2 C}+R^2\frac{{\left(Q_1 + Q_2 \right)}^{2}}{2 L} \,,
\end{align}
while
the dissipation function is, in these coordinates,
\begin{align}
  \label{eq:rayleighRLC}
  \mathcal{F}=\frac{R}{2}\left(\dot{Q}_1+\dot{Q}_2\right)^2\,.
\end{align}
The equations of motion, as there are no sources, are in matrix form
\begin{align}
  \label{eq:eomsmatrixRLC}
  \mathsf{E}\dot{z}= \nabla_{z} H+\nabla_{\dot{z}}\mathcal{F}\,,
\end{align}
and thus one obtains
\begin{align}
  \label{eq:badeomsRLC}
  -R\dot{Q}_2 &=
                \frac{Q_1}{C}+\frac{R^2}{L}\left(Q_1+Q_2\right)+ R \left(\dot{Q}_1+\dot{Q}_2\right)\,,\\
  R\dot{Q}_1&= \frac{R^2}{L}\left(Q_1+Q_2\right) + R \left(\dot{Q}_1+\dot{Q}_2\right)\,.  
\end{align}
\\
  To reach this point, a number of choices have been made,
  namely for the ordering of nodes and  branches, and the basis for
the kernel of $\mathsf{F}$. It should
be clear that the actual dynamics would be recovered under different
ones, as the prescription is univocal in that regard. Indeed, that is
precisely the main reason to use a Hamiltonian description, namely to
increase the set of possible variables so as to have more tools to understand
and, where possible, solve the dynamics. To make this
point more apparent, let us carry out an additional change of variables,
\begin{align}
  \label{eq:chofvarsdiagdissip}
  \begin{pmatrix}
    Q_1\\ Q_2
  \end{pmatrix}=
  \begin{pmatrix}
    1&0\\-1&-1/R
  \end{pmatrix}
             \begin{pmatrix}
               Q\\ \Phi
             \end{pmatrix}\,.
\end{align}
In terms of these new variables, we read directly the harmonic
oscillator Hamiltonian
\begin{align}
	H=\frac{Q^{2}}{2 C}+\frac{{\Phi}^{2}}{2 L},
\end{align}
the canonical commutator $\left\{\Phi,Q\right\}=1$, and the dissipation
function
\begin{align}
  \mathcal{F}= \frac{\dot{\Phi}^2}{2R}=&\,\frac{1}{2}
  \begin{pmatrix}
    \dot{Q}&\dot{\Phi}
  \end{pmatrix}
             \begin{pmatrix}
               1&-1\\0&-\frac{1}{R}
             \end{pmatrix}\times\dots\\&\dots\times
             \begin{pmatrix}
              R&R\\ R&R
            \end{pmatrix}
            \begin{pmatrix}
            1&0\\-1&-1/R
            \end{pmatrix}
             \begin{pmatrix}
               \dot{Q}\\ \dot{\Phi}
             \end{pmatrix},\nonumber
\end{align}
such that the equations of motion, Eq.~(\ref{eq:constmatrix}), are directly
computed from the final Hamiltonian, Poisson bracket, and dissipation
function,
\begin{align}
  \label{eq:directRLC}
  \dot{\Phi}&=\frac{\partial H}{\partial
              Q}+\frac{\partial\mathcal{F}}{\partial\dot{Q}}=\frac{Q}{C}\,,\\
  \dot{Q}&= -\frac{\partial H}{\partial
           \Phi}-\frac{\partial\mathcal{F}}{\partial\dot{\Phi}}=-\frac{\Phi}{L}-\frac{\dot{\Phi}}{R}\,. \nonumber
\end{align}
Naturally, we recover the elementary RLC description of a damped
harmonic oscillator, as was only to be expected. Analogous to the analysis of the circuit in Fig.~\ref{fig:NL_LC_oscillators}(a), under the assumptions made in Fig.~\ref{fig:Conjectured_M2B}, and independent of any topological assignments to the branch variables in the resistor, the final submanifold is $\mathbbm{R}^2$.

\subsection{The star circuit}
\label{sec:star_circuit}
A cutset matrix for the star circuit in
Fig.~\ref{fig:Circuit_examples}(a) with $N=4$ nodes and $B=6$ branches is
\begin{align}
	\msF_C= \begin{pmatrix}
		1 & 0 & -1 & 1 & 0 & 0 \\ -1 & 1 & 0 & 0 & 1 & 0 \\ 0 & -1 & 1 & 0 & 0 & 1
	\end{pmatrix},
\end{align}
where the first (last) three columns correspond to the capacitor (inductor) branches with kernel 
\begin{align}
	\msK=\begin{pmatrix}
		1 & 0 & 0 & 0 & 0 & 0 \\ 0 & 1 & 0 & 0 & 0 & 0 \\ 0 & 0 & 1 & 0 & 0 & 0 \\ -1 & 0 & 1 & 0 & 0 & 0 \\ 1 & -1 & 0 & 0 & 0 & 0 \\ 0 & 1 & -1 & 0 & 0 & 0 \\ 0 & 0 & 0 & 1 & -1 & 0 \\ 0 & 0 & 0 & 0 & 1 & -1 \\ 0 & 0 & 0 & -1 & 0 & 1 \\ 0 & 0 & 0 & 1 & 0 & 0 \\ 0 & 0 & 0 & 0 & 1 & 0 \\ 0 & 0 & 0 & 0 & 0 & 1 
	\end{pmatrix},
\end{align}
such that the two-form matrix representation is 
\begin{align}
	\msE=\msK^T\msE_{2B}\msK=&\begin{pmatrix}
		0 & 0 & 0 & 1 & -1 & 0 \\ 0 & 0 & 0 & 0 & 1 & -1 \\ 0 & 0 & 0 & -1 & 0 & 1 \\  -1 & 0 & 1 & 0 & 0 & 0 \\ 1 & -1 & 0 & 0 & 0 & 0 \\ 0 & 1 & -1 & 0 & 0 & 0
	\end{pmatrix}\\
	&\sim\begin{pmatrix}
		0 & 0 & 1 & 0 & 0 & 0 \\ 0 & 0 & 0 & 1 & 0 & 0 \\ -1 & 0 & 0 & 0 & 0 & 0 \\ 0 & -1 & 0 & 0 & 0 & 0 \\ 0 & 0 & 0 & 0 & 0 & 0 \\ 0 & 0 & 0 & 0 & 0 & 0
	\end{pmatrix},
\end{align}
where the second line is written in the Darboux basis under a linear (point) transformation of coordinates. The null eigenspace of $\msE$ corresponding to the voltage and current constraints in the outer loop and inner node, respectively, is expanded by
\begin{align}
	\mbf{W}_l=\begin{pmatrix}
		0\\0\\0\\1\\1\\1
	\end{pmatrix},\quad\mbf{W}_c=\begin{pmatrix}
		1\\1\\1\\0\\0\\0
	\end{pmatrix}.
\end{align}
Alternatively, we could start from the loop charges and the node fluxes depicted in Fig.~\ref{fig:Circuit_examples}(a), since we know that they suffice to express all branch charges and fluxes, and obtain the two-form
\begin{align}\label{eq:omegastargood}
    \omega &= \dd Q_1\wedge\left(\dd \Phi_1-\dd \Phi_2\right)\nonumber\\
    &\quad +\dd Q_2\wedge\left(\dd \Phi_2-\dd \Phi_3\right)\\
    &\quad +\dd Q_3\wedge\left(\dd \Phi_3-\dd \Phi_1\right)\,,\nonumber
\end{align}
with zero vectors
\begin{subequations}\label{eq:zerovectorsstar}
    \begin{align}
        \mbf{W}_l&= \frac{\partial}{\partial \Phi_1}+\frac{\partial}{\partial \Phi_2}+\frac{\partial}{\partial \Phi_3}\,,\\
        \mbf{W}_c&= \frac{\partial}{\partial Q_1}+\frac{\partial}{\partial Q_2}+\frac{\partial}{\partial Q_3}\,.
    \end{align}
\end{subequations}

The Darboux basis above corresponds to the change of variables
\begin{align}
\begin{aligned}
    Q_1 &= Q_a + w_c,&\quad \Phi_1 &= -(w_l + \Phi_b) \\
    Q_2 &= w_c,&\quad \Phi_2 &= -(w_l + \Phi_a + \Phi_b) \\
    Q_3 &= Q_b + w_c,&\quad \Phi_3 &= -w_l,
\end{aligned}
\end{align}
such that $\omega= \dd Q_a\wedge\dd \Phi_a+\dd Q_b\wedge\dd \Phi_b$. Considering, as in the main text, that all the elements are linear, the Hamiltonian in this Darboux basis $\{\xi^\mu\}\cup\{w^I\}\equiv(Q_a,Q_b,\Phi_a,\Phi_b,w_c,w_l)$ is written as 
\begin{align}
\begin{aligned}
    H=&\,\frac{{\left(Q_a + w_c	\right)}^{2}}{2 C_{1}} + \frac{w_c^{2}}{2 C_{2}} + \frac{{\left(Q_b + w_c\right)}^{2}}{2 C_{3}} \\
	&+ \frac{{\left(w_l+ \Phi_{b}\right)}^{2}}{2 L_{1}} + \frac{{\left(w_l + \Phi_a + \Phi_b\right)}^{2}}{2 L_{2}} + \frac{w_l^{2}}{2 L_{3}}.
\end{aligned}
\end{align}

We can now solve the zero mode constraints (\ref{eq:rewriteconstraint}) $w_l$ and $w_c$, and arrive to the quantizable Hamiltonian 
\begin{align}
	H&=\frac{1}{2}\left(\bQ^T\msC^{-1}\bQ+\bPhi^T\msL^{-1}\bPhi\right),\\
	\msC^{-1}&=\frac{1}{C_*^2}\begin{pmatrix}
		{C_{2} + C_{3}}& -{C_{2}}\\ -{C_{2}}& {C_{1} + C_{2}}
	\end{pmatrix},\\
	\msL^{-1}&=\frac{1}{L_*^2}\begin{pmatrix}
		{L_{1} + L_{3}}& {L_{1}}\\ 
		{L_{1}}& {L_{1} + L_{2}}
	\end{pmatrix},
\end{align}
where $C_*^2={C_{1} C_{2} + {\left(C_{1} + C_{2}\right)} C_{3}}$ and $L_*^2={L_{1} L_{2} + {\left(L_{1} + L_{2}\right)} L_{3}}$, with two canonically-quantizable pairs of coordinates $\{\Phi_\alpha,Q_\beta\}=\delta_{\alpha\beta}$, and positive-definite kinetic and potential matrices, i.e., there are two normal frequencies. Observe that there are only two dynamical degrees of freedom, even though the Kirchhoff constraints allow up to three. With a capacitive/inductive partition of the form being portrayed, one of those disappears, as there exists the nondynamical possibility of a steady current in the capacitor perimeter, together with a symmetry of global displacement of the external node fluxes. 

Even though at this point we have only presented the final result for a linear circuit, these statements hold more generally. Namely, let one capacitor, say the third one, and one inductor, say the first one, be nonlinear, with energy functions $h_c(Q_3)=h_c(Q_b+w_c)$ and $h_l(\Phi_1)=h_l(-(w_l+\Phi_b))$, see Fig.~\ref{fig:Circuit_examples}(a). Then the constraints read
\begin{align}\label{eq:nonlinstarconst}
    h_c'(Q_b+w_c)+ \frac{w_c}{C_2} +\frac{Q_a+w_c}{C_1}&=0\,,\\
    -h_l'(-(w_l+\Phi_b))+ \frac{w_l+\Phi_a+\Phi_b}{L_2}+\frac{w_l}{L_3}&=0\,.\nonumber
\end{align}
Under the conditions
\begin{align}\label{eq:invcondstar}
\begin{aligned}
    h_c''(x)+\frac{1}{C_1}+\frac{1}{C_2}&>0\qquad \mathrm{and}\\
    h_l''(-y)+\frac{1}{L_2}+\frac{1}{L_3}&>0,
\end{aligned}
\end{align}
for all values of the argument of $h_c''$ and $h_l''$, we can solve explicitly $w_c$ and $w_l$ from Eqs. \eqref{eq:nonlinstarconst}, and insert the solution in the Hamiltonian, obtaining a reduced Hamiltonian and canonical two-form, with two degrees of freedom. In fact we impose greater than zero sign in Eqs. \eqref{eq:invcondstar} for energy stability reasons, but in order to  ensure invertibility simple constancy of sign would be enough. Observe that this nonlinear example is not amenable to the Y-$\triangle$ transformation that can help in the solution of the linear one in terms of fluxes.
Even if, however, the conditions \eqref{eq:invcondstar} do not hold, the constraints \eqref{eq:nonlinstarconst} can be solved parametrically. In that case we would be led possibly to a non-homogeneous two-form, and a singularity in the form of non-homogeneous rank might arise. We shall see similar situations in  subsection~\ref{sec:nonl-sing-circ}.

Before setting aside this example, let us point out that the dual circuit, in which the capacitors are all three on the inner branches while the inductors are on the perimeter, is a good illustration of the purely charge and purely flux gauge vector case. Namely, the two-form $\omega$ for that case would be $(-1)$ times that of Eq. 
\eqref{eq:omegastargood}, and the zero vectors are exactly as in \eqref{eq:zerovectorsstar}  . However, they give rise now to gauge constraints, exactly because of the purely inductive loop and the purely capacitive cutset.

Finally, we note that the previous analysis has set aside the explicit integration of the manifold $\mcl{M}_{2B}$. We remark here that, under the assumptions in Fig.~\ref{fig:Conjectured_M2B}, the integrated submanifolds $\mcl{M}$ would be $\mathbbm{R}^6$ for both the linear case and the nonlinear case, if the nonlinear inductor and capacitor were specifically a Josephson junction and a phase-slip junction, respectively. It is evident that this result is compatible with the fact that there is neither an island nor a closed superconducting loop in either described case.
\subsection{Admittance black-box circuit}
\label{sec:Y-bb-circuit}
We now examine the circuit depicted in Fig. 3(a) of the main text.
The circuit constraints are encoded in the $\msF$ matrix in
Eq. (\ref{eq:fullPfaff}). We order the branch elements (columns) in
the following sequence: the phase-slip $E_P$ with its series inductance $L_P$, the Josephson element $E_J$, capacitors at the ports of the matrix $C_i$, left branches of the Belevitch transformer $T_{\mathrm{left,i}}$, right branches of the Belevitch transformer $T_{\mathrm{right,i}}$, the two internal inductors $L$ and the gyrator branches $G_i$. We have (in non-canonical form)
\onecolumn
\setcounter{MaxMatrixCols}{20}
\begin{align}
	\msF_C&=	\begin{pmatrix} 
		-1 & 1 & 0 & 0 & 0 & 0 & 0 & 0 & 0 & 0 & 0 & 0 & 0 \\ 
		0 & 1 & 0 & 1 & 0 & 1 & 0 & 0 & 0 & 0 & 0 & 0 & 0 \\ 
		0 & 0 & 1 & 0 & 1 & 0 & 1 & 0 & 0 & 0 & 0 & 0 & 0 \\ 
		0 & 0 & 0 & 0 & 0 & 0 & 0 & 1 & 0 & 1 & 0 & 0 & 0 \\ 
		0 & 0 & 0 & 0 & 0 & 0 & 0 & 0 & 1 & 0 & 1 & 0 & 0 \\ 
		0 & 0 & 0 & 0 & 0 & 0 & 0 & 0 & 0 & 1 & 0 & -1 & 0 \\ 
		0 & 0 & 0 & 0 & 0 & 0 & 0 & 0 & 0 & 0 & 1 & 0 & -1 
	\end{pmatrix},\\
	\msF_L&=\begin{pmatrix}
		1 & 1 & 0 & 0 & 0 & -1 & 0 & 0 & 0 & 0 & 0 & 0 & 0 \\ 0 & 0 & 1 & 0 & 0 & 0 & -1 & 0 & 0 & 0 & 0 & 0 & 0 \\ 0 & 0 & 0 & 1 & 0 & -1 & 0 & 0 & 0 & 0 & 0 & 0 & 0 \\ 0 & 0 & 0 & 0 & 1 & 0 & -1 & 0 & 0 & 0 & 0 & 0 & 0 \\ 0 & 0 & 0 & 0 & 0 & 0 & 0 & 1 & 0 & -1 & 0 & -1 & 0 \\ 0 & 0 & 0 & 0 & 0 & 0 & 0 & 0 & 1 & 0 & -1 & 0 & -1
	\end{pmatrix},
\end{align}
\twocolumn
The transformer constraints are written as
\begin{align}
	\msF_T=&\left(\begin{array}{r|r}
		\msF_{T,\mathrm{cut}} &0\\
		0&\msF_{T,\mathrm{loop}}
	\end{array}\right),\\
	\msF_{T,\mathrm{cut}}=&\begin{pmatrix}
		0&\dots&\mone&\msN&\dots&0
	\end{pmatrix},\\
	\msF_{T,\mathrm{loop}}=&\begin{pmatrix}
		0&\dots&-\msN^T&\mone&\dots&0
	\end{pmatrix},
\end{align}
with \begin{align}
		\msN=&\begin{pmatrix}
		n_{11}&n_{12}\\n_{21}&n_{22}
	\end{pmatrix},
\end{align}
the Belevitch transformer ratios, while the gyrator ones as  
\begin{align}
	\msF_G=\left(\begin{array}{rrr|rrr}
		0&\dots& \mathbbm{1} & 0&\dots &-\msY_G
	\end{array}\right),
\end{align}
with $\msY_G=(1/R) i \sigma_y$ the pure gyrator matrix. The fundamental cutset and loop matrices have not been written at this point in
a canonical form, precisely to put to good use the canonical forms of the
transformer and gyrator constraints. The local integration of the Pfaff equations with all the constraints ($\msF \dd{\bzeta}=0$) results in a complete basis for the state of the circuit composed of four charges and three fluxes, $\bz=(Q_1,w,Q_a,Q_2,\Phi_1,\Phi_a,\Phi_2)^T$. We leave the more involved analysis on the topology of this sub-manifold (global integration) based on the assumptions in Sec.~\ref{sec:quantization_superc_circuits} for future work. Explicitly, the branch variables of the energy-storing elements are expressed in terms of this basis as 
\begin{align}
    \bq=&\begin{pmatrix}
        q_P\\
        q_{L_P}\\
        q_{J}\\
        q_{C_1}\\
        q_{C_2}\\
        q_{L_1}\\
        q_{L_2}
    \end{pmatrix}=\begin{pmatrix}
Q_{1} \\
Q_{1} \\
w \\
Q_{a} \\
Q_{2} \\
\frac{(Q_{2} + w)n_{12}-(Q_{1}+ Q_{a}) n_{22}}{\Delta_{\msN}} \\
\frac{(Q_{1} + Q_{a}) n_{21}-(Q_{2}+w) n_{11}}{\Delta_{\msN}} \\
 \end{pmatrix},\\
 \bphi=&\begin{pmatrix}
        \phi_P\\
        \phi_{L_P}\\
        \phi_{J}\\
        \phi_{C_1}\\
        \phi_{C_2}\\
        \phi_{L_1}\\
        \phi_{L_2}
    \end{pmatrix}=\begin{pmatrix}
 \Phi_{1} \\
\Phi_{a} \\
\Phi_{2} \\
(\Phi_{1} + \Phi_{a})\\
\Phi_{2}\\
\Phi_{1} + \Phi_{a}\\
\Phi_{2}
\end{pmatrix},
\end{align}
with $\Delta_{\msN}=\mathrm{det}(\msN)=n_{11} n_{22}-n_{12} n_{21}$. Following the method  described here, we obtain the two form 
\begin{align}
\begin{aligned}
\omega=&\, \dd{Q}_1 \wedge \dd(\Phi_1-R/\Delta_{\msN}(Q_2+w))\\
    &+\dd{Q_a}\wedge\dd(\Phi_a+\Phi_1-R/\Delta_{\msN}(Q_2+w))\\
    &+\dd{Q_2}\wedge\dd{\Phi_2}.  
\end{aligned}  
\end{align}

To bring it to canonical form, we perform a transformation of the variables $\Phi_1$ and $\Phi_a$,
\begin{subequations}
\begin{align}
    \tilde{\Phi}_1&=\Phi_1-R/\Delta_{\msN}(Q_2+w),\\
    \tilde{\Phi}_a&=\Phi_a+\Phi_1-R/\Delta_{\msN}(Q_2+w),
\end{align}   
\end{subequations}
while the rest of the variables remain unchanged,  to obtain 
\begin{align}
\begin{aligned}
\omega=&\, \dd{Q}_1 \wedge \dd\tilde{\Phi}_1+\dd{Q_a}\wedge\dd\tilde{\Phi}_a+\dd{Q_2}\wedge\dd{\Phi_2}. 
\end{aligned}  
\end{align}
Here, the separation between dynamical conjugate pairs and the zero mode $w$ is made evident. This is indeed a gauge mode that can be fixed to an arbitrary value since it does not appear in  Eq.~(\ref{eq:H_final_Ycircuit}). Writing the energy function in terms of the natural state circuit variables, we arrive at the Hamiltonian 
\begin{align}
	H=&\frac{(\tilde{\Phi}_{1}-\tilde{\Phi}_a)^{2}}{2 L_{\mathit{PS}}} +h_P(Q_{1})+\frac{Q_{2}^{2}}{2 C_{2}} + h_J(\Phi_2)+ \frac{Q_a^2}{2 C_{1}}\nonumber\\
	&+\frac{{\left(n_{11}\tilde{\Phi}_a + n_{21}\Phi_{2} + n_{21} \frac{R}{\Delta_{\msN}}(Q_1+Q_a)\right)}^{2}}{2 L}\nonumber\\
	&+ \frac{{\left(n_{12}\tilde{\Phi}_{a} +  n_{22}\Phi_{2} + n_{22} \frac{R}{\Delta_{\msN}}(Q_1+Q_a)\right)}^{2}}{2 L},\label{eq:H_final_Ycircuit}
\end{align}
with the phase-slip and Josephson energy functions $h_P(Q_1) =-E_P \cos(2 \pi Q_1/(2e))$, and  $h_J(\Phi_{2}) =-E_J \cos(2 \pi \Phi_2/\Phi_Q)$, such that it can be recast in the form
\begin{align}
	H=&\frac{1}{2}\left(\bQ^T\msC^{-1}\bQ+\bPhi^T\msL^{-1}\bPhi+\bQ^T\msG\bPhi\right)\nonumber\\&+h_P(Q_1) + h_J(\Phi_2).\label{eq:H_Ycircuit_app}
\end{align}

\subsection{Time-dependent external flux in the SQUID}\label{sec:time-depend-extern}
\begin{figure}[!ht]
	\includegraphics[width=1\linewidth]{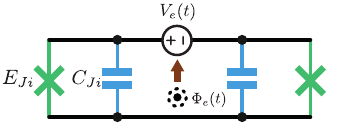}
	\caption{A SQUID loop threaded by a time dependent flux that	is replaced by a time dependent voltage source (emf).}
	\label{fig:time-dep-ext-flux-SQUID}
\end{figure}

Here we rederive the well-known result for the quantization of (lumped) superconducting loops threaded by time-dependent fluxes, see further details in~
\cite{You:2019,Riwar:2022,Bryon:2023}. We shall examine the circuit
depicted in Fig. \ref{fig:time-dep-ext-flux-SQUID}. We map a
time-dependent external flux threading only the internal loop to an
{\it emf} source $V_e(t)= \dot{\Phi}_e(t)$, as previously discussed
(section \ref{sec:time-depend-magn}). Next we carry out the by now standard constraint matrix $\mathsf{F}$/embedding matrix $\mathsf{K}$/two-form matrix $\mathsf{E}$ (or $\omega$) sequence,
\begin{align}
	\msF_C= \begin{pmatrix}
		1 & 0 & -1 & 1 & 0 & 0 \\ -1 & 1 & 0 & 0 & 1 & 0 \\ 0 & -1 & 1 & 0 & 0 & 1
	\end{pmatrix},
\end{align}
\begin{align}
	\msK=\begin{pmatrix}
		1 & 0 & 0 & 0 & 0 \\ 
		0 & 1 & 0 & 0 & 0 \\
	    0 & 0 & 1 & 0 & 0 \\ 
	    -1 & -1 & -1 & 0 & 0 \\ 
	    -1 & 0 & -1 & 0 & 0 \\ 
	    0 & 0 & 0 & 1 & 0 \\ 
	    0 & 0 & 0 & 0 & 1 \\ 
	    0 & 0 & 0 & 1 & 0 \\ 
	    0 & 0 & 0 & 0 & 1 \\ 
	    0 & 0 & 0 & 1 & -1
	\end{pmatrix}\,,
\end{align}

\begin{align}
	\msE&=\begin{pmatrix}
		0 & 0 & 0 & -1 & 0 \\ 0 & 0 & 0 & 0 & -1 \\ 0 & 0 & 0 & 0 & 0 \\ 1 & 0 & 0 & 0 & 0 \\ 0 & 1 & 0 & 0 & 0
	\end{pmatrix}\sim\begin{pmatrix}
		0 & 0 & 1 & 0 & 0 \\ 0 & 0 & 0 & 1 & 0 \\ -1 & 0 & 0 & 0 & 0 \\ 0 & -1 & 0 & 0 & 0 \\ 0 & 0 & 0 & 0 & 0
	\end{pmatrix}.
\end{align}
After this process, the total energy function reads
\begin{align}
	H_T(t)=&H+S_\alpha(t) z^{\alpha}=\frac{{\left(Q_{1} + Q_{2} + w_{1}\right)}^{2}}{2 \, C_{J2}} \nonumber\\
	&+h_1(\Phi_1) +h_2(\Phi_2)+ \frac{w_{1}^{2}}{2 \, C_{J1}}-(Q_{1} + w_{1})V_{e} \,,\label{eq:energy_f_SQUID}
\end{align}
and the constraint $\partial H_T/\partial w_1=0$ is
actually the voltage constraint around the loop,
\begin{align}
	-V_{e} + \frac{Q_{1} + Q_{2} + w_{1}}{C_{J2}} + \frac{w_{1}}{C_{J1}} = 0.
\end{align}
Solving $w_1$ from the constraint, substituting in the total energy function~\eqref{eq:energy_f_SQUID}, and making the change of variables 
\begin{align}
		\bq=\frac{1}{\sqrt{2}}\begin{pmatrix}
			1&1\\1&-1
		\end{pmatrix}\bQ=\mathsf{O}\bQ,
\end{align}
and, dually, $\bphi=\mathsf{O}\bPhi$, we derive the energy function
\begin{align}
	H_T(t)=&\frac{1}{2}(\bq^T\msC^{-1}\bq+2\bq^T\bsb{s}(t))\nonumber\\
	&+h_1(Q_1(\bq))+h_2(Q_2(\bq)),
\end{align}
with \begin{align}
	\msC^{-1}=\frac{2}{C_\Sigma}\begin{pmatrix}
		1&0\\0&0
	\end{pmatrix},
\end{align}
where $C_\Sigma=C_{J1}+C_{J2}$. Note that extra terms not contributing to the dynamics have been neglected.

At this stage, the evolution is hamiltonian with two canonical pairs, and the total energy function is the Hamiltonian. However, the equations of motion are such that the really dynamical part is of one single degree of freedom, with  effective Hamiltonian
\begin{align}
	\tilde{H}_T(t)=&\,\frac{q_1^2}{C_\Sigma}+q_1 V_e(t) \left(\frac{2\tilde{C}-1}{\sqrt{2}}\right)\nonumber\\
	&+h_1\left(\frac{\phi_1+\Phi_e/\sqrt{2}}{\sqrt{2}}\right)+h_2\left(\frac{\phi_1-\Phi_e/\sqrt{2}}{\sqrt{2}}\right),\label{eq:H_eff_SQUID}
\end{align}
with $\tilde{C}=C_{J1}/C_\Sigma$. The reader may note the equivalence of this Eq.~(\ref{eq:H_eff_SQUID}) with that in Eq.~(14) in Ref.~\cite{You:2019} under the particular choice of parameters $m_l=1/\sqrt{2}=-m_r$, and thus $m_\triangle=m_l-m_r=\sqrt{2}$, together with the correspondance of junctions $J_l\leftrightarrow J_1$ and $J_r\leftrightarrow J_2$. Explicitly, the equations for the $\{\phi_{2},q_2\}$ pair are 
\begin{align}
	\dot{\phi}_2&=\frac{V_e(t)}{\sqrt{2}},\\
	\sqrt{2}\dot{q}_2&=h'_2\left(\frac{\phi_1-\phi_2}{\sqrt{2}}\right)-h'_1\left(\frac{\phi_1+\phi_2}{\sqrt{2}}\right).
\end{align}
These equations show that these two variables are slaved to the external flux and the other pair $\{\phi_{1},q_1\}$. We can integrate the first one, and substitute $\phi_2$ in total energy function. After this substitution, there are time dependent terms in the total energy function that no effect on the dynamics of $\phi_1$ and $q_1$. We discard those, and in this manner we arrive at the effective Hamiltonian (\ref{eq:H_eff_SQUID}).

Observe, that under constant external flux, the emf $V_e$ is zero, and there is an ignorable coordinate, and therefore a conserved quantity in the total energy function (Eq.~(\ref{eq:energy_f_SQUID}) with the substitution of $w_1$). In the presence of time dependent external flux, this ignorable coordinate and conserved quantity couple become the slaved pair. We finally observe that, under the assumption for the initial manifold $\mcl{M}_{2B}$ in Sec.~\ref{sec:quantization_superc_circuits}, the final reduced submanifold $\mcl{M}_S=\{q_1,\phi_1\}$ has, as usually phenomenologically posited, an $S^1\times \mathbbm{R}$ topology.

\section{Nonlinear singular circuits}
\label{sec:nonl-sing-circ}
In this work we present a systematic method for deriving the Lagrangian and Hamiltonian dynamics of lumped element circuits. Our approach is rooted in a geometric description and incorporates a Faddeev-Jackiw analysis to uncover potential implicit constraints. In the preceding section, we offered concrete examples of circuits, both linear and nonlinear, along with explicit solutions for cases where linear constraints were in play. However, in dealing with the non-linear star circuit, we introduced the overarching challenge of devising reduced models in the presence of nonlinear constraints.

In this section, we conduct a more comprehensive examination of circuits exhibiting nonlinear singularities. These are circuits where nonlinear constraints can manifest, e.g., when non-homogeneous rank (singular) two-forms are involved. We take two recent works focused on finding quantum mechanical models for (approximately) singular circuits, readily, Refs.~\cite{Rymarz:2023} and~\cite{Miano:2023}, and scrutinize some of their findings through the lens of our approach. 

\subsection{Revisiting Rymarz-DiVincenzo's singular circuit}
\begin{figure}[!ht]
\includegraphics[width=1\linewidth]{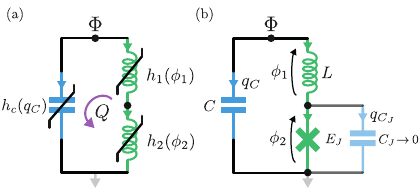}
	\caption{(a) A circuit containing two nonlinear inductive elements and a nonlinear capacitive element. Potential singularities arise in obtaining a Hamiltonian description starting with second-order Lagrangians with only flux-node/loop-charge variables. (b) A particular instance of the singular circuit in (a), studied in Ref.~\cite{Rymarz:2023}. A Josephson junction with a capacitor $C_J\ll C$ is sometimes approximated by an open circuit ($C_J\rightarrow 0$)~\cite{Rymarz:2023,Miano:2023}.}
	\label{fig:Circuit_singular_NL_L}
\end{figure}

Let us now start the analysis of the circuit in
Fig.~\ref{fig:Circuit_singular_NL_L}(a), whose constraints we already solved in
Eq.~(\ref{eq:explicitkkir}).  There we observed that the system is
generally described by one loop charge and two node fluxes. Therefore
the two-form $\omega$ is necessarily degenerate and we have to
consider either reduction or Faddeev--Jackiw increase in the number of dimensions. Notice that this system could be understood as a limit of the one in Fig.~\ref{fig:Circuit_singular_NL_L}(b) when a capacitor shunting the second (NL) inductor tends to zero.

Following the steps above, we compute the kernel of the $\msF$ matrix, where the elements are ordered such that the projectors are written as 
\begin{align}
  \mathsf{P}^{\mathcal{C}}=
  \begin{pmatrix}
    0&0&0\\0&0&0\\0&0&1
  \end{pmatrix}\qquad
  \mathrm{and}\quad
  \mathsf{P}^{\mathcal{L}}=
  \begin{pmatrix}
    1&0&0\\0&1&0\\0&0&0
    \end{pmatrix}\,,
\end{align}
which provides us with the two-form matrix  
\begin{align}
[\omega_{\alpha\beta}]=\msK^T\msE_{2B}\msK=\begin{pmatrix}
    0&1&1\\-1&0&0\\ -1&0&0
  \end{pmatrix},
\end{align}
with the ordering $Q$, $\phi_1$ and $\phi_2$. 

Observe that we could have obtained the same result writing directly the Kirchhoff's constraints in differential form 
\begin{align}
\begin{aligned}
    \dd Q&=\dd q_C=-\dd q_1=-\dd q_2\nonumber\\
    \dd \Phi&=\dd \phi_C=\dd \phi_1+\dd \phi_2,
\end{aligned}
\end{align}
to compute $\omega=\dd Q\wedge(\dd \phi_1+\dd \phi_2)$, from which one can read the matrix $\omega_{\alpha\beta}$; recall the $2$ factor in the definition of the matrix with respect to the two-form expression. The single zero mode is 
\begin{align}
  \mathbf{W}=
  \begin{pmatrix}
    0\\1\\-1
  \end{pmatrix}.
\end{align}
The corresponding constraint function, $\mathbf{W}(H)=W(\Phi_1,\Phi_2)$, is
\begin{subequations}\label{eq:cosntrfunctfig5}
  \begin{align}
    W(\phi_1,\phi_2)&=\frac{\partial H}{\partial \phi_1}-\frac{\partial H}{\partial \phi_2}\label{eq:wwithpartials}\\
    &=h_1'(\phi_1)-h_2'(\phi_2)=0\,,\label{eq:wexplicit}
  \end{align}
\end{subequations}
since the Hamiltonian in terms of the loop-charge and the node-fluxes
is $H(Q,\phi_1,\phi_2)=h_1(\phi_1)+h_2(\phi_2)+h_3(Q)$.

This constraint is a useful example to better understand a type of constraints that arise from the formalism. Namely, it is in this case simply the condition that the current through both inductors has to be the same. This contrasts with the constraints in the example of \ref{sec:star_circuit}, that reflected the fact that there can be a nondynamical current through the capacitors, for instance.

We shall use this example to illustrate obstructions to a Hamiltonian description due to the interplay of energy functions and topology. To do so we have to examine the process of reduction by the dynamical constraint \eqref{eq:wexplicit} in more detail.

From the structure of the initial manifold of branch variables and from the first reduction being just Kirchhoff, we know that the three dimensional state manifold is $\mathbb{R}^{3-k}\times\mathbbm{T}^k$, i.e. $\mathbb{R}^3$, $\mathbb{R}^2\times S^1$,  $\mathbb{R}\times\mathbbm{T}^2$ or $\mathbbm{T}^3$. We shall only consider the first two, that correspond \emph{classically} to an extended ($\mathbb{R}$) variable for the loop charge and for the branch flux of one of the inductors (the linear one in the detailed analysis that follows) and a compact ($S^1$) variable for the other branch flux (the Josephson junction in the particularization below). In all cases, because the only constraints to this point were Kirchhoff, the total manifold is a product of the charge and fluxes submanifolds.  Thus the dynamical constraint of eq. \eqref{eq:wexplicit} is an implicit definition of a  curve in the two dimensional fluxes manifold: we are systematically assuming that the branch energy functions are smooth, and therefore so is $W(\phi_1,\phi_2)$ in Eq. \eqref{eq:cosntrfunctfig5}, so by the implicit function theorem at all points except at flux energy extrema there will local expressions of one variable in terms of the other. If there is only one connected component of the constraint we can solve it parametrically, and we will have a description of the reduced two dimensional dynamical manifold. The remaining task is to determine the restricted two-form, and to check whether it is symplectic or not.

 If we take $h_1(\phi_1)=\phi_1^2/2L$ we can make a stronger assertion. As $h_1''$ never vanishes in this case, we can solve $\phi_1$ in terms of $\phi_2$, so we have an explicit expression for the curve satisfying the constraint, and $\phi_2$ can be used to describe the reduced two dimensional manifold of interest, with coordinates $(Q,\phi_2)$. Explicitly,  we can solve Eq. \eqref{eq:wexplicit} as
 \begin{align}
     \phi_1=L h'_2(\phi_2)\,,
 \end{align}
 and the reduced elements for the equations of motion are 
 \begin{subequations}\label{eq:reducedsystem}
\begin{align}
     \omega &= \left[1+L h_2''(\phi_2)\right]\,\dd Q\wedge\dd \phi_2\,,\label{eq:reduced2form}\\
     H &= h_c(Q)+ h_2(\phi_2)+\frac{L}{2}\left[h_2'(\phi_2)\right]^2\,.\label{eq:reducedham}
 \end{align}
 \end{subequations}
The equations of motion,
\begin{align}\label{eq:eomsimpler}
\left[1+L h_2''(\phi_2)\right]\dot{\phi}_2 &= h'_c(Q)\\
\dot{Q} &= -h_2'(\phi_2)\nonumber
\end{align}
are a two-dimensional autonomous dynamical system, for which an integrating factor always exists \cite{ince:1956}, and one readily computes the first integral
 \begin{align}
    E(Q,\phi_2)= h_c(Q)+ h_2(\phi_2)+\frac{L}{2} \left[h_2'(\phi_2)\right]^2\,.
 \end{align}
 The level sets of this first integral are the trajectories in phase space $(\phi_2,Q)$ of the solutions of the dynamical system \eqref{eq:eomsimpler}. Therefore, there exists a classically equivalent Hamiltonian system with canonical Poisson bracket, such that the equations of motion are 
 \begin{subequations}\label{eq:alteomsham}
     \begin{align}
         \frac{\dd \phi_2}{\dd \tau}&=\frac{\partial E}{\partial Q}= h_c'(Q)\,,\label{eq:phi2eqham}\\
         \frac{\dd Q}{\dd \tau}&= -\frac{\partial E}{\partial\phi_2}=-h_2'(\phi_2)\left[1+L h_2''(\phi_2)\right]\,.\label{eq:Qeqham2}
     \end{align}
 \end{subequations}

 We now have to take stock of this process. The strong statement is that  the classical dynamics of the system has been written in  Hamiltonian form, as  shown in Eqs. \eqref{eq:alteomsham}.  Nonetheless, there are subtleties to be considered. First and foremost, systematic reduction has taken us to Eqs. \eqref{eq:reducedham}. For $\omega$ in Eq. \eqref{eq:reduced2form} to be a symplectic form we require that it be nondegenerate. However,  it will be degenerate if there are lines for which $1+L h''_2(\phi_2)=0$. The integrating factor \emph{resolves} that classical singularity if it exists.  The manifold described by coordinates $\phi_2$ and $Q$ does admit a canonical symplectic form, and the level sets of $E(Q,\phi_2)$ are the phase space portrait. Even so, in many cases we do not simply want a Hamiltonian description, but rather a Hamiltonian description in terms of some preferred variables, and that is a completely different issue.

\subsubsection{Singularities and otherwise}
\label{sec:further}
As stated above, for the case $h_1(\phi_1)=\phi_1^2/2L$ and smoothness conditions we have a Hamiltonian description as in Eqs. \eqref{eq:alteomsham}, in terms of the loop charge $Q$ and the branch flux of the possibly nonlinear inductor $\phi_2$. However, in most cases of interest we would rather have a Hamiltonian description in terms of the total flux $\Phi=\phi_1+\phi_2$. As $\phi_1$ has been solved in terms of $\phi_2$, we have the relation $\Phi=\phi_2+L h'_2(\phi_2)$, which slaves the evolution of $\Phi$ to $\phi_2$. However, if we want a description exclusively in terms of $\Phi$ this relation has to be inverted. The invertibility condition, given the smoothness of the energy functions, is $1+L h_2''(\phi_2)\neq0$ in the range of $\phi_2$. i.e. monotony. Notice that this means that the two-form $\omega$ in Eq. \eqref{eq:reduced2form} is nondegenerate. In that case it is readily seen that eqs. \eqref{eq:eomsimpler} are equivalent to  the Hamiltonian equations
\begin{align}\label{eq:formalhamiltonian}
    \dot{\Phi} &= h_c'(Q)\,,\nonumber\\
    \dot{Q} &= -V'(\Phi)\,,\\
    V(\Phi)&= h_2\left[\phi_2(\Phi)\right]+ \frac{L}{2}\left\{h_2'\left[\phi_2(\Phi)\right]\right\}^2\,,\nonumber
\end{align}
where $V(\Phi)$ is the   energy for the equivalent inductor.
We can recognize this result from Eqs. \eqref{eq:reducedsystem}, with the observation that $\omega= \dd Q\wedge\dd \Phi$. 

Consider now what happens  if the invertibility condition is not satisfied. Formally, 
inspecting Eqs. \eqref{eq:formalhamiltonian},  we see that the evolution is Hamiltonian with conjugate $\Phi$ and $Q$, and an effective potential that has to be expressed parameterically. Much has been made of the fact that this effective potential is multibranched as a function of $\Phi$ if the invertibility condition does not hold. However, the classical motion is completely deterministic, and the phase portrait is unequivocal. Even so, the quantization of a system described parametrically deserves further investigation. We shall now give concrete examples of invertible and noninvertible situations, restricting the analysis always to the classical models.
\subsubsection{Case 1: Direct reduction with two linear inductors}
\label{sec:direct-reduction}
We
now consider that the two
inductors are linear, such that $h_i(\phi_i)=\phi^2_i/2L_i$. Then the
constraint can be solved, and we coordinatize the constrained surface as
\begin{subequations}\label{eq:linearindfig5}
  \begin{align}
  \phi_1&= \frac{L_1}{L_1+L_2}\Phi\,,\label{eq:phi1fig5}\\
  \phi_2&= \frac{L_2}{L_1+L_2}\Phi\,,\label{eq:phi2fig5}
  \end{align}
\end{subequations}
so the equations of motion for $Q$ and $\Phi$ are derived from
the first order Lagrangian
\begin{align}
  \label{eq:explicitfirstorderfinalfig5linear}
  L=Q\dot{\Phi}-h_3(Q) -\frac{\Phi^2}{2(L_1+L_2)}\,.
\end{align}
Naturally, this procedure has provided us with the
standard reduction of series linear inductors. 

To focus our mind for what follows, let us rephrase what we have done here: we have applied our procedure and obtained and \emph{effective} Hamiltonian in a reduced phase space. The standard reduction formulae for linear inductors and capacitors are exactly that, a recipe for the construction of effective energy functions in fewer variables, and flow immediately from the formalism.

\subsubsection{Case 2: Direct reduction with linear inductor and Josephson junction}
As we have stated repeatedly, the formalism we propose here encompasses the traditional approaches to circuit analysis. Thus, the dynamical constraints of the form $\mathbf{W}(H)=0$ for zero vectors of the two-form will be readily understood in many cases as flowing from current or energy conservation. The same can be said of a dual method for obtaining Hamiltonian systems,  the Dirac-Bergmann~\cite{Rymarz:2023}. Therefore the issue of invertibility of the function that expresses the total flux $\Phi$ as a function of a parameter has already been addressed in the literature \cite{Rymarz:2023,Miano:2023}. The idea of regularization to start from a non-defective quadratic Lagrangian has also been put forward in this regard \cite{Rymarz:2023}, with special emphasis on the derivation of an effective Lagrangian both under invertibility or in its absence.

For concreteness let us set the first inductor to be linear ($h_1(\phi_1)=\phi_1^2/2 L$)
and the second one a Josephson junction with energy function $h_2(\phi_{2})=-
E_J\cos(2 \pi \phi_{2}/\Phi_0)$, where $\Phi_0$ the standard quantum of flux, see Fig.~\ref{fig:Circuit_singular_NL_L}(b), where the limit of $C_J\rightarrow 0$ has been taken before the analysis.  In order to ease comparison with previous   literature, we perform a standard transformation of the flux and charge variables to phase and Cooper-pair number variables, respectively, i.e., $n=Q/2e$, and $\varphi_\alpha=2\pi \phi_\alpha/\Phi_0$ , a transformation that is symplectic if one sets $\hbar=1$. The classical Lagrangian reads now
\begin{align}
    L&=\hbar n(\dot{\varphi}_1+\dot{\varphi}_2)-H(n,\bsb{\varphi}),\label{eq:L_LJJ_singular}\\
    H(n,\bsb{\varphi})&=4E_C n^2-\frac{E_L}{2}\varphi_1^2-E_J\cos(\varphi_2),\label{eq:H_LJJ_singular}
\end{align}
with $E_C=(e)^2/2C$,  $E_L=(\Phi_0/2\pi)^2/L$, and $\bsb{\varphi}=(\varphi_1,\varphi_2)$. Applying again the zero-mode vector to the energy term, we  solve $\varphi_1$ from the constraint as
\begin{align}\label{eq:linearjosephsonsolv}
  \varphi_1= \beta \sin\left(\varphi_2\right),
\end{align}
where $\beta=L_1 E_J(2\pi/\Phi_0)^2=E_J/E_L$, always assumed positive in what follows. The total phase (at the capacitor), $\varphi=\Phi/\Phi_0=\varphi_1+\varphi_2$, is expressed after the reduction as 
\begin{align}
  \varphi=\varphi_{2}+\beta\sin\left(\varphi_2\right)\,.
\end{align}
We can only invert this function $\varphi(\varphi_2)$ if
 \begin{align}\label{eq:criticalcondition}
  \beta < 1\,.
\end{align}
In this particular case, the Lagrangian in Eq.~(\ref{eq:L_LJJ_singular}) becomes~\cite{Rymarz:2023}
\begin{align}
    L=\hbar n \dot{\varphi}-H(n,\bsb{\varphi}(\varphi)),
\end{align}
amenable to quantization, with the only caveats arising for compact $\varphi_2$, as inequivalent quantizations will appear~\cite{Egusquiza0pi:2022,galindo:2012,Reed:1975}. In other words one must then consider the possibility of a gate charge parameter.

\subsubsection{Multivaluedness and singularity of case 2}
\label{sec:multival-sing}
In spite of or because of  the issues with the regime $\beta\geq1$, that we shall now examine, it has picqued the curiosity of researchers \cite{Rymarz:2023,Miano:2023}, and the connection to a regularized model is under active discussion. Now we study this regime intrisically, i.e. without connection to a regularized circuit.  

We shall consider the two cases of $\varphi_2$ being: i) an extended variable ($\varphi_2\in\mathbb{R}$), and ii) a compact variable ($\varphi_2\in[0,2\pi)$ for definiteness). Observe that the total flux $\varphi$ inherits this extended or compact characteristic, although not its range.

We restate the system of Eqs. \eqref{eq:eomsimpler} for the aforementioned elements:
\begin{subequations}
  \label{eq:eomsafterreduction}
  \begin{align}
   \hbar\left[1+\beta\cos\left(\varphi_2\right)\right]\dot{\varphi}_2
    &= h_3'(n)=8E_C n,\label{eq:firstreducedeom}\\
    \hbar\dot{n} &= E_J\sin\left(\varphi_{2}\right).\label{eq:secondreducedeom}
  \end{align}
\end{subequations}
This system has a first integral
\begin{align}\label{eq:firstintegralbetalarg1}
E_1= 4 E_C n^2 -E_J\left[\cos\varphi_2+\frac{\beta}{4}\cos(2\varphi_2)\right]    \,.
\end{align}
This first integral is periodic in $\varphi_2$ with period $2\pi$, for the extended case. The extrema in the interval $[0,2\pi)$ are at 0, $\varphi_*$  (the solution of $1+\beta\cos\varphi_2=0$ in the interval $(\pi/2,\pi)$), $\pi$, and $2\pi-\varphi_*$, when $\beta>1$. $\varphi_2=0$ is always a minimum of the potential. However, $\varphi_2=\pi$ is  a maximum when $\beta<1$ and becomes a minimum for $\beta>1$. $\varphi_*$ and $2\pi-\varphi_*$ enter the stage only for $\beta>1$, and they are maxima in that case. The parameter $\beta$ is a classical bifurcation parameter.

For later use, we study $\varphi_*$ as a function of $\beta$. Clearly, $\varphi_*(1)=\pi$, and $\varphi_*\to\pi/2$ as $\beta\to\infty$. One readily computes
\begin{align}
    \frac{\dd \varphi_*}{\dd \beta}=\frac{-1}{\beta\sqrt{\beta^2-1}}\,.
\end{align}
Notice that  $\varphi_*$ is a monotonous decreasing function of $\beta$.
Direct computation also shows that close to $\beta=1$ one has $\varphi_*=\pi-\sqrt{2(\beta-1)}+O\left[\left(\beta-1\right)^2\right]$.   Consider now $\varphi_2$  compact with support in $[0,2\pi)$. The extrema for $\varphi(\varphi_2)=\varphi_2+\beta\sin(\varphi_2)$ will be reached either at the endpoints of the interval for $\varphi_2$ or at the solutions of $1+\beta\cos(\varphi_2)=0$, if they exist. Thus, for $\beta<1$ the range of $\varphi$ will also be $(0,2\pi)$. On the other hand, for $\beta\geq1$ we have to assess $\varphi(\varphi_*)=\varphi_*+\sqrt{\beta^2-1}$ and $\varphi(2\pi-\varphi_*)=2\pi-\varphi_*-\sqrt{\beta^2-1}$. For values of $\beta$ slightly larger than 1, the range of $\varphi$ will be again $(0,2\pi)$, until a critical value $\beta_c\approx4.6033$ is reached, such that $2\pi=\varphi_*(\beta_c)+\sqrt{\beta_c^2-1}$. From that point on, the range of $\varphi$ will be $(\varphi(2\pi-\varphi_*),\varphi(\varphi_*))$. Asymptotically, $(-\beta,\beta)$ for large $\beta$.

Let us examine qualitatively the system from the perspective of the total phase $\varphi$. The current is always bounded, by Eq. \eqref{eq:secondreducedeom}, and as $\dot\varphi= 8E_C n$ the evolution is not singular. The issue is that if $\varphi_2$ reaches $\varphi_*$ or $2\pi-\varphi_*$ (modulo $2\pi$ in the extended case) the velocity of $\varphi_2$ is required to be infinite, to compensate that $\dot{\varphi}=8 E_C n/\hbar$ is finite. This corresponds to  infinite voltage drops in both inductor branches, even though the total voltage drop is finite. We can describe the motion  with a  parametric expression for potential energy and  the total phase, 
\begin{align}
    \varphi&=\sigma+ \beta\sin\sigma\,,\nonumber\\
    V&= -E_J\left[\cos\sigma+\frac{\beta}{4}\cos(2\sigma)\right]\,.
\end{align}
 There has been previous discussion about this kind of system and its quantization (for instance by Shapere and Wilczeck \cite{Shapere:2012}), and we defer to future work this aspect of the model.

The parametric presentation of the potential allows a qualitative analysis for small and large $\beta$, that will guide that quantum analysis. Small $\beta$ corresponds to the inductance of the linear inductor going to 0, which corresponds to there being no phase drop, and the system is simply the capacitor and the nonlinear inductor. This physical description matches the limit of the effective potential. Large $\beta$, physically, means that the linear inductor has no current flowing through it, so the system is no longer dynamical, and all the energy will be the electrostatic energy of the capacitor. I.e., the physical expectation is that the effective potential will be flat in that limit. Consider the case of compact $\varphi_2$. Then so is $\varphi$, with range (in the large $\beta$ case) $(\varphi(2\pi-\varphi_*),\varphi(\varphi_*))$. Therefore $\varphi/\beta$ is uniformly bounded as $\beta\to\infty$, $|\varphi|/\beta<1+\pi/\beta$. Inversion is not possible in the full range. However, given a large fixed $\Phi$, for all $|\varphi|<\Phi$ , as $\beta\to\infty$, inversion is possible, $\sigma=\varphi/\beta+ O(\beta^{-2})$. It follows that 
\begin{align}
    V\approx -E_J\left[1+\frac{\beta}{4}-\frac{1}{2}\frac{\varphi^2}{\beta}+ O(\beta^{-2})\right]\,,
\end{align}
effectively flat as expected, for a fixed range $(-\Phi,\Phi)$. 
For extended $\varphi_2$, the argument is slightly different. There will exist an integer $N$ such that $\sigma=2\pi N+\zeta$, with $\zeta$ in $[0,2\pi)$. Then $\zeta$ will be small, $\zeta= \left(\varphi-2\pi/N\right)/\beta+O(\beta^{-2})$, and applying the same argument as before we see again that $V$ flattens out as $\beta\to\infty$, matching the physical argument.

It is important to observe that in the large $\beta$ limit the variable $\varphi$ effectively decompactifies, if it started out with a finite range, at the same time that its potential flattens out away from the boundaries. Therefore, when it comes to quantization, where the behaviour of compact variables is radically different from extended ones, we will in both cases have an approximate continuum spectrum.  This argument can be presented in an alternative way: in this large $\beta$ limit, all the energy is in the capacitor, due to its charge, and that charge need no longer be discretized.


\subsubsection{Regularized problem of case 2}

In order to have a better handle of the quantization of these circuits, Rymarz and DiVincenzo \cite{Rymarz:2023} start their analysis with the regularized classical circuit Hamiltonian (obtained from  Fig.~\ref{fig:Circuit_singular_NL_L}(b) with finite Josephson capacitance $C_J$). After  canonical quantization, they carry out a Born--Oppenheimer analysis as $C_J\rightarrow 0$ with the expectation that there will be a clear separation of scales.

Using the same redefinitions as above and the notation in the corresponding figure, and following our method (observe that here the standard flux-node analysis~\cite{Devoret:1997} provides us with the same result), one obtains the Hamiltonian  
\begin{align}\label{eq:reghammartin}
    H=4 E_C(n_C^2+r n_{C_J}^2)+\frac{E_J}{2\beta} (\varphi_C-\varphi_2)^2-E_J\cos{\varphi_2},
\end{align}
with the additional parameter $r=C/C_J$, as well as the canonical brackets $\{\varphi_C,n_C\}=\{\varphi_2,n_{C_J}\}=1$. Recall the definition $\beta=E_J/E_L$. The main goal in the analysis of \cite{Rymarz:2023}  is its quantization and separation of scales. Here we shall just analyze it classically, including adiabatic expansions, i.e. classical separation of scales.
\begin{table*}[]
    \centering
    
    \begin{tabular}{|c|c|c|}\hline
         & $0<r<\infty$ &$r\rightarrow\infty$\\
           \hline

         $0<\beta<\infty$&2 d.o.f., $H$ in Eq. \eqref{eq:reghammartin}& 1 d.o.f. $H$ in \eqref{eq:H_LJJ_singular}\\
           \hline
$\beta\rightarrow0$&1 d.o.f. ($H=4 E_C r n_a^2/(r+1) -E_J\cos\varphi_a$)& 1. d.o.f ($H=4 E_C  n_a^2 -E_J\cos\varphi_a$)\\
           \hline
$\beta\rightarrow\infty$&2 d.o.f. (1 free) ($H=4 E_Cn_C^2+\left[E_Cr n_{C_J}^2-E_J\cos{\varphi_2}\right]$)& 1 (free) d.o.f. ($H=4 E_Cn_C^2$)\\
\hline
    \end{tabular}
    \caption{Summary table for the different classical Hamiltonian models of circuit in Fig.~\ref{fig:Circuit_singular_NL_L}(b). We stress the number of effective degrees of freedom, and the appearance of a free particle.}
    \label{tab:my_label}
\end{table*}
In particular, we want to address the exchangeability of the limits $\beta\to\infty$ or $\beta\to0$, on one hand, and $r\to\infty$ on the other. Physically, as $\beta\to0$ the argument above holds: there will be no phase drop in the strict limit through the linear inductor. Therefore the circuit will be equivalent to that of a JJ with a parallel total equivalent capacitance $\tilde{C}=C+C_J$, independent of the value of $C_J$. Let us now check this intuition against the equations of motion, after a symplectic change of variables
\begin{align}\label{eq:chagneofvarslimiting}
    \begin{aligned}
        n_C&= \frac{1}{2} n_a+ n_b,\qquad
        n_{C_J}= \frac{1}{2}n_a-n_b,\\
        \varphi_C&= \varphi_a+\frac{1}{2} \varphi_b,\qquad
        \varphi_2= \varphi_a -\frac{1}{2}\varphi_b,
     \end{aligned}
\end{align}
with obvious notation, $\varphi_a$ and $n_a$ being a conjugate pair and $\varphi_b$ and $n_b$ a second one,
such that the mass matrix is no longer diagonal, 
\begin{align}\label{eq:otherhamiltonianforslow}
    H=&\, E_C\left[4(1+r)n_b^2 + 4(1-r) n_a n_b +(1+r)n_a^2\right]\nonumber\\
    &+ \frac{E_J}{2\beta}\varphi_b^2 - E_J\cos\left(\varphi_a-\varphi_b/2\right)\,.
\end{align}
This means that the only equation of motion in which the  $\beta$ parameter appears is $\dot{n}_b=-\partial H/\partial\varphi_b$, and we conclude that
\begin{align}
    \varphi_b =\beta\left[\frac{1}{2}\sin\left(\varphi_a-\frac{\varphi_b}{2}\right)-\frac{1}{E_J}\dot{n}_b\right]\,.
\end{align}
Adiabatic elimination to first order is therefore $\varphi_b=0$, and for consistency $\dot{\varphi}_b=\partial H/\partial n_b=0$ that fixes $n_b=(r-1)n_a/2(r+1)$, providing us with a \emph{reduced} system, i.e. of one fewer degree of freedom, with total charge number $n$ conjugate to the phase of the JJ, and potential energy that of the JJ. Explicitly, $H=4 E_C r n_a^2/(r+1) -E_J\cos\varphi_a$, which corresponds to the equivalent capacitor with total capacity $C+ C_J$. Furthermore, notice that this system of equations is amenable to a systematic expansion in $\beta$, giving rise perturbatively to a reduced dynamics. Explicitly, in a $\beta$ expansion of the form $\varphi_j=\sum_{m=0}^\infty \beta^m \varphi_j^{(m)}$, with $j$ being either $a$ or $b$, and correspondingly for $n_j$, order by order $\varphi_b^{(m)}$ and $n_b^{(m)}$ are completely determined by $\left\{\varphi_a^{(m')}\right\}$ and $\left\{n_a^{(m')}\right\}$, with $m'\leq m$. This result is independent of the value of $r$, so the limit $r\to\infty$ is immediate.

Now we start from the limit $r\to\infty$ first with Hamiltonian \eqref{eq:reghammartin}. From the equations of motion we see that $n_2$ has to be zero. Rather that direct substitution in the Hamiltonian, one should stick to the equations of motion to carry out this approximation. Thus $n_2=0$ demands for consistency $\dot{n}_2=0$, whence $\varphi_C=\varphi_2+\beta\sin\varphi_2$, and the approximate equations of motion are exactly the system we analyzed above. Again the limit is singular in that there is only one remaining degree of freedom. It has to be pointed out that this appears to be in contradiction with the result in \cite{Rymarz:2023}, where it was suggested that in the quantum mechanical context the effective potential in the limit $r\to\infty$ is always zero. Clearly the issue deserves further investigation, that we defer to future work.

Finally we check again the limit $\beta\to\infty$, now in this regularized model. The physical intuition is as above, that there will be no current through the linear inductor. Consequently, we shall have electrostatic energy in the $C$ capacitor, and a dynamical system composed exclusively of the JJ and the regularization capacitor $C_J$. This intuition is read off directly from the equations of motion, in fact, since one sees the requirement for $\dot{n}_C=0$ in the limit, separating off the $n_2$ and $\varphi_2$ variables. However there are two degrees of freedom, one of which is a free particle.

As $\beta\to\infty$  gives us two degrees of freedom and $r\to\infty$ one, we now check whether these limits commute. Starting from the $\beta\to\infty$ limit Hamiltonian, $H=4 E_Cn_C^2+\left[E_Cr n_{C_J}^2-E_J\cos{\varphi_2}\right]$, and considering $r\to\infty$, $n_{{C_J}}$ must be frozen to zero and $\varphi_2$ has no dynamics, thus having an effective Hamiltonian $H=4E_C n_C^2$. Interchanging the order, we begin with $H=4 E_C n^2-E_J\varphi_1^2/2\beta-E_J\cos\varphi_2$ and take $\beta\to\infty$, as we analyzed above, with the effective potential flattening out and leaving us again with one free particle.

In these classical analyses the issue of compactness or otherwise of the variables has not played a crucial role. On the other hand, were we to fully address quantization (which we defer to later work), it would be crucial to have this aspect in mind. For instance, even if the effective potential for a compact variable were flat, the corresponding spectrum would be discrete. On the other hand, in a system with two degrees of freedom with an extended variable and a bounded potential in some direction that partakes of this extended aspect, the spectrum will be continuous. 
The models we have been looking at here are debated currently as extremely idealized ones for simple circuits in which a superconducting island appears.  In this situation, the assumption of a compact phase  ($\varphi_2\in S^1$) matches the experimental data, e.g., the discrete spectra, on \emph{charge qubits} in any coupling regime, including the Cooper-Pair Box ($E_J\sim E_C$) and \emph{transmon} regimes ($E_J\gg E_C$). Thus, even deferring to future work the detailed study of quantization under regularization, we point out that the classical analysis we put forward here, together with general considerations on the quantum models that flow from it, is in accordance with experiments~\cite{Nakamura:1999,Schreier:2008}.

\subsection{Singular multi-junction branches}
\label{sec:miano}
In this section we will apply our method to reproduce, contextualize, and expand on the results presented in Ref.~\cite{Miano:2023}, as an example of the versatility of this formalism. The objective of Miano et al. is to provide a 
method to identify minima in a superconducting circuit context and thus allow for an expansion around those minima. Miano et al. consider a very particular class of circuits, in which there exists a loop of Josephson junctions and inductors. They want to study  low energy physics, in the sense that they want to characterize the minima of the potential energy, that is a function of the node fluxes, as they assume that a node-flux description of the larger circuit is not defective. 
To do so, they ask what current intensity is being injected into the inductive loop at each node between inductive elements in a general state, and then they observe that in a minimum of inductive energy, and thus in a possible steady state, this current injection must be zero. They arrive to that result by a Tellegen like argument. They next study the equivalent inductive element that would describe the same low energy phenomena, if the adjoining capacitors are negligible.   

\begin{figure}[h]
\centering
\includegraphics[width=.85\linewidth]{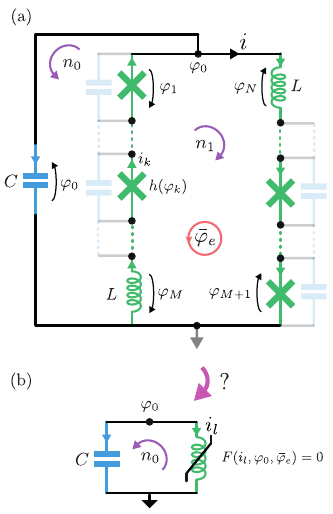}
	\caption{A nonlinear dipole, investigated in Ref.~\cite{Miano:2023}, is constructed using a series of Josephson junctions and linear inductors. A persistent inquiry in the literature revolves around the challenge of mapping the response of a multi-dipole problem to a single element.}
    \label{fig:NL_inductors_chain}
\end{figure}

\subsubsection{Effective Hamiltonian}\label{sec:effectivemiano}
The main objective of Miano et al. is the study of the effective Hamiltonian for a configuration included in those presented in Fig. \ref{fig:NL_inductors_chain}. We shall compute that effective Hamiltonian straightforwardly using our method. In particular, we will make use of the differential form presentation rather than the equivalent matricial one we have insisted on above, as it is more convenient for the treatment of general models.  We intend this material  to be pedagogical, so we shall be very explicit in the construction. It should be pointed out that, as the system consists purely of ideal reactances, the result can be derived as a direct application of Kirchhoff's laws. However, we contend that the generality and systematics of our proposal are useful in identifying correctly the relevant quantities. In particular, the goal of \cite{Miano:2023} is to evidence that the full system is well described by 
\begin{align}
    H= 4 E_C n_0^2+ U_{\mathrm{eff}}(\varphi_0,\varphi_e)\,,
\end{align}
where $\varphi_0$ is conjugate to $n_0$ and dynamical, and $\varphi_e$ a (constant) external flux controls the inner current.

In a nutshell, we have three effective branches, the capacitor one and two inductive series, that we shall refer to respectively as the outer and inner edges of the inductive loop. We shall use $n_0$ and $n_1$ charge number  and $\left\{\varphi_k\right\}_{k=0}^{N+1}$ branch phase variables. From the figure, we see that $\varphi_0$ is also the top node phase. The energy function is simply
\begin{align}
    H\left(n_0,\left\{\varphi_k\right\}_{k=0}^{N+1}\right) = h_c(n_0)+\sum_{k=1}^{N+1}h_k(\varphi_k)\,.
\end{align}
Prima facie, it seems that neither $\varphi_0$ nor $\varphi_e$ play a role. However, we have yet to determine all the constraints. To do so, we observe that the KVL forces us to have 
\begin{subequations}\label{eq:KVLMiano}
    \begin{align}
\dd \varphi_e&=\sum_{k=1}^{N+1}\dd \varphi_k\,,\label{eq:phieMiano}\\
    \dd \varphi_0 &= -\sum_{k=1}^M\dd \varphi_k\label{eq:firstphi0miano}\\
    &= -\dd \varphi_e+ \sum_{k=M+1}^{N+1}\dd \varphi_k\,.\label{eq:secondphi0miano}
\end{align}
\end{subequations}
We shall make use of these in the two form, that, following our general constructions, reads
\begin{subequations}\label{eq:Mianotwoform}
    \begin{align}
    \omega &= \frac{1}{2}\dd n_0\wedge\dd \varphi_0\label{eq:capacitive2formMiano}\\
    &\,+\frac{1}{2}\sum_{k=M+1}^{N+1}\dd \varphi_k\wedge\dd  n_1\label{eq:outerinductive2formMiano}\\
    &\,+\frac{1}{2}\sum_{k=1}^{M}\dd \varphi_k\wedge\left(\dd n_0+\dd n_1\right)\label{eq:innerinductive2formMiano}\\
    &=\dd n_0\wedge\dd \varphi_0+\frac{1}{2}\dd \varphi_e\wedge\dd n_1\,.
\end{align}
\end{subequations}
For simplicity we have set $\hbar=1$. The first line \eqref{eq:capacitive2formMiano} corresponds to the capacitive branch, the second one \eqref{eq:outerinductive2formMiano} to the outer inductive branch, and finally the third one \eqref{eq:innerinductive2formMiano} to the inner inductive branch. 

As was to be expected, there can be only two degrees of freedom, and that just if $\varphi_e$ is time dependent. Otherwise only $n_0$ and its conjugate will be dynamical, and we have to carry out symplectic reduction to make that fact explicit.
For definiteness we shall only consider fixed $\varphi_e$, so we have the simple two-form $\omega=\dd n_0\wedge\dd \varphi_0$.

It is clear that this two form is defective: we have $N+2$ branches in total, so before imposing Kirchhoff's laws we had $2N+4$ branch variables, and after imposing Kirchhoff these must be reduced to $N+2$. These are in fact the  the two charges and $N$ phases. One of the charges, $n_1$, is directly eliminated, as it appears neither in the two-form nor in the Hamiltonian, so the associated zero vector $\partial/\partial n_1$ is gauge. In other words, there can exist a nondynamical current around the inductive loop. If $\varphi_e$ were time-dependent it would drive this current $\dot{n}_1$. We fix it to zero, with constant $\varphi_e$.

Therefore, we are left with one charge, $n_0$, and $N$ phases. However the two form, $\omega=\dd n_0\wedge\dd \varphi_0$, is obviously rank two. Therefore there must exist $N-1$ independent zero vectors. Now, because of Eq. \eqref{eq:firstphi0miano} we have
\begin{align}
    \left\langle\dd \varphi_0,\frac{\partial}{\partial\varphi_k}\right\rangle=-1
\end{align}
for $k\in\left\{1,\ldots,M\right\}$, and from Eq. \eqref{eq:secondphi0miano}, together with $\dd \varphi_e=0$,
\begin{align}
    \left\langle\dd \varphi_0,\frac{\partial}{\partial\varphi_k}\right\rangle=1
\end{align}
for $k\in\{M+1,\ldots,N+1\}$.
Therefore we have 
\begin{align}
    \mathbf{W}_k=\frac{\partial}{\partial\varphi_{k+1}}-\frac{\partial}{\partial\varphi_{k}}
\end{align}
with $k$ from $1$ to $M-1$ and from $M+1$ to $N$ are zero vectors. They are in fact independent, and they number a total of $M-1+\left(N-(M+1)+1\right)=N-1$. We have a complete independent set.

These zero vectors provide us with the consistency constraints, $\mathbf{W}_k(H)=0$, which are in fact current conservation constraints in this case, here
\begin{align}\label{eq:currentconstraintsMiano}
    h'_{k+1}(\varphi_{k+1})=h'_k(\varphi_k)
\end{align} 
for $k\in\left\{1,\ldots,M-1\right\}\cup \{M+1,\ldots,N\}$. Essentially, for $k=0$ and $k=M$ there can be current extraction or injection, and therefore those nodes are missing from the sequence of Eq. \eqref{eq:currentconstraintsMiano}.

Putting everything together, we have an expression for the inductive energy in terms of the $N+1$ inductive branch phases, on the one hand, 2 KVL constraint equations in which the additional $\varphi_0$ and $\varphi_e$ are introduced (raising the number of total phase variables to $N+3$), and an additional $N-1$ current conservation constraints. All in all, $N+3$ phases and $1+2+N-1=N+2$ relations, which, after elimination of the $N+1$ loop inductive branch phases, leaves us with an expression of the inductive energy in terms of $\varphi_0$ and $\varphi_e$. In other words, this system will always present with an inductive energy depending on those variables. The only caveat to provide an explicit form is that it might well be the case, as we will now show, that it is not possible to reexpress all $\varphi_k$ from 1 to $N+1$ in terms of $\varphi_0$ and $\varphi_e$. At any rate, if not we would still have the inductive energy as a function of $\varphi_0$ and $\varphi_e$, albeit in  implicit form.
\subsubsection{Example}
Let us apply the previous computation to the case where all the inductive elements, in both the inner and the outer edges of the loop, are either linear inductances or JJ inductances. I.e. where all the inductive energies $h_k$ are of either the form
\begin{align}
    h_k(\varphi_k)= \frac{E_k}{2}\varphi_k^2
\end{align}
or
\begin{align}
    h_k(\varphi_k)=- E_k\cos(\varphi_k)\,.
\end{align}
If all the inductors in one of the edges of the loop are linear, they can all be parameterized by $\varphi_{\mathrm{linedge}}= \sum_{k\in\mathrm{edge}}\varphi_k$ as $\varphi_k= E_{\mathrm{linedge}}\varphi/E_k$, with $E_{\mathrm{linedge}}=\left(\sum_{k\in\mathrm{edge}}E_k^{-1}\right)^{-1}$, and we have the standard equivalent inductor of the linear case. Let us therefore assume that there is at least one JJ inductance in the edge. Without loss of generality, let the JJ inductance with the \emph{smallest} $E$ coefficient amongst the JJ inductances be the first one of the edge of the loop. We shall first take the inner edge for definiteness, so $E_1$ is the smallest energy coefficient of all the JJ inductances on that loop. Define $\beta_k=E_1/E_k$ for all inductors on the inner edge. Then the constraint equations \eqref{eq:currentconstraintsMiano} are either
\begin{align}
\varphi_k = \beta_k\sin\varphi_1\,,
\end{align}
for the linear inductances of the edge, or
\begin{align}\label{eq:JJparameter}
    \varphi_k=\arcsin{\left(\beta_k\sin\varphi_1\right)}
\end{align}
for the JJ inductances. Observe that for all JJ inductances $\beta_k\leq1$ by construction, so Eq. \eqref{eq:JJparameter} can be followed consistently. At any rate, we have an explicit parameterization of the solution of the constraint equations, $\varphi_k=f_k(\varphi_1)$, for the inner edges. In exactly the same manner we also have $\varphi_k=f_k(\varphi_{M+1})$ for the outer edge.

Now we express the final variables $\varphi_0$ and $\varphi_e$ in terms of the two parameters $\varphi_1$ and $\varphi_{M+1}$,
\begin{subequations}\label{eq:phi0andphiefromparam}
    \begin{align}
        \varphi_0 &= -\sum_{k=1}^M f_k(\varphi_1)\,,\label{eq:phi0fromparam}\\
        \varphi_e &= -\varphi_0+\sum_{k=M+1}^{N+1} f_k(\varphi_{M+1})\,.\label{eq:phiefromparam}
    \end{align}
\end{subequations}
The relevant question at this point is whether these relations can be inverted, so that we can express $\varphi_1= g_1(\varphi_0)$ and $\varphi_{M+1}= g_2(\varphi_0+\varphi_e)$. As all $f_k$ functions are smooth, the question is whether the sums in eqs. \eqref{eq:phi0andphiefromparam} are monotonous functions of $\varphi_1$ and $\varphi_{M+1}$. In other words, we have to investigate whether the equations
\begin{align}
    \sum_{k=1}^M f_k'(\sigma_1) &=0\qquad \mathrm{and}\nonumber\\
    \sum_{k=M+1}^{N+1} f_k'(\sigma_2) &=0\nonumber
\end{align}
have roots. We defer to Ref. \cite{Miano:2023} for an analysis of some examples.
\subsubsection{Stationary state and inductive loop.}\label{sec:mianofirstresult}
In the previous \ref{sec:effectivemiano} we have addressed the construction of the principal object of interest in \cite{Miano:2023}, the effective potential of an inductive dipole. Another aspect that Miano et al addressed was the identification of minima for expansions. 
Let us give a direct proof of their first result at this point, concerning the steady state of an inductive loop connected to a network through capacitive elements. We use an index $k$ to identify individual inductive elements in the loop, and denote branch currents (in fact, \emph{number} currents, as we are normalizing charges with the Cooper pair charge, but we maintain the notation of current intensity) and branch phases as $i_k$ and $\varphi_k$. The energy function for each inductive element is $h_k(\varphi_k)$. The total inductive energy of the loop is the sum of these energy functions, $U_{\mathrm{ind}}=\sum_k h_k$.  In fact, a reduction of variables due to the Kirchhoff constraints in the circuits they consider is equivalent to considering node phases. We number the inductive elements sequentially, and choose the numbering of the nodes correspondingly, such that the $k$-th element connects the $k-1$-th node to the $k$-th node. The $0$th node is the reference node. Thus the branch phase of the $k$-th element, $\varphi_k$, equals the difference of the $k$-th and the $k-1$-th node phases, $\varphi_k=\phi_k-\phi_{k-1}$. The inductive energy of the loop is presented as a function of the node phases, $U_{\mathrm{ind}}(\boldsymbol{\phi})$.

The constitutive equation for the inductive elements is
\begin{align}
  \label{eq:const}
  {i}_k=\frac{\partial h_k}{\partial {\varphi}_k}\,. 
\end{align}
Now, by construction, ${\varphi}_k=\phi_k-\phi_{k-1}$, so
\begin{align}
  \label{eq:chofvar}
  \frac{\partial}{\partial\phi_k}=-\frac{\partial}{\partial{\varphi}_k}+\frac{\partial}{\partial{\varphi}_{k+1}}\,.
\end{align}
Therefore,
\begin{align}
  \label{eq:ik}
  \frac{\partial U_{\mathrm{ind}}}{\partial\phi_k}= -{i}_{k+1}+{i}_{k}=j_k\,,
\end{align}
where we have introduced the symbol $j_k$ for the current injected at the $k$-th node. This result (Eq. (5) of Ref.~\cite{Miano:2023}) holds
for any dynamical situation, be it a steady state or not. Even more, this holds even if the inductive loop is connected to other elements, not solely capacitors providing one loop per inductive element, as in the description by Miano et al. For instance, this holds true even if the node-flux Lagrangian is defective. The derivation here  flows from the Kirchhoff constraints. In Ref.~\cite{Miano:2023} it was obtained from a Tellegen type argument. As Tellegen's global property flows from the Kirchhoff laws, and, furthermore, any one of the Kirchhoff laws together with Tellegen's property implies the other one, these derivations are equivalent. 

The next statement in the search of minima that are suitable for expansion and posterior quantization is that, in a classical stationary state, $\nabla_\phi
U_{\mathrm{ind}}(\boldsymbol{\phi})=0$. This is not as general as the previous result. For this to hold, the only inductive elements connected to the nodes of interest must be those in the loop. Under this condition, the  next result by Miano et al. is that in a stationary
state  the current in all  branches of a purely inductive loop
has to be the same. In other words, for all elements $j$, $k$ in the inductive loop in a stationary state we must have
\begin{align} \label{eq:Mianostatcondition}
    h'_k(\varphi_k)=h'_j(\varphi_j)\,.
\end{align}
This more concrete result is the starting point in the analysis of Miano et al to look at an effective inductive dipole, along the lines of subsection  \ref{sec:effectivemiano}. Notice that \eqref{eq:Mianostatcondition} are in direct correspondence with \eqref{eq:currentconstraintsMiano}.

\bibliographystyle{apsrev4-2}
\bibliography{bibliography}

\begin{thebibliography}{141}%
\makeatletter
\providecommand \@ifxundefined [1]{%
 \@ifx{#1\undefined}
}%
\providecommand \@ifnum [1]{%
 \ifnum #1\expandafter \@firstoftwo
 \else \expandafter \@secondoftwo
 \fi
}%
\providecommand \@ifx [1]{%
 \ifx #1\expandafter \@firstoftwo
 \else \expandafter \@secondoftwo
 \fi
}%
\providecommand \natexlab [1]{#1}%
\providecommand \enquote  [1]{``#1''}%
\providecommand \bibnamefont  [1]{#1}%
\providecommand \bibfnamefont [1]{#1}%
\providecommand \citenamefont [1]{#1}%
\providecommand \href@noop [0]{\@secondoftwo}%
\providecommand \href [0]{\begingroup \@sanitize@url \@href}%
\providecommand \@href[1]{\@@startlink{#1}\@@href}%
\providecommand \@@href[1]{\endgroup#1\@@endlink}%
\providecommand \@sanitize@url [0]{\catcode `\\12\catcode `\$12\catcode `\&12\catcode `\#12\catcode `\^12\catcode `\_12\catcode `\%12\relax}%
\providecommand \@@startlink[1]{}%
\providecommand \@@endlink[0]{}%
\providecommand \url  [0]{\begingroup\@sanitize@url \@url }%
\providecommand \@url [1]{\endgroup\@href {#1}{\urlprefix }}%
\providecommand \urlprefix  [0]{URL }%
\providecommand \Eprint [0]{\href }%
\providecommand \doibase [0]{https://doi.org/}%
\providecommand \selectlanguage [0]{\@gobble}%
\providecommand \bibinfo  [0]{\@secondoftwo}%
\providecommand \bibfield  [0]{\@secondoftwo}%
\providecommand \translation [1]{[#1]}%
\providecommand \BibitemOpen [0]{}%
\providecommand \bibitemStop [0]{}%
\providecommand \bibitemNoStop [0]{.\EOS\space}%
\providecommand \EOS [0]{\spacefactor3000\relax}%
\providecommand \BibitemShut  [1]{\csname bibitem#1\endcsname}%
\let\auto@bib@innerbib\@empty
\bibitem [{\citenamefont {Jackson}(1999)}]{Jackson:1999}%
  \BibitemOpen
  \bibfield  {author} {\bibinfo {author} {\bibfnamefont {J.~D.}\ \bibnamefont {Jackson}},\ }\href {https://doi.org/10.1002/3527600434.eap109} {\emph {\bibinfo {title} {Classical electrodynamics}}},\ \bibinfo {edition} {3rd}\ ed.\ (\bibinfo  {publisher} {Wiley},\ \bibinfo {address} {New York},\ \bibinfo {year} {1999})\BibitemShut {NoStop}%
\bibitem [{\citenamefont {Feynman}\ \emph {et~al.}(2010)\citenamefont {Feynman}, \citenamefont {Leighton},\ and\ \citenamefont {Sands}}]{Feynman:2010}%
  \BibitemOpen
  \bibfield  {author} {\bibinfo {author} {\bibfnamefont {R.}~\bibnamefont {Feynman}}, \bibinfo {author} {\bibfnamefont {R.}~\bibnamefont {Leighton}},\ and\ \bibinfo {author} {\bibfnamefont {M.}~\bibnamefont {Sands}},\ }\href {https://www.feynmanlectures.caltech.edu/II_toc.html} {\emph {\bibinfo {title} {The Feynman Lectures on Physics, Vol. II: Mainly Electromagnetism and Matter}}},\ \bibinfo {edition} {new millennium}\ ed.\ (\bibinfo  {publisher} {Basic Books},\ \bibinfo {address} {New York},\ \bibinfo {year} {2010})\BibitemShut {NoStop}%
\bibitem [{\citenamefont {Caldeira}\ and\ \citenamefont {Leggett}(1981)}]{CaldeiraLeggett:1981}%
  \BibitemOpen
  \bibfield  {author} {\bibinfo {author} {\bibfnamefont {A.~O.}\ \bibnamefont {Caldeira}}\ and\ \bibinfo {author} {\bibfnamefont {A.~J.}\ \bibnamefont {Leggett}},\ }\href {https://doi.org/10.1103/PhysRevLett.46.211} {\bibfield  {journal} {\bibinfo  {journal} {Physical Review Letters}\ }\textbf {\bibinfo {volume} {46}},\ \bibinfo {pages} {211} (\bibinfo {year} {1981})}\BibitemShut {NoStop}%
\bibitem [{\citenamefont {Yurke}\ and\ \citenamefont {Denker}(1984)}]{YurkeDenker:1984}%
  \BibitemOpen
  \bibfield  {author} {\bibinfo {author} {\bibfnamefont {B.}~\bibnamefont {Yurke}}\ and\ \bibinfo {author} {\bibfnamefont {J.~S.}\ \bibnamefont {Denker}},\ }\href {https://doi.org/10.1103/PhysRevA.29.1419} {\bibfield  {journal} {\bibinfo  {journal} {Physical Review A}\ }\textbf {\bibinfo {volume} {29}},\ \bibinfo {pages} {1419} (\bibinfo {year} {1984})}\BibitemShut {NoStop}%
\bibitem [{\citenamefont {Blais}\ \emph {et~al.}(2021)\citenamefont {Blais}, \citenamefont {Grimsmo}, \citenamefont {Girvin},\ and\ \citenamefont {Wallraff}}]{Blais:2021}%
  \BibitemOpen
  \bibfield  {author} {\bibinfo {author} {\bibfnamefont {A.}~\bibnamefont {Blais}}, \bibinfo {author} {\bibfnamefont {A.~L.}\ \bibnamefont {Grimsmo}}, \bibinfo {author} {\bibfnamefont {S.~M.}\ \bibnamefont {Girvin}},\ and\ \bibinfo {author} {\bibfnamefont {A.}~\bibnamefont {Wallraff}},\ }\href {https://doi.org/10.1103/RevModPhys.93.025005} {\bibfield  {journal} {\bibinfo  {journal} {Review Modern Physics}\ }\textbf {\bibinfo {volume} {93}},\ \bibinfo {pages} {025005} (\bibinfo {year} {2021})}\BibitemShut {NoStop}%
\bibitem [{\citenamefont {Parra-Rodriguez}(2021)}]{ParraRodriguezPhD:2021}%
  \BibitemOpen
  \bibfield  {author} {\bibinfo {author} {\bibfnamefont {A.}~\bibnamefont {Parra-Rodriguez}},\ }\href {http://hdl.handle.net/10810/51132} {\emph {\bibinfo {title} {PhD Thesis: Canonical Quantization of Superconducting Circuits}}}\ (\bibinfo  {publisher} {Universidad del Pais Vasco},\ \bibinfo {address} {Leioa},\ \bibinfo {year} {2021})\BibitemShut {NoStop}%
\bibitem [{\citenamefont {Foster}(1924)}]{Foster:1924}%
  \BibitemOpen
  \bibfield  {author} {\bibinfo {author} {\bibfnamefont {R.~M.}\ \bibnamefont {Foster}},\ }\href@noop {} {\bibfield  {journal} {\bibinfo  {journal} {Bell Systems Technical Journal}\ }\textbf {\bibinfo {volume} {6}},\ \bibinfo {pages} {259} (\bibinfo {year} {1924})}\BibitemShut {NoStop}%
\bibitem [{\citenamefont {Cauer}(1926)}]{Cauer:1926}%
  \BibitemOpen
  \bibfield  {author} {\bibinfo {author} {\bibfnamefont {W.}~\bibnamefont {Cauer}},\ }\href@noop {} {\emph {\bibinfo {title} {Doktorarbeit: Die Verwirklichung der Wechselstrom- widerst\"ande vorgeschriebener Frequenzabh\"angigkeit.}}}\ (\bibinfo  {publisher} {TH Berlin},\ \bibinfo {address} {Berlin},\ \bibinfo {year} {1926})\BibitemShut {NoStop}%
\bibitem [{\citenamefont {Cauer}(1929)}]{Cauer:1929}%
  \BibitemOpen
  \bibfield  {author} {\bibinfo {author} {\bibfnamefont {W.}~\bibnamefont {Cauer}},\ }\href@noop {} {\bibfield  {journal} {\bibinfo  {journal} {Elektrische Nachrichtentechnik (ENT)}\ }\textbf {\bibinfo {volume} {6}},\ \bibinfo {pages} {272} (\bibinfo {year} {1929})}\BibitemShut {NoStop}%
\bibitem [{\citenamefont {Brune}(1931)}]{Brune:1931}%
  \BibitemOpen
  \bibfield  {author} {\bibinfo {author} {\bibfnamefont {O.}~\bibnamefont {Brune}},\ }\href@noop {} {\emph {\bibinfo {title} {PhD Thesis: Synthesis of a finite two-terminal network whose driving-point impedance is a prescribed function of frequency}}}\ (\bibinfo  {publisher} {M.I.T.},\ \bibinfo {address} {Boston},\ \bibinfo {year} {1931})\BibitemShut {NoStop}%
\bibitem [{\citenamefont {Tellegen}(1948{\natexlab{a}})}]{Tellegen:1948a}%
  \BibitemOpen
  \bibfield  {author} {\bibinfo {author} {\bibfnamefont {B.~D.~H.}\ \bibnamefont {Tellegen}},\ }\href {{https://web.archive.org/web/20140423045739/http://techpreservation.dyndns.org/beitman/abpr/newfiles/The%20Gyrator.pdf}} {\bibfield  {journal} {\bibinfo  {journal} {Philips Research Reports}\ }\textbf {\bibinfo {volume} {3}},\ \bibinfo {pages} {81} (\bibinfo {year} {1948}{\natexlab{a}})}\BibitemShut {NoStop}%
\bibitem [{\citenamefont {Belevitch}(1950)}]{Belevitch:1950}%
  \BibitemOpen
  \bibfield  {author} {\bibinfo {author} {\bibfnamefont {V.}~\bibnamefont {Belevitch}},\ }\href@noop {} {\bibfield  {journal} {\bibinfo  {journal} {Electrical Communication}\ }\textbf {\bibinfo {volume} {27}},\ \bibinfo {pages} {231} (\bibinfo {year} {1950})}\BibitemShut {NoStop}%
\bibitem [{\citenamefont {Cauer}(1954)}]{Cauer:1954}%
  \BibitemOpen
  \bibfield  {author} {\bibinfo {author} {\bibfnamefont {W.}~\bibnamefont {Cauer}},\ }\href@noop {} {\emph {\bibinfo {title} {Theorie der linearen Wechselstromschaltungen}}},\ \bibinfo {edition} {2nd}\ ed.\ (\bibinfo  {publisher} {Akademie-Verlag GmbH, Berlin},\ \bibinfo {year} {1954})\BibitemShut {NoStop}%
\bibitem [{\citenamefont {Newcomb}(1966)}]{Newcomb:1966}%
  \BibitemOpen
  \bibfield  {author} {\bibinfo {author} {\bibfnamefont {R.~W.}\ \bibnamefont {Newcomb}},\ }\href@noop {} {\emph {\bibinfo {title} {Linear Multiport Synthesis}}}\ (\bibinfo  {publisher} {McGraw-Hill},\ \bibinfo {address} {New York},\ \bibinfo {year} {1966})\BibitemShut {NoStop}%
\bibitem [{\citenamefont {Anderson}\ and\ \citenamefont {Moylan}(1975)}]{Anderson:1975}%
  \BibitemOpen
  \bibfield  {author} {\bibinfo {author} {\bibfnamefont {B.}~\bibnamefont {Anderson}}\ and\ \bibinfo {author} {\bibfnamefont {P.}~\bibnamefont {Moylan}},\ }\href {https://doi.org/10.1002/cta.4490030209} {\bibfield  {journal} {\bibinfo  {journal} {International Journal of Circuit Theory and Applications}\ }\textbf {\bibinfo {volume} {3}},\ \bibinfo {pages} {193} (\bibinfo {year} {1975})}\BibitemShut {NoStop}%
\bibitem [{\citenamefont {Wells}(1938)}]{Wells:1938}%
  \BibitemOpen
  \bibfield  {author} {\bibinfo {author} {\bibfnamefont {D.~A.}\ \bibnamefont {Wells}},\ }\href {https://doi.org/10.1063/1.1710422} {\bibfield  {journal} {\bibinfo  {journal} {Journal of Applied Physics}\ }\textbf {\bibinfo {volume} {9}},\ \bibinfo {pages} {312} (\bibinfo {year} {1938})}\BibitemShut {NoStop}%
\bibitem [{\citenamefont {Wells}(1945)}]{Wells:1945}%
  \BibitemOpen
  \bibfield  {author} {\bibinfo {author} {\bibfnamefont {D.~A.}\ \bibnamefont {Wells}},\ }\href {https://doi.org/10.1063/1.1707623} {\bibfield  {journal} {\bibinfo  {journal} {Journal of Applied Physics}\ }\textbf {\bibinfo {volume} {16}},\ \bibinfo {pages} {535} (\bibinfo {year} {1945})}\BibitemShut {NoStop}%
\bibitem [{\citenamefont {MacFarlane}(1969)}]{MacFarlane:1969}%
  \BibitemOpen
  \bibfield  {author} {\bibinfo {author} {\bibfnamefont {A.}~\bibnamefont {MacFarlane}},\ }\href {https://doi.org/10.1049/piee.1969.0262} {\bibfield  {journal} {\bibinfo  {journal} {Proceedings of the Institution of Electrical Engineers}\ }\textbf {\bibinfo {volume} {116}},\ \bibinfo {pages} {1453(4)} (\bibinfo {year} {1969})}\BibitemShut {NoStop}%
\bibitem [{\citenamefont {{Chua}}\ and\ \citenamefont {{McPherson}}(1974)}]{Chua:1974}%
  \BibitemOpen
  \bibfield  {author} {\bibinfo {author} {\bibfnamefont {L.}~\bibnamefont {{Chua}}}\ and\ \bibinfo {author} {\bibfnamefont {J.}~\bibnamefont {{McPherson}}},\ }\href {https://doi.org/10.1109/TCS.1974.1083849} {\bibfield  {journal} {\bibinfo  {journal} {IEEE Transactions on Circuits and Systems}\ }\textbf {\bibinfo {volume} {21}},\ \bibinfo {pages} {277} (\bibinfo {year} {1974})}\BibitemShut {NoStop}%
\bibitem [{\citenamefont {{Kwatny}}\ \emph {et~al.}(1982)\citenamefont {{Kwatny}}, \citenamefont {{Massimo}},\ and\ \citenamefont {{Bahar}}}]{Kwatny:1982}%
  \BibitemOpen
  \bibfield  {author} {\bibinfo {author} {\bibfnamefont {H.}~\bibnamefont {{Kwatny}}}, \bibinfo {author} {\bibfnamefont {F.}~\bibnamefont {{Massimo}}},\ and\ \bibinfo {author} {\bibfnamefont {L.}~\bibnamefont {{Bahar}}},\ }\href {https://doi.org/10.1109/TCS.1982.1085140} {\bibfield  {journal} {\bibinfo  {journal} {IEEE Transactions on Circuits and Systems}\ }\textbf {\bibinfo {volume} {29}},\ \bibinfo {pages} {220} (\bibinfo {year} {1982})}\BibitemShut {NoStop}%
\bibitem [{\citenamefont {Shragowitz}\ and\ \citenamefont {Gerlovin}(1988)}]{Shragowitz:1988}%
  \BibitemOpen
  \bibfield  {author} {\bibinfo {author} {\bibfnamefont {E.}~\bibnamefont {Shragowitz}}\ and\ \bibinfo {author} {\bibfnamefont {E.}~\bibnamefont {Gerlovin}},\ }\href {https://doi.org/10.1002/cta.4490160203} {\bibfield  {journal} {\bibinfo  {journal} {International Journal of Circuit Theory and Applications}\ }\textbf {\bibinfo {volume} {16}},\ \bibinfo {pages} {129} (\bibinfo {year} {1988})}\BibitemShut {NoStop}%
\bibitem [{\citenamefont {Bernstein}\ and\ \citenamefont {Lieberman}(1989)}]{Bernstein:1989}%
  \BibitemOpen
  \bibfield  {author} {\bibinfo {author} {\bibfnamefont {G.}~\bibnamefont {Bernstein}}\ and\ \bibinfo {author} {\bibfnamefont {M.}~\bibnamefont {Lieberman}},\ }\href {https://doi.org/10.1109/31.17588} {\bibfield  {journal} {\bibinfo  {journal} {{IEEE} Transactions on Circuits and Systems}\ }\textbf {\bibinfo {volume} {36}},\ \bibinfo {pages} {411} (\bibinfo {year} {1989})}\BibitemShut {NoStop}%
\bibitem [{\citenamefont {Maschke}\ \emph {et~al.}(1995)\citenamefont {Maschke}, \citenamefont {van~der Schaft},\ and\ \citenamefont {Breedveld}}]{Maschke:1995}%
  \BibitemOpen
  \bibfield  {author} {\bibinfo {author} {\bibfnamefont {B.}~\bibnamefont {Maschke}}, \bibinfo {author} {\bibfnamefont {A.}~\bibnamefont {van~der Schaft}},\ and\ \bibinfo {author} {\bibfnamefont {P.}~\bibnamefont {Breedveld}},\ }\href {https://doi.org/10.1109/81.372847} {\bibfield  {journal} {\bibinfo  {journal} {IEEE Transactions on Circuits and Systems I: Fundamental Theory and Applications}\ }\textbf {\bibinfo {volume} {42}},\ \bibinfo {pages} {73} (\bibinfo {year} {1995})}\BibitemShut {NoStop}%
\bibitem [{\citenamefont {Weiss}\ and\ \citenamefont {Mathis}(1997)}]{Weiss:1997}%
  \BibitemOpen
  \bibfield  {author} {\bibinfo {author} {\bibfnamefont {L.}~\bibnamefont {Weiss}}\ and\ \bibinfo {author} {\bibfnamefont {W.}~\bibnamefont {Mathis}},\ }\href {https://doi.org/10.1109/81.622990} {\bibfield  {journal} {\bibinfo  {journal} {IEEE Transactions on Circuits and Systems I: Fundamental Theory and Applications}\ }\textbf {\bibinfo {volume} {44}},\ \bibinfo {pages} {843} (\bibinfo {year} {1997})}\BibitemShut {NoStop}%
\bibitem [{\citenamefont {Van Der~Schaft}(2006)}]{VanDerSchaft:2006}%
  \BibitemOpen
  \bibfield  {author} {\bibinfo {author} {\bibfnamefont {A.}~\bibnamefont {Van Der~Schaft}},\ }in\ \href {https://doi.org/10.4171/022-3/65} {\emph {\bibinfo {booktitle} {Proceedings of the international congress of mathematicians}}},\ Vol.~\bibinfo {volume} {3},\ \bibinfo {organization} {Marta Sanz-Sole, Javier Soria, Juan Luis Verona, Joan Verdura, Madrid, Spain}\ (\bibinfo  {publisher} {European Mathematical Society Publishing House},\ \bibinfo {year} {2006})\ pp.\ \bibinfo {pages} {1339--1365}\BibitemShut {NoStop}%
\bibitem [{\citenamefont {Van Der~Schaft}\ and\ \citenamefont {Jeltsema}(2014)}]{VanDerSchaft:2014}%
  \BibitemOpen
  \bibfield  {author} {\bibinfo {author} {\bibfnamefont {A.}~\bibnamefont {Van Der~Schaft}}\ and\ \bibinfo {author} {\bibfnamefont {D.}~\bibnamefont {Jeltsema}},\ }\href {https://doi.org/10.1561/2600000002} {\bibfield  {journal} {\bibinfo  {journal} {Foundations and Trends{\textregistered} in Systems and Control}\ }\textbf {\bibinfo {volume} {1}},\ \bibinfo {pages} {173} (\bibinfo {year} {2014})}\BibitemShut {NoStop}%
\bibitem [{\citenamefont {Devoret}(1995)}]{Devoret:1997}%
  \BibitemOpen
  \bibfield  {author} {\bibinfo {author} {\bibfnamefont {M.~H.}\ \bibnamefont {Devoret}},\ }in\ \href@noop {} {\emph {\bibinfo {booktitle} {Proceedings of the Les Houches Summer School, Session LXIII}}}\ (\bibinfo  {publisher} {Elsevier, edited by S. Reynaud, E. Giacobino, and J. Zinn-Justin},\ \bibinfo {address} {New York},\ \bibinfo {year} {1995})\BibitemShut {NoStop}%
\bibitem [{\citenamefont {Burkard}\ \emph {et~al.}(2004)\citenamefont {Burkard}, \citenamefont {Koch},\ and\ \citenamefont {DiVincenzo}}]{Burkard:2004}%
  \BibitemOpen
  \bibfield  {author} {\bibinfo {author} {\bibfnamefont {G.}~\bibnamefont {Burkard}}, \bibinfo {author} {\bibfnamefont {R.~H.}\ \bibnamefont {Koch}},\ and\ \bibinfo {author} {\bibfnamefont {D.~P.}\ \bibnamefont {DiVincenzo}},\ }\href {https://doi.org/10.1103/PhysRevB.69.064503} {\bibfield  {journal} {\bibinfo  {journal} {Physical Review B}\ }\textbf {\bibinfo {volume} {69}},\ \bibinfo {pages} {064503} (\bibinfo {year} {2004})}\BibitemShut {NoStop}%
\bibitem [{\citenamefont {Burkard}(2005)}]{Burkard:2005}%
  \BibitemOpen
  \bibfield  {author} {\bibinfo {author} {\bibfnamefont {G.}~\bibnamefont {Burkard}},\ }\href {https://doi.org/10.1103/PhysRevB.71.144511} {\bibfield  {journal} {\bibinfo  {journal} {Physical Review B}\ }\textbf {\bibinfo {volume} {71}},\ \bibinfo {pages} {144511} (\bibinfo {year} {2005})}\BibitemShut {NoStop}%
\bibitem [{\citenamefont {Ulrich}\ and\ \citenamefont {Hassler}(2016)}]{Ulrich:2016}%
  \BibitemOpen
  \bibfield  {author} {\bibinfo {author} {\bibfnamefont {J.}~\bibnamefont {Ulrich}}\ and\ \bibinfo {author} {\bibfnamefont {F.}~\bibnamefont {Hassler}},\ }\href {https://doi.org/10.1103/PhysRevB.94.094505} {\bibfield  {journal} {\bibinfo  {journal} {Physical Review B}\ }\textbf {\bibinfo {volume} {94}},\ \bibinfo {pages} {094505} (\bibinfo {year} {2016})}\BibitemShut {NoStop}%
\bibitem [{\citenamefont {Parra-Rodriguez}\ and\ \citenamefont {Egusquiza}(2022{\natexlab{a}})}]{ParraRodriguez:2022}%
  \BibitemOpen
  \bibfield  {author} {\bibinfo {author} {\bibfnamefont {A.}~\bibnamefont {Parra-Rodriguez}}\ and\ \bibinfo {author} {\bibfnamefont {I.~L.}\ \bibnamefont {Egusquiza}},\ }\href {https://doi.org/10.22331/q-2022-04-04-681} {\bibfield  {journal} {\bibinfo  {journal} {{Quantum}}\ }\textbf {\bibinfo {volume} {6}},\ \bibinfo {pages} {681} (\bibinfo {year} {2022}{\natexlab{a}})}\BibitemShut {NoStop}%
\bibitem [{\citenamefont {Egusquiza}\ and\ \citenamefont {Parra-Rodriguez}(2022)}]{Egusquiza:2022}%
  \BibitemOpen
  \bibfield  {author} {\bibinfo {author} {\bibfnamefont {I.~L.}\ \bibnamefont {Egusquiza}}\ and\ \bibinfo {author} {\bibfnamefont {A.}~\bibnamefont {Parra-Rodriguez}},\ }\href {https://doi.org/10.1103/PhysRevB.106.024510} {\bibfield  {journal} {\bibinfo  {journal} {Physical Review B}\ }\textbf {\bibinfo {volume} {106}},\ \bibinfo {pages} {024510} (\bibinfo {year} {2022})}\BibitemShut {NoStop}%
\bibitem [{\citenamefont {Chakravarty}\ and\ \citenamefont {Schmid}(1986)}]{Chakravarty:1986}%
  \BibitemOpen
  \bibfield  {author} {\bibinfo {author} {\bibfnamefont {S.}~\bibnamefont {Chakravarty}}\ and\ \bibinfo {author} {\bibfnamefont {A.}~\bibnamefont {Schmid}},\ }\href {https://doi.org/10.1103/PhysRevB.33.2000} {\bibfield  {journal} {\bibinfo  {journal} {Physical Review B}\ }\textbf {\bibinfo {volume} {33}},\ \bibinfo {pages} {2000} (\bibinfo {year} {1986})}\BibitemShut {NoStop}%
\bibitem [{\citenamefont {Yurke}(1987)}]{Yurke:1987}%
  \BibitemOpen
  \bibfield  {author} {\bibinfo {author} {\bibfnamefont {B.}~\bibnamefont {Yurke}},\ }\href {https://doi.org/10.1364/JOSAB.4.001551} {\bibfield  {journal} {\bibinfo  {journal} {Journal of the Optical Society of America B}\ }\textbf {\bibinfo {volume} {4}},\ \bibinfo {pages} {1551} (\bibinfo {year} {1987})}\BibitemShut {NoStop}%
\bibitem [{\citenamefont {Paladino}\ \emph {et~al.}(2003)\citenamefont {Paladino}, \citenamefont {Taddei}, \citenamefont {Giaquinta},\ and\ \citenamefont {Falci}}]{Paladino:2003}%
  \BibitemOpen
  \bibfield  {author} {\bibinfo {author} {\bibfnamefont {E.}~\bibnamefont {Paladino}}, \bibinfo {author} {\bibfnamefont {F.}~\bibnamefont {Taddei}}, \bibinfo {author} {\bibfnamefont {G.}~\bibnamefont {Giaquinta}},\ and\ \bibinfo {author} {\bibfnamefont {G.}~\bibnamefont {Falci}},\ }\href {https://doi.org/10.1016/S1386-9477(02)00948-7} {\bibfield  {journal} {\bibinfo  {journal} {Physica E: Low-Dimensional Systems and Nanostructures}\ }\textbf {\bibinfo {volume} {18}},\ \bibinfo {pages} {39} (\bibinfo {year} {2003})}\BibitemShut {NoStop}%
\bibitem [{\citenamefont {Blais}\ \emph {et~al.}(2004)\citenamefont {Blais}, \citenamefont {Huang}, \citenamefont {Wallraff}, \citenamefont {Girvin},\ and\ \citenamefont {Schoelkopf}}]{Blais:2004}%
  \BibitemOpen
  \bibfield  {author} {\bibinfo {author} {\bibfnamefont {A.}~\bibnamefont {Blais}}, \bibinfo {author} {\bibfnamefont {R.-S.}\ \bibnamefont {Huang}}, \bibinfo {author} {\bibfnamefont {A.}~\bibnamefont {Wallraff}}, \bibinfo {author} {\bibfnamefont {S.~M.}\ \bibnamefont {Girvin}},\ and\ \bibinfo {author} {\bibfnamefont {R.~J.}\ \bibnamefont {Schoelkopf}},\ }\href {https://doi.org/10.1103/PhysRevA.69.062320} {\bibfield  {journal} {\bibinfo  {journal} {Physical ReviewA}\ }\textbf {\bibinfo {volume} {69}},\ \bibinfo {pages} {062320} (\bibinfo {year} {2004})}\BibitemShut {NoStop}%
\bibitem [{\citenamefont {Houck}\ \emph {et~al.}(2008)\citenamefont {Houck}, \citenamefont {Schreier}, \citenamefont {Johnson}, \citenamefont {Chow}, \citenamefont {Koch}, \citenamefont {Gambetta}, \citenamefont {Schuster}, \citenamefont {Frunzio}, \citenamefont {Devoret}, \citenamefont {Girvin},\ and\ \citenamefont {Schoelkopf}}]{Houck:2008}%
  \BibitemOpen
  \bibfield  {author} {\bibinfo {author} {\bibfnamefont {A.~A.}\ \bibnamefont {Houck}}, \bibinfo {author} {\bibfnamefont {J.~A.}\ \bibnamefont {Schreier}}, \bibinfo {author} {\bibfnamefont {B.~R.}\ \bibnamefont {Johnson}}, \bibinfo {author} {\bibfnamefont {J.~M.}\ \bibnamefont {Chow}}, \bibinfo {author} {\bibfnamefont {J.}~\bibnamefont {Koch}}, \bibinfo {author} {\bibfnamefont {J.~M.}\ \bibnamefont {Gambetta}}, \bibinfo {author} {\bibfnamefont {D.~I.}\ \bibnamefont {Schuster}}, \bibinfo {author} {\bibfnamefont {L.}~\bibnamefont {Frunzio}}, \bibinfo {author} {\bibfnamefont {M.~H.}\ \bibnamefont {Devoret}}, \bibinfo {author} {\bibfnamefont {S.~M.}\ \bibnamefont {Girvin}},\ and\ \bibinfo {author} {\bibfnamefont {R.~J.}\ \bibnamefont {Schoelkopf}},\ }\href {https://doi.org/10.1103/PhysRevLett.101.080502} {\bibfield  {journal} {\bibinfo  {journal} {Physical Review Letters}\ }\textbf {\bibinfo {volume} {101}},\ \bibinfo {pages} {080502} (\bibinfo {year} {2008})}\BibitemShut {NoStop}%
\bibitem [{\citenamefont {Bourassa}\ \emph {et~al.}(2009)\citenamefont {Bourassa}, \citenamefont {Gambetta}, \citenamefont {Abdumalikov}, \citenamefont {Astafiev}, \citenamefont {Nakamura},\ and\ \citenamefont {Blais}}]{Bourassa:2009}%
  \BibitemOpen
  \bibfield  {author} {\bibinfo {author} {\bibfnamefont {J.}~\bibnamefont {Bourassa}}, \bibinfo {author} {\bibfnamefont {J.~M.}\ \bibnamefont {Gambetta}}, \bibinfo {author} {\bibfnamefont {A.~A.}\ \bibnamefont {Abdumalikov}}, \bibinfo {author} {\bibfnamefont {O.}~\bibnamefont {Astafiev}}, \bibinfo {author} {\bibfnamefont {Y.}~\bibnamefont {Nakamura}},\ and\ \bibinfo {author} {\bibfnamefont {A.}~\bibnamefont {Blais}},\ }\href {https://doi.org/10.1103/PhysRevA.80.032109} {\bibfield  {journal} {\bibinfo  {journal} {Physical Review A}\ }\textbf {\bibinfo {volume} {80}},\ \bibinfo {pages} {032109} (\bibinfo {year} {2009})}\BibitemShut {NoStop}%
\bibitem [{\citenamefont {Clerk}\ \emph {et~al.}(2010)\citenamefont {Clerk}, \citenamefont {Devoret}, \citenamefont {Girvin}, \citenamefont {Marquardt},\ and\ \citenamefont {Schoelkopf}}]{Clerk:2010}%
  \BibitemOpen
  \bibfield  {author} {\bibinfo {author} {\bibfnamefont {A.~A.}\ \bibnamefont {Clerk}}, \bibinfo {author} {\bibfnamefont {M.~H.}\ \bibnamefont {Devoret}}, \bibinfo {author} {\bibfnamefont {S.~M.}\ \bibnamefont {Girvin}}, \bibinfo {author} {\bibfnamefont {F.}~\bibnamefont {Marquardt}},\ and\ \bibinfo {author} {\bibfnamefont {R.~J.}\ \bibnamefont {Schoelkopf}},\ }\href {https://doi.org/10.1103/RevModPhys.82.1155} {\bibfield  {journal} {\bibinfo  {journal} {Reviews of Modern Physics}\ }\textbf {\bibinfo {volume} {82}},\ \bibinfo {pages} {1155} (\bibinfo {year} {2010})}\BibitemShut {NoStop}%
\bibitem [{\citenamefont {Koch}\ \emph {et~al.}(2010)\citenamefont {Koch}, \citenamefont {Houck}, \citenamefont {Hur},\ and\ \citenamefont {Girvin}}]{Koch:2010}%
  \BibitemOpen
  \bibfield  {author} {\bibinfo {author} {\bibfnamefont {J.}~\bibnamefont {Koch}}, \bibinfo {author} {\bibfnamefont {A.~A.}\ \bibnamefont {Houck}}, \bibinfo {author} {\bibfnamefont {K.~L.}\ \bibnamefont {Hur}},\ and\ \bibinfo {author} {\bibfnamefont {S.~M.}\ \bibnamefont {Girvin}},\ }\href {https://doi.org/10.1103/PhysRevA.82.043811} {\bibfield  {journal} {\bibinfo  {journal} {Physical Review A}\ }\textbf {\bibinfo {volume} {82}},\ \bibinfo {pages} {043811} (\bibinfo {year} {2010})}\BibitemShut {NoStop}%
\bibitem [{\citenamefont {Filipp}\ \emph {et~al.}(2011)\citenamefont {Filipp}, \citenamefont {G\"oppl}, \citenamefont {Fink}, \citenamefont {Baur}, \citenamefont {Bianchetti}, \citenamefont {Steffen},\ and\ \citenamefont {Wallraff}}]{Filipp:2011}%
  \BibitemOpen
  \bibfield  {author} {\bibinfo {author} {\bibfnamefont {S.}~\bibnamefont {Filipp}}, \bibinfo {author} {\bibfnamefont {M.}~\bibnamefont {G\"oppl}}, \bibinfo {author} {\bibfnamefont {J.~M.}\ \bibnamefont {Fink}}, \bibinfo {author} {\bibfnamefont {M.}~\bibnamefont {Baur}}, \bibinfo {author} {\bibfnamefont {R.}~\bibnamefont {Bianchetti}}, \bibinfo {author} {\bibfnamefont {L.}~\bibnamefont {Steffen}},\ and\ \bibinfo {author} {\bibfnamefont {A.}~\bibnamefont {Wallraff}},\ }\href {https://doi.org/10.1103/PhysRevA.83.063827} {\bibfield  {journal} {\bibinfo  {journal} {Physical Review A}\ }\textbf {\bibinfo {volume} {83}},\ \bibinfo {pages} {063827} (\bibinfo {year} {2011})}\BibitemShut {NoStop}%
\bibitem [{\citenamefont {Bourassa}\ \emph {et~al.}(2012)\citenamefont {Bourassa}, \citenamefont {Beaudoin}, \citenamefont {Gambetta},\ and\ \citenamefont {Blais}}]{Bourassa:2012}%
  \BibitemOpen
  \bibfield  {author} {\bibinfo {author} {\bibfnamefont {J.}~\bibnamefont {Bourassa}}, \bibinfo {author} {\bibfnamefont {F.}~\bibnamefont {Beaudoin}}, \bibinfo {author} {\bibfnamefont {J.~M.}\ \bibnamefont {Gambetta}},\ and\ \bibinfo {author} {\bibfnamefont {A.}~\bibnamefont {Blais}},\ }\href {https://doi.org/10.1103/PhysRevA.86.013814} {\bibfield  {journal} {\bibinfo  {journal} {Physical Review A}\ }\textbf {\bibinfo {volume} {86}},\ \bibinfo {pages} {013814} (\bibinfo {year} {2012})}\BibitemShut {NoStop}%
\bibitem [{\citenamefont {Bergenfeldt}\ and\ \citenamefont {Samuelsson}(2012)}]{Bergenfeldt:2012}%
  \BibitemOpen
  \bibfield  {author} {\bibinfo {author} {\bibfnamefont {C.}~\bibnamefont {Bergenfeldt}}\ and\ \bibinfo {author} {\bibfnamefont {P.}~\bibnamefont {Samuelsson}},\ }\href {https://doi.org/10.1103/PhysRevB.85.045446} {\bibfield  {journal} {\bibinfo  {journal} {Physical Review B}\ }\textbf {\bibinfo {volume} {85}},\ \bibinfo {pages} {045446} (\bibinfo {year} {2012})}\BibitemShut {NoStop}%
\bibitem [{\citenamefont {Peropadre}\ \emph {et~al.}(2013)\citenamefont {Peropadre}, \citenamefont {Lindkvist}, \citenamefont {Hoi}, \citenamefont {Wilson}, \citenamefont {Garcia-Ripoll}, \citenamefont {Delsing},\ and\ \citenamefont {Johansson}}]{Peropadre:2013}%
  \BibitemOpen
  \bibfield  {author} {\bibinfo {author} {\bibfnamefont {B.}~\bibnamefont {Peropadre}}, \bibinfo {author} {\bibfnamefont {J.}~\bibnamefont {Lindkvist}}, \bibinfo {author} {\bibfnamefont {I.-C.}\ \bibnamefont {Hoi}}, \bibinfo {author} {\bibfnamefont {C.~M.}\ \bibnamefont {Wilson}}, \bibinfo {author} {\bibfnamefont {J.~J.}\ \bibnamefont {Garcia-Ripoll}}, \bibinfo {author} {\bibfnamefont {P.}~\bibnamefont {Delsing}},\ and\ \bibinfo {author} {\bibfnamefont {G.}~\bibnamefont {Johansson}},\ }\href {https://doi.org/10.1088/1367-2630/15/3/035009} {\bibfield  {journal} {\bibinfo  {journal} {New Journal of Physics}\ }\textbf {\bibinfo {volume} {15}},\ \bibinfo {pages} {035009} (\bibinfo {year} {2013})}\BibitemShut {NoStop}%
\bibitem [{\citenamefont {Nigg}\ \emph {et~al.}(2012)\citenamefont {Nigg}, \citenamefont {Paik}, \citenamefont {Vlastakis}, \citenamefont {Kirchmair}, \citenamefont {Shankar}, \citenamefont {Frunzio}, \citenamefont {Devoret}, \citenamefont {Schoelkopf},\ and\ \citenamefont {Girvin}}]{Nigg:2012}%
  \BibitemOpen
  \bibfield  {author} {\bibinfo {author} {\bibfnamefont {S.~E.}\ \bibnamefont {Nigg}}, \bibinfo {author} {\bibfnamefont {H.}~\bibnamefont {Paik}}, \bibinfo {author} {\bibfnamefont {B.}~\bibnamefont {Vlastakis}}, \bibinfo {author} {\bibfnamefont {G.}~\bibnamefont {Kirchmair}}, \bibinfo {author} {\bibfnamefont {S.}~\bibnamefont {Shankar}}, \bibinfo {author} {\bibfnamefont {L.}~\bibnamefont {Frunzio}}, \bibinfo {author} {\bibfnamefont {M.~H.}\ \bibnamefont {Devoret}}, \bibinfo {author} {\bibfnamefont {R.~J.}\ \bibnamefont {Schoelkopf}},\ and\ \bibinfo {author} {\bibfnamefont {S.~M.}\ \bibnamefont {Girvin}},\ }\href {https://doi.org/10.1103/PhysRevLett.108.240502} {\bibfield  {journal} {\bibinfo  {journal} {Physical Review Letters}\ }\textbf {\bibinfo {volume} {108}},\ \bibinfo {pages} {240502} (\bibinfo {year} {2012})}\BibitemShut {NoStop}%
\bibitem [{\citenamefont {Devoret}\ and\ \citenamefont {Schoelkopf}(2013)}]{Devoret:2013}%
  \BibitemOpen
  \bibfield  {author} {\bibinfo {author} {\bibfnamefont {M.}~\bibnamefont {Devoret}}\ and\ \bibinfo {author} {\bibfnamefont {R.}~\bibnamefont {Schoelkopf}},\ }\href {https://doi.org/10.1126/science.1231930} {\bibfield  {journal} {\bibinfo  {journal} {Science}\ }\textbf {\bibinfo {volume} {339}},\ \bibinfo {pages} {1169} (\bibinfo {year} {2013})}\BibitemShut {NoStop}%
\bibitem [{\citenamefont {Solgun}\ \emph {et~al.}(2014)\citenamefont {Solgun}, \citenamefont {Abraham},\ and\ \citenamefont {DiVincenzo}}]{Solgun:2014}%
  \BibitemOpen
  \bibfield  {author} {\bibinfo {author} {\bibfnamefont {F.}~\bibnamefont {Solgun}}, \bibinfo {author} {\bibfnamefont {D.~W.}\ \bibnamefont {Abraham}},\ and\ \bibinfo {author} {\bibfnamefont {D.~P.}\ \bibnamefont {DiVincenzo}},\ }\href {https://doi.org/10.1103/PhysRevB.90.134504} {\bibfield  {journal} {\bibinfo  {journal} {Physical Review B}\ }\textbf {\bibinfo {volume} {90}},\ \bibinfo {pages} {134504} (\bibinfo {year} {2014})}\BibitemShut {NoStop}%
\bibitem [{\citenamefont {Sundaresan}\ \emph {et~al.}(2015)\citenamefont {Sundaresan}, \citenamefont {Liu}, \citenamefont {Sadri}, \citenamefont {Sz\ifmmode~\mbox{\H{o}}\else \H{o}\fi{}cs}, \citenamefont {Underwood}, \citenamefont {Malekakhlagh}, \citenamefont {T\"ureci},\ and\ \citenamefont {Houck}}]{Sundaresan:2015}%
  \BibitemOpen
  \bibfield  {author} {\bibinfo {author} {\bibfnamefont {N.~M.}\ \bibnamefont {Sundaresan}}, \bibinfo {author} {\bibfnamefont {Y.}~\bibnamefont {Liu}}, \bibinfo {author} {\bibfnamefont {D.}~\bibnamefont {Sadri}}, \bibinfo {author} {\bibfnamefont {L.~J.}\ \bibnamefont {Sz\ifmmode~\mbox{\H{o}}\else \H{o}\fi{}cs}}, \bibinfo {author} {\bibfnamefont {D.~L.}\ \bibnamefont {Underwood}}, \bibinfo {author} {\bibfnamefont {M.}~\bibnamefont {Malekakhlagh}}, \bibinfo {author} {\bibfnamefont {H.~E.}\ \bibnamefont {T\"ureci}},\ and\ \bibinfo {author} {\bibfnamefont {A.~A.}\ \bibnamefont {Houck}},\ }\href {https://doi.org/10.1103/PhysRevX.5.021035} {\bibfield  {journal} {\bibinfo  {journal} {Physical Review X}\ }\textbf {\bibinfo {volume} {5}},\ \bibinfo {pages} {021035} (\bibinfo {year} {2015})}\BibitemShut {NoStop}%
\bibitem [{\citenamefont {Solgun}\ and\ \citenamefont {DiVincenzo}(2015)}]{Solgun:2015}%
  \BibitemOpen
  \bibfield  {author} {\bibinfo {author} {\bibfnamefont {F.}~\bibnamefont {Solgun}}\ and\ \bibinfo {author} {\bibfnamefont {D.}~\bibnamefont {DiVincenzo}},\ }\href {https://doi.org/10.1016/j.aop.2015.07.005} {\bibfield  {journal} {\bibinfo  {journal} {Annals of Physics}\ }\textbf {\bibinfo {volume} {361}},\ \bibinfo {pages} {605} (\bibinfo {year} {2015})}\BibitemShut {NoStop}%
\bibitem [{\citenamefont {Malekakhlagh}\ and\ \citenamefont {T\"ureci}(2016)}]{Malekakhlagh:2016}%
  \BibitemOpen
  \bibfield  {author} {\bibinfo {author} {\bibfnamefont {M.}~\bibnamefont {Malekakhlagh}}\ and\ \bibinfo {author} {\bibfnamefont {H.~E.}\ \bibnamefont {T\"ureci}},\ }\href {https://doi.org/10.1103/PhysRevA.93.012120} {\bibfield  {journal} {\bibinfo  {journal} {Physical Review A}\ }\textbf {\bibinfo {volume} {93}},\ \bibinfo {pages} {012120} (\bibinfo {year} {2016})}\BibitemShut {NoStop}%
\bibitem [{\citenamefont {Mortensen}\ \emph {et~al.}(2016)\citenamefont {Mortensen}, \citenamefont {M\o{}lmer},\ and\ \citenamefont {Andersen}}]{Mortensen:2016}%
  \BibitemOpen
  \bibfield  {author} {\bibinfo {author} {\bibfnamefont {H.~L.}\ \bibnamefont {Mortensen}}, \bibinfo {author} {\bibfnamefont {K.}~\bibnamefont {M\o{}lmer}},\ and\ \bibinfo {author} {\bibfnamefont {C.~K.}\ \bibnamefont {Andersen}},\ }\href {https://doi.org/10.1103/PhysRevA.94.053817} {\bibfield  {journal} {\bibinfo  {journal} {Physical Review A}\ }\textbf {\bibinfo {volume} {94}},\ \bibinfo {pages} {053817} (\bibinfo {year} {2016})}\BibitemShut {NoStop}%
\bibitem [{\citenamefont {Roy}\ and\ \citenamefont {Devoret}(2016)}]{Roy:2016}%
  \BibitemOpen
  \bibfield  {author} {\bibinfo {author} {\bibfnamefont {A.}~\bibnamefont {Roy}}\ and\ \bibinfo {author} {\bibfnamefont {M.}~\bibnamefont {Devoret}},\ }\href {https://doi.org/https://doi.org/10.1016/j.crhy.2016.07.012} {\bibfield  {journal} {\bibinfo  {journal} {Comptes Rendus Physique}\ }\textbf {\bibinfo {volume} {17}},\ \bibinfo {pages} {740} (\bibinfo {year} {2016})}\BibitemShut {NoStop}%
\bibitem [{\citenamefont {Vool}\ and\ \citenamefont {Devoret}(2017)}]{Vool:2017}%
  \BibitemOpen
  \bibfield  {author} {\bibinfo {author} {\bibfnamefont {U.}~\bibnamefont {Vool}}\ and\ \bibinfo {author} {\bibfnamefont {M.}~\bibnamefont {Devoret}},\ }\href {https://doi.org/10.1002/cta.2359} {\bibfield  {journal} {\bibinfo  {journal} {International Journal of Circuit Theory and Applications}\ }\textbf {\bibinfo {volume} {45}},\ \bibinfo {pages} {897} (\bibinfo {year} {2017})}\BibitemShut {NoStop}%
\bibitem [{\citenamefont {Malekakhlagh}\ \emph {et~al.}(2017)\citenamefont {Malekakhlagh}, \citenamefont {Petrescu},\ and\ \citenamefont {T\"ureci}}]{Malekakhlagh:2017}%
  \BibitemOpen
  \bibfield  {author} {\bibinfo {author} {\bibfnamefont {M.}~\bibnamefont {Malekakhlagh}}, \bibinfo {author} {\bibfnamefont {A.}~\bibnamefont {Petrescu}},\ and\ \bibinfo {author} {\bibfnamefont {H.~E.}\ \bibnamefont {T\"ureci}},\ }\href {https://doi.org/10.1103/PhysRevLett.119.073601} {\bibfield  {journal} {\bibinfo  {journal} {Physical Review Letters}\ }\textbf {\bibinfo {volume} {119}},\ \bibinfo {pages} {073601} (\bibinfo {year} {2017})}\BibitemShut {NoStop}%
\bibitem [{\citenamefont {Roy}\ and\ \citenamefont {Devoret}(2018)}]{Roy:2018}%
  \BibitemOpen
  \bibfield  {author} {\bibinfo {author} {\bibfnamefont {A.}~\bibnamefont {Roy}}\ and\ \bibinfo {author} {\bibfnamefont {M.}~\bibnamefont {Devoret}},\ }\href {https://doi.org/10.1103/PhysRevB.98.045405} {\bibfield  {journal} {\bibinfo  {journal} {Physical Review B}\ }\textbf {\bibinfo {volume} {98}},\ \bibinfo {pages} {045405} (\bibinfo {year} {2018})}\BibitemShut {NoStop}%
\bibitem [{\citenamefont {Parra-Rodriguez}\ \emph {et~al.}(2018)\citenamefont {Parra-Rodriguez}, \citenamefont {Rico}, \citenamefont {Solano},\ and\ \citenamefont {Egusquiza}}]{ParraRodriguez:2018}%
  \BibitemOpen
  \bibfield  {author} {\bibinfo {author} {\bibfnamefont {A.}~\bibnamefont {Parra-Rodriguez}}, \bibinfo {author} {\bibfnamefont {E.}~\bibnamefont {Rico}}, \bibinfo {author} {\bibfnamefont {E.}~\bibnamefont {Solano}},\ and\ \bibinfo {author} {\bibfnamefont {I.~L.}\ \bibnamefont {Egusquiza}},\ }\href {https://doi.org/10.1088/2058-9565/aab1ba} {\bibfield  {journal} {\bibinfo  {journal} {Quantum Science and Technology}\ }\textbf {\bibinfo {volume} {3}},\ \bibinfo {pages} {024012} (\bibinfo {year} {2018})}\BibitemShut {NoStop}%
\bibitem [{\citenamefont {Jeltsema}\ and\ \citenamefont {{Van Der Schaft}}(2009)}]{Jeltsema:2009}%
  \BibitemOpen
  \bibfield  {author} {\bibinfo {author} {\bibfnamefont {D.}~\bibnamefont {Jeltsema}}\ and\ \bibinfo {author} {\bibfnamefont {A.~J.}\ \bibnamefont {{Van Der Schaft}}},\ }\href {https://doi.org/https://doi.org/10.1016/S0034-4877(09)00009-3} {\bibfield  {journal} {\bibinfo  {journal} {Reports on Mathematical Physics}\ }\textbf {\bibinfo {volume} {63}},\ \bibinfo {pages} {55} (\bibinfo {year} {2009})}\BibitemShut {NoStop}%
\bibitem [{\citenamefont {Mariantoni}(2020)}]{Mariantoni:2020}%
  \BibitemOpen
  \bibfield  {author} {\bibinfo {author} {\bibfnamefont {M.}~\bibnamefont {Mariantoni}},\ }\Eprint {https://arxiv.org/abs/2007.08519} {arXiv:2007.08519 [class-ph]}  (\bibinfo {year} {2020})\BibitemShut {NoStop}%
\bibitem [{\citenamefont {Parra-Rodriguez}\ \emph {et~al.}(2019)\citenamefont {Parra-Rodriguez}, \citenamefont {Egusquiza}, \citenamefont {DiVincenzo},\ and\ \citenamefont {Solano}}]{ParraRodriguez:2019}%
  \BibitemOpen
  \bibfield  {author} {\bibinfo {author} {\bibfnamefont {A.}~\bibnamefont {Parra-Rodriguez}}, \bibinfo {author} {\bibfnamefont {I.~L.}\ \bibnamefont {Egusquiza}}, \bibinfo {author} {\bibfnamefont {D.~P.}\ \bibnamefont {DiVincenzo}},\ and\ \bibinfo {author} {\bibfnamefont {E.}~\bibnamefont {Solano}},\ }\href {https://doi.org/10.1103/PhysRevB.99.014514} {\bibfield  {journal} {\bibinfo  {journal} {Physical Review B}\ }\textbf {\bibinfo {volume} {99}},\ \bibinfo {pages} {014514} (\bibinfo {year} {2019})}\BibitemShut {NoStop}%
\bibitem [{\citenamefont {Rymarz}\ \emph {et~al.}(2021)\citenamefont {Rymarz}, \citenamefont {Bosco}, \citenamefont {Ciani},\ and\ \citenamefont {DiVincenzo}}]{Rymarz:2021}%
  \BibitemOpen
  \bibfield  {author} {\bibinfo {author} {\bibfnamefont {M.}~\bibnamefont {Rymarz}}, \bibinfo {author} {\bibfnamefont {S.}~\bibnamefont {Bosco}}, \bibinfo {author} {\bibfnamefont {A.}~\bibnamefont {Ciani}},\ and\ \bibinfo {author} {\bibfnamefont {D.~P.}\ \bibnamefont {DiVincenzo}},\ }\href {https://doi.org/10.1103/PhysRevX.11.011032} {\bibfield  {journal} {\bibinfo  {journal} {Physical Review X}\ }\textbf {\bibinfo {volume} {11}},\ \bibinfo {pages} {011032} (\bibinfo {year} {2021})}\BibitemShut {NoStop}%
\bibitem [{\citenamefont {Parra-Rodriguez}\ and\ \citenamefont {Egusquiza}(2022{\natexlab{b}})}]{ParraRodriguez:2022b}%
  \BibitemOpen
  \bibfield  {author} {\bibinfo {author} {\bibfnamefont {A.}~\bibnamefont {Parra-Rodriguez}}\ and\ \bibinfo {author} {\bibfnamefont {I.~L.}\ \bibnamefont {Egusquiza}},\ }\href {https://doi.org/10.1103/PhysRevB.106.054504} {\bibfield  {journal} {\bibinfo  {journal} {Physical Review B}\ }\textbf {\bibinfo {volume} {106}},\ \bibinfo {pages} {054504} (\bibinfo {year} {2022}{\natexlab{b}})}\BibitemShut {NoStop}%
\bibitem [{\citenamefont {Kerckhoff}\ \emph {et~al.}(2015)\citenamefont {Kerckhoff}, \citenamefont {Lalumi\`ere}, \citenamefont {Chapman}, \citenamefont {Blais},\ and\ \citenamefont {Lehnert}}]{Kerckhoff:2015}%
  \BibitemOpen
  \bibfield  {author} {\bibinfo {author} {\bibfnamefont {J.}~\bibnamefont {Kerckhoff}}, \bibinfo {author} {\bibfnamefont {K.}~\bibnamefont {Lalumi\`ere}}, \bibinfo {author} {\bibfnamefont {B.~J.}\ \bibnamefont {Chapman}}, \bibinfo {author} {\bibfnamefont {A.}~\bibnamefont {Blais}},\ and\ \bibinfo {author} {\bibfnamefont {K.~W.}\ \bibnamefont {Lehnert}},\ }\href {https://doi.org/10.1103/PhysRevApplied.4.034002} {\bibfield  {journal} {\bibinfo  {journal} {Physical Review Applied}\ }\textbf {\bibinfo {volume} {4}},\ \bibinfo {pages} {034002} (\bibinfo {year} {2015})}\BibitemShut {NoStop}%
\bibitem [{\citenamefont {Sliwa}\ \emph {et~al.}(2015)\citenamefont {Sliwa}, \citenamefont {Hatridge}, \citenamefont {Narla}, \citenamefont {Shankar}, \citenamefont {Frunzio}, \citenamefont {Schoelkopf},\ and\ \citenamefont {Devoret}}]{Sliwa:2015}%
  \BibitemOpen
  \bibfield  {author} {\bibinfo {author} {\bibfnamefont {K.~M.}\ \bibnamefont {Sliwa}}, \bibinfo {author} {\bibfnamefont {M.}~\bibnamefont {Hatridge}}, \bibinfo {author} {\bibfnamefont {A.}~\bibnamefont {Narla}}, \bibinfo {author} {\bibfnamefont {S.}~\bibnamefont {Shankar}}, \bibinfo {author} {\bibfnamefont {L.}~\bibnamefont {Frunzio}}, \bibinfo {author} {\bibfnamefont {R.~J.}\ \bibnamefont {Schoelkopf}},\ and\ \bibinfo {author} {\bibfnamefont {M.~H.}\ \bibnamefont {Devoret}},\ }\href {https://doi.org/10.1103/PhysRevX.5.041020} {\bibfield  {journal} {\bibinfo  {journal} {Physical Review X}\ }\textbf {\bibinfo {volume} {5}},\ \bibinfo {pages} {041020} (\bibinfo {year} {2015})}\BibitemShut {NoStop}%
\bibitem [{\citenamefont {Lecocq}\ \emph {et~al.}(2017)\citenamefont {Lecocq}, \citenamefont {Ranzani}, \citenamefont {Peterson}, \citenamefont {Cicak}, \citenamefont {Simmonds}, \citenamefont {Teufel},\ and\ \citenamefont {Aumentado}}]{Lecocq:2017}%
  \BibitemOpen
  \bibfield  {author} {\bibinfo {author} {\bibfnamefont {F.}~\bibnamefont {Lecocq}}, \bibinfo {author} {\bibfnamefont {L.}~\bibnamefont {Ranzani}}, \bibinfo {author} {\bibfnamefont {G.~A.}\ \bibnamefont {Peterson}}, \bibinfo {author} {\bibfnamefont {K.}~\bibnamefont {Cicak}}, \bibinfo {author} {\bibfnamefont {R.~W.}\ \bibnamefont {Simmonds}}, \bibinfo {author} {\bibfnamefont {J.~D.}\ \bibnamefont {Teufel}},\ and\ \bibinfo {author} {\bibfnamefont {J.}~\bibnamefont {Aumentado}},\ }\href {https://doi.org/10.1103/PhysRevApplied.7.024028} {\bibfield  {journal} {\bibinfo  {journal} {Physical Review Applied}\ }\textbf {\bibinfo {volume} {7}},\ \bibinfo {pages} {024028} (\bibinfo {year} {2017})}\BibitemShut {NoStop}%
\bibitem [{\citenamefont {Chapman}\ \emph {et~al.}(2017)\citenamefont {Chapman}, \citenamefont {Rosenthal}, \citenamefont {Kerckhoff}, \citenamefont {Moores}, \citenamefont {Vale}, \citenamefont {Mates}, \citenamefont {Hilton}, \citenamefont {Lalumi\`ere}, \citenamefont {Blais},\ and\ \citenamefont {Lehnert}}]{Chapman:2017}%
  \BibitemOpen
  \bibfield  {author} {\bibinfo {author} {\bibfnamefont {B.~J.}\ \bibnamefont {Chapman}}, \bibinfo {author} {\bibfnamefont {E.~I.}\ \bibnamefont {Rosenthal}}, \bibinfo {author} {\bibfnamefont {J.}~\bibnamefont {Kerckhoff}}, \bibinfo {author} {\bibfnamefont {B.~A.}\ \bibnamefont {Moores}}, \bibinfo {author} {\bibfnamefont {L.~R.}\ \bibnamefont {Vale}}, \bibinfo {author} {\bibfnamefont {J.~A.~B.}\ \bibnamefont {Mates}}, \bibinfo {author} {\bibfnamefont {G.~C.}\ \bibnamefont {Hilton}}, \bibinfo {author} {\bibfnamefont {K.}~\bibnamefont {Lalumi\`ere}}, \bibinfo {author} {\bibfnamefont {A.}~\bibnamefont {Blais}},\ and\ \bibinfo {author} {\bibfnamefont {K.~W.}\ \bibnamefont {Lehnert}},\ }\href {https://doi.org/10.1103/PhysRevX.7.041043} {\bibfield  {journal} {\bibinfo  {journal} {Physical Review X}\ }\textbf {\bibinfo {volume} {7}},\ \bibinfo {pages} {041043} (\bibinfo {year} {2017})}\BibitemShut {NoStop}%
\bibitem [{\citenamefont {Mahoney}\ \emph {et~al.}(2017)\citenamefont {Mahoney}, \citenamefont {Colless}, \citenamefont {Pauka}, \citenamefont {Hornibrook}, \citenamefont {Watson}, \citenamefont {Gardner}, \citenamefont {Manfra}, \citenamefont {Doherty},\ and\ \citenamefont {Reilly}}]{Mahoney:2017}%
  \BibitemOpen
  \bibfield  {author} {\bibinfo {author} {\bibfnamefont {A.~C.}\ \bibnamefont {Mahoney}}, \bibinfo {author} {\bibfnamefont {J.~I.}\ \bibnamefont {Colless}}, \bibinfo {author} {\bibfnamefont {S.~J.}\ \bibnamefont {Pauka}}, \bibinfo {author} {\bibfnamefont {J.~M.}\ \bibnamefont {Hornibrook}}, \bibinfo {author} {\bibfnamefont {J.~D.}\ \bibnamefont {Watson}}, \bibinfo {author} {\bibfnamefont {G.~C.}\ \bibnamefont {Gardner}}, \bibinfo {author} {\bibfnamefont {M.~J.}\ \bibnamefont {Manfra}}, \bibinfo {author} {\bibfnamefont {A.~C.}\ \bibnamefont {Doherty}},\ and\ \bibinfo {author} {\bibfnamefont {D.~J.}\ \bibnamefont {Reilly}},\ }\href {https://doi.org/10.1103/PhysRevX.7.011007} {\bibfield  {journal} {\bibinfo  {journal} {Physical Review X}\ }\textbf {\bibinfo {volume} {7}},\ \bibinfo {pages} {011007} (\bibinfo {year} {2017})}\BibitemShut {NoStop}%
\bibitem [{\citenamefont {Barzanjeh}\ \emph {et~al.}(2017)\citenamefont {Barzanjeh}, \citenamefont {Wulf}, \citenamefont {Peruzzo}, \citenamefont {Kalaee}, \citenamefont {Dieterle}, \citenamefont {Painter},\ and\ \citenamefont {Fink}}]{Barzanjeh:2017}%
  \BibitemOpen
  \bibfield  {author} {\bibinfo {author} {\bibfnamefont {S.}~\bibnamefont {Barzanjeh}}, \bibinfo {author} {\bibfnamefont {M.}~\bibnamefont {Wulf}}, \bibinfo {author} {\bibfnamefont {M.}~\bibnamefont {Peruzzo}}, \bibinfo {author} {\bibfnamefont {M.}~\bibnamefont {Kalaee}}, \bibinfo {author} {\bibfnamefont {P.}~\bibnamefont {Dieterle}}, \bibinfo {author} {\bibfnamefont {O.}~\bibnamefont {Painter}},\ and\ \bibinfo {author} {\bibfnamefont {J.}~\bibnamefont {Fink}},\ }\href {https://doi.org/10.1038/s41467-017-01304-x} {\bibfield  {journal} {\bibinfo  {journal} {Nature Communications}\ }\textbf {\bibinfo {volume} {8}},\ \bibinfo {pages} {953} (\bibinfo {year} {2017})}\BibitemShut {NoStop}%
\bibitem [{\citenamefont {Rosario~Hamann}\ \emph {et~al.}(2018)\citenamefont {Rosario~Hamann}, \citenamefont {M\"uller}, \citenamefont {Jerger}, \citenamefont {Zanner}, \citenamefont {Combes}, \citenamefont {Pletyukhov}, \citenamefont {Weides}, \citenamefont {Stace},\ and\ \citenamefont {Fedorov}}]{Rosario:2018}%
  \BibitemOpen
  \bibfield  {author} {\bibinfo {author} {\bibfnamefont {A.}~\bibnamefont {Rosario~Hamann}}, \bibinfo {author} {\bibfnamefont {C.}~\bibnamefont {M\"uller}}, \bibinfo {author} {\bibfnamefont {M.}~\bibnamefont {Jerger}}, \bibinfo {author} {\bibfnamefont {M.}~\bibnamefont {Zanner}}, \bibinfo {author} {\bibfnamefont {J.}~\bibnamefont {Combes}}, \bibinfo {author} {\bibfnamefont {M.}~\bibnamefont {Pletyukhov}}, \bibinfo {author} {\bibfnamefont {M.}~\bibnamefont {Weides}}, \bibinfo {author} {\bibfnamefont {T.~M.}\ \bibnamefont {Stace}},\ and\ \bibinfo {author} {\bibfnamefont {A.}~\bibnamefont {Fedorov}},\ }\href {https://doi.org/10.1103/PhysRevLett.121.123601} {\bibfield  {journal} {\bibinfo  {journal} {Physical Review Letters}\ }\textbf {\bibinfo {volume} {121}},\ \bibinfo {pages} {123601} (\bibinfo {year} {2018})}\BibitemShut {NoStop}%
\bibitem [{\citenamefont {Navarathna}\ \emph {et~al.}(2023)\citenamefont {Navarathna}, \citenamefont {Le}, \citenamefont {Hamann}, \citenamefont {Nguyen}, \citenamefont {Stace},\ and\ \citenamefont {Fedorov}}]{Navarathna:2023}%
  \BibitemOpen
  \bibfield  {author} {\bibinfo {author} {\bibfnamefont {R.}~\bibnamefont {Navarathna}}, \bibinfo {author} {\bibfnamefont {D.~T.}\ \bibnamefont {Le}}, \bibinfo {author} {\bibfnamefont {A.~R.}\ \bibnamefont {Hamann}}, \bibinfo {author} {\bibfnamefont {H.~D.}\ \bibnamefont {Nguyen}}, \bibinfo {author} {\bibfnamefont {T.~M.}\ \bibnamefont {Stace}},\ and\ \bibinfo {author} {\bibfnamefont {A.}~\bibnamefont {Fedorov}},\ }\href {https://doi.org/10.1103/PhysRevLett.130.037001} {\bibfield  {journal} {\bibinfo  {journal} {Physical Review Letters}\ }\textbf {\bibinfo {volume} {130}},\ \bibinfo {pages} {037001} (\bibinfo {year} {2023})}\BibitemShut {NoStop}%
\bibitem [{\citenamefont {Brayton}\ and\ \citenamefont {Moser}(1964{\natexlab{a}})}]{Brayton:1964a}%
  \BibitemOpen
  \bibfield  {author} {\bibinfo {author} {\bibfnamefont {R.~K.}\ \bibnamefont {Brayton}}\ and\ \bibinfo {author} {\bibfnamefont {J.~K.}\ \bibnamefont {Moser}},\ }\href {https://doi.org/10.1090/qam/169746} {\bibfield  {journal} {\bibinfo  {journal} {Quarterly of Applied Mathematics}\ }\textbf {\bibinfo {volume} {22}},\ \bibinfo {pages} {1} (\bibinfo {year} {1964}{\natexlab{a}})}\BibitemShut {NoStop}%
\bibitem [{\citenamefont {Brayton}\ and\ \citenamefont {Moser}(1964{\natexlab{b}})}]{Brayton:1964b}%
  \BibitemOpen
  \bibfield  {author} {\bibinfo {author} {\bibfnamefont {R.~K.}\ \bibnamefont {Brayton}}\ and\ \bibinfo {author} {\bibfnamefont {J.~K.}\ \bibnamefont {Moser}},\ }\href {https://doi.org/10.1090/qam/169747} {\bibfield  {journal} {\bibinfo  {journal} {Quarterly of applied mathematics}\ }\textbf {\bibinfo {volume} {22}},\ \bibinfo {pages} {81} (\bibinfo {year} {1964}{\natexlab{b}})}\BibitemShut {NoStop}%
\bibitem [{\citenamefont {Smale}(1972)}]{Smale:1972}%
  \BibitemOpen
  \bibfield  {author} {\bibinfo {author} {\bibfnamefont {S.}~\bibnamefont {Smale}},\ }\href@noop {} {\bibfield  {journal} {\bibinfo  {journal} {Journal of Differential Geometry}\ }\textbf {\bibinfo {volume} {7}},\ \bibinfo {pages} {193} (\bibinfo {year} {1972})}\BibitemShut {NoStop}%
\bibitem [{\citenamefont {Weinstein}(1983)}]{Weinstein:1983}%
  \BibitemOpen
  \bibfield  {author} {\bibinfo {author} {\bibfnamefont {A.}~\bibnamefont {Weinstein}},\ }\href {https://doi.org/10.4310/jdg/1214437787} {\bibfield  {journal} {\bibinfo  {journal} {Journal of Differential Geometry}\ }\textbf {\bibinfo {volume} {18}},\ \bibinfo {pages} {523 } (\bibinfo {year} {1983})}\BibitemShut {NoStop}%
\bibitem [{\citenamefont {Courant}(1990)}]{Courant:1990}%
  \BibitemOpen
  \bibfield  {author} {\bibinfo {author} {\bibfnamefont {T.~J.}\ \bibnamefont {Courant}},\ }\href@noop {} {\bibfield  {journal} {\bibinfo  {journal} {Transactions of the American Mathematical Society}\ }\textbf {\bibinfo {volume} {319}},\ \bibinfo {pages} {631} (\bibinfo {year} {1990})}\BibitemShut {NoStop}%
\bibitem [{\citenamefont {Yoshimura}\ and\ \citenamefont {Marsden}(2006)}]{Yoshimura:2006}%
  \BibitemOpen
  \bibfield  {author} {\bibinfo {author} {\bibfnamefont {H.}~\bibnamefont {Yoshimura}}\ and\ \bibinfo {author} {\bibfnamefont {J.~E.}\ \bibnamefont {Marsden}},\ }in\ \href@noop {} {\emph {\bibinfo {booktitle} {Proceedings of the 17th International Symposium on Mathematical Theory of Networks and Systems}}}\ (\bibinfo {address} {Kyoto},\ \bibinfo {year} {2006})\ pp.\ \bibinfo {pages} {Paper WeA08.5, pp 1--6}\BibitemShut {NoStop}%
\bibitem [{\citenamefont {Kirchhoff}(1847)}]{Kirchhoff:1847}%
  \BibitemOpen
  \bibfield  {author} {\bibinfo {author} {\bibfnamefont {G.}~\bibnamefont {Kirchhoff}},\ }\href {https://doi.org/10.1002/andp.18471481202} {\bibfield  {journal} {\bibinfo  {journal} {Annalen der Physik}\ }\textbf {\bibinfo {volume} {148}},\ \bibinfo {pages} {497} (\bibinfo {year} {1847})}\BibitemShut {NoStop}%
\bibitem [{\citenamefont {Gely}\ and\ \citenamefont {Steele}(2020)}]{Gely:2020}%
  \BibitemOpen
  \bibfield  {author} {\bibinfo {author} {\bibfnamefont {M.~F.}\ \bibnamefont {Gely}}\ and\ \bibinfo {author} {\bibfnamefont {G.~A.}\ \bibnamefont {Steele}},\ }\href {https://doi.org/10.1088/1367-2630/ab60f6} {\bibfield  {journal} {\bibinfo  {journal} {New Journal of Physics}\ }\textbf {\bibinfo {volume} {22}},\ \bibinfo {pages} {013025} (\bibinfo {year} {2020})}\BibitemShut {NoStop}%
\bibitem [{\citenamefont {Minev}\ \emph {et~al.}(2021)\citenamefont {Minev}, \citenamefont {McConkey}, \citenamefont {Takita}, \citenamefont {Corcoles},\ and\ \citenamefont {Gambetta}}]{Minev:2021}%
  \BibitemOpen
  \bibfield  {author} {\bibinfo {author} {\bibfnamefont {Z.~K.}\ \bibnamefont {Minev}}, \bibinfo {author} {\bibfnamefont {T.~G.}\ \bibnamefont {McConkey}}, \bibinfo {author} {\bibfnamefont {M.}~\bibnamefont {Takita}}, \bibinfo {author} {\bibfnamefont {A.~D.}\ \bibnamefont {Corcoles}},\ and\ \bibinfo {author} {\bibfnamefont {J.~M.}\ \bibnamefont {Gambetta}},\ }\Eprint {https://arxiv.org/abs/2103.10344} {arXiv:2103.10344 [quant-ph]}  (\bibinfo {year} {2021})\BibitemShut {NoStop}%
\bibitem [{\citenamefont {Chitta}\ \emph {et~al.}(2022)\citenamefont {Chitta}, \citenamefont {Zhao}, \citenamefont {Huang}, \citenamefont {Mondragon-Shem},\ and\ \citenamefont {Koch}}]{Chitta:2022}%
  \BibitemOpen
  \bibfield  {author} {\bibinfo {author} {\bibfnamefont {S.~P.}\ \bibnamefont {Chitta}}, \bibinfo {author} {\bibfnamefont {T.}~\bibnamefont {Zhao}}, \bibinfo {author} {\bibfnamefont {Z.}~\bibnamefont {Huang}}, \bibinfo {author} {\bibfnamefont {I.}~\bibnamefont {Mondragon-Shem}},\ and\ \bibinfo {author} {\bibfnamefont {J.}~\bibnamefont {Koch}},\ }\href {https://doi.org/10.1088/1367-2630/ac94f2} {\bibfield  {journal} {\bibinfo  {journal} {New Journal of Physics}\ }\textbf {\bibinfo {volume} {24}},\ \bibinfo {pages} {103020} (\bibinfo {year} {2022})}\BibitemShut {NoStop}%
\bibitem [{\citenamefont {Rajabzadeh}\ \emph {et~al.}(2023)\citenamefont {Rajabzadeh}, \citenamefont {Zhaoyou~Wang}, \citenamefont {Makihara}, \citenamefont {Guo},\ and\ \citenamefont {Safavi-Naeini}}]{Rajabzadeh:2023}%
  \BibitemOpen
  \bibfield  {author} {\bibinfo {author} {\bibfnamefont {T.}~\bibnamefont {Rajabzadeh}}, \bibinfo {author} {\bibfnamefont {N.~L.}\ \bibnamefont {Zhaoyou~Wang}}, \bibinfo {author} {\bibfnamefont {T.}~\bibnamefont {Makihara}}, \bibinfo {author} {\bibfnamefont {Y.}~\bibnamefont {Guo}},\ and\ \bibinfo {author} {\bibfnamefont {A.~H.}\ \bibnamefont {Safavi-Naeini}},\ }\href {https://doi.org/10.22331/q-2023-09-25-1118} {\bibfield  {journal} {\bibinfo  {journal} {{Quantum}}\ }\textbf {\bibinfo {volume} {7}},\ \bibinfo {pages} {1118} (\bibinfo {year} {2023})}\BibitemShut {NoStop}%
\bibitem [{\citenamefont {Faddeev}(1969)}]{Faddeev:1969}%
  \BibitemOpen
  \bibfield  {author} {\bibinfo {author} {\bibfnamefont {L.~D.}\ \bibnamefont {Faddeev}},\ }\href {https://doi.org/10.1007/BF01028566} {\bibfield  {journal} {\bibinfo  {journal} {Theoretical and Mathematical Physics}\ }\textbf {\bibinfo {volume} {1}},\ \bibinfo {pages} {1} (\bibinfo {year} {1969})}\BibitemShut {NoStop}%
\bibitem [{\citenamefont {Faddeev}\ and\ \citenamefont {Jackiw}(1988)}]{Faddeev:1988}%
  \BibitemOpen
  \bibfield  {author} {\bibinfo {author} {\bibfnamefont {L.}~\bibnamefont {Faddeev}}\ and\ \bibinfo {author} {\bibfnamefont {R.}~\bibnamefont {Jackiw}},\ }\href {https://doi.org/10.1103/PhysRevLett.60.1692} {\bibfield  {journal} {\bibinfo  {journal} {Physical Review Letters}\ }\textbf {\bibinfo {volume} {60}},\ \bibinfo {pages} {1692} (\bibinfo {year} {1988})}\BibitemShut {NoStop}%
\bibitem [{\citenamefont {{Jackiw}}(1993)}]{Jackiw:1993}%
  \BibitemOpen
  \bibfield  {author} {\bibinfo {author} {\bibfnamefont {R.}~\bibnamefont {{Jackiw}}},\ }\Eprint {https://arxiv.org/abs/hep-th/9306075} {arXiv:hep-th/9306075 [hep-th]}  (\bibinfo {year} {1993})\BibitemShut {NoStop}%
\bibitem [{\citenamefont {Jackiw}(1994)}]{Jackiw:1994}%
  \BibitemOpen
  \bibfield  {author} {\bibinfo {author} {\bibfnamefont {R.}~\bibnamefont {Jackiw}},\ }in\ \href {https://doi.org/10.1142/2194} {\emph {\bibinfo {booktitle} {Constraint Theory and Quantization Methods: From Relativistic Particles to Field Theory and General Relativity}}},\ \bibinfo {editor} {edited by\ \bibinfo {editor} {\bibfnamefont {F.}~\bibnamefont {Colomo}}, \bibinfo {editor} {\bibfnamefont {L.}~\bibnamefont {Lusanna}},\ and\ \bibinfo {editor} {\bibfnamefont {G.}~\bibnamefont {Marmo}}}\ (\bibinfo  {publisher} {World Scientific},\ \bibinfo {year} {1994})\BibitemShut {NoStop}%
\bibitem [{\citenamefont {Rymarz}\ and\ \citenamefont {DiVincenzo}(2023)}]{Rymarz:2023}%
  \BibitemOpen
  \bibfield  {author} {\bibinfo {author} {\bibfnamefont {M.}~\bibnamefont {Rymarz}}\ and\ \bibinfo {author} {\bibfnamefont {D.~P.}\ \bibnamefont {DiVincenzo}},\ }\href {https://doi.org/10.1103/PhysRevX.13.021017} {\bibfield  {journal} {\bibinfo  {journal} {Physical Review X}\ }\textbf {\bibinfo {volume} {13}},\ \bibinfo {pages} {021017} (\bibinfo {year} {2023})}\BibitemShut {NoStop}%
\bibitem [{\citenamefont {Miano}\ \emph {et~al.}(2023)\citenamefont {Miano}, \citenamefont {Joshi}, \citenamefont {Liu}, \citenamefont {Dai}, \citenamefont {Parakh}, \citenamefont {Frunzio},\ and\ \citenamefont {Devoret}}]{Miano:2023}%
  \BibitemOpen
  \bibfield  {author} {\bibinfo {author} {\bibfnamefont {A.}~\bibnamefont {Miano}}, \bibinfo {author} {\bibfnamefont {V.~R.}\ \bibnamefont {Joshi}}, \bibinfo {author} {\bibfnamefont {G.}~\bibnamefont {Liu}}, \bibinfo {author} {\bibfnamefont {W.}~\bibnamefont {Dai}}, \bibinfo {author} {\bibfnamefont {P.~D.}\ \bibnamefont {Parakh}}, \bibinfo {author} {\bibfnamefont {L.}~\bibnamefont {Frunzio}},\ and\ \bibinfo {author} {\bibfnamefont {M.~H.}\ \bibnamefont {Devoret}},\ }\href {https://doi.org/10.1103/PRXQuantum.4.030324} {\bibfield  {journal} {\bibinfo  {journal} {PRX Quantum}\ }\textbf {\bibinfo {volume} {4}},\ \bibinfo {pages} {030324} (\bibinfo {year} {2023})}\BibitemShut {NoStop}%
\bibitem [{\citenamefont {Manucharyan}\ \emph {et~al.}(2009)\citenamefont {Manucharyan}, \citenamefont {Koch}, \citenamefont {Glazman},\ and\ \citenamefont {Devoret}}]{Manucharyan:2009}%
  \BibitemOpen
  \bibfield  {author} {\bibinfo {author} {\bibfnamefont {V.~E.}\ \bibnamefont {Manucharyan}}, \bibinfo {author} {\bibfnamefont {J.}~\bibnamefont {Koch}}, \bibinfo {author} {\bibfnamefont {L.~I.}\ \bibnamefont {Glazman}},\ and\ \bibinfo {author} {\bibfnamefont {M.~H.}\ \bibnamefont {Devoret}},\ }\href {https://doi.org/10.1126/science.1175552} {\bibfield  {journal} {\bibinfo  {journal} {Science}\ }\textbf {\bibinfo {volume} {326}},\ \bibinfo {pages} {113} (\bibinfo {year} {2009})}\BibitemShut {NoStop}%
\bibitem [{\citenamefont {B\"uttiker}(1987)}]{Buettiker:1987}%
  \BibitemOpen
  \bibfield  {author} {\bibinfo {author} {\bibfnamefont {M.}~\bibnamefont {B\"uttiker}},\ }\href {https://doi.org/10.1103/PhysRevB.36.3548} {\bibfield  {journal} {\bibinfo  {journal} {Physical Review B}\ }\textbf {\bibinfo {volume} {36}},\ \bibinfo {pages} {3548} (\bibinfo {year} {1987})}\BibitemShut {NoStop}%
\bibitem [{\citenamefont {Thanh~Le}\ \emph {et~al.}(2019)\citenamefont {Thanh~Le}, \citenamefont {Grimsmo}, \citenamefont {Müller},\ and\ \citenamefont {Stace}}]{ThanhLe:2019}%
  \BibitemOpen
  \bibfield  {author} {\bibinfo {author} {\bibfnamefont {D.}~\bibnamefont {Thanh~Le}}, \bibinfo {author} {\bibfnamefont {A.}~\bibnamefont {Grimsmo}}, \bibinfo {author} {\bibfnamefont {C.}~\bibnamefont {Müller}},\ and\ \bibinfo {author} {\bibfnamefont {T.~M.}\ \bibnamefont {Stace}},\ }\href {https://doi.org/10.1103/PhysRevA.100.062321} {\bibfield  {journal} {\bibinfo  {journal} {Physical Review A}\ }\textbf {\bibinfo {volume} {100}},\ \bibinfo {pages} {062321} (\bibinfo {year} {2019})}\BibitemShut {NoStop}%
\bibitem [{\citenamefont {Guillemin}(1953)}]{Guillemin:1953}%
  \BibitemOpen
  \bibfield  {author} {\bibinfo {author} {\bibfnamefont {E.~A.}\ \bibnamefont {Guillemin}},\ }\href@noop {} {\emph {\bibinfo {title} {Introductory circuit theory}}}\ (\bibinfo  {publisher} {John Wiley \& Sons},\ \bibinfo {address} {New York},\ \bibinfo {year} {1953})\BibitemShut {NoStop}%
\bibitem [{\citenamefont {Chua}(1980)}]{Chua:1980}%
  \BibitemOpen
  \bibfield  {author} {\bibinfo {author} {\bibfnamefont {L.}~\bibnamefont {Chua}},\ }\href {https://doi.org/10.1109/TCS.1980.1084742} {\bibfield  {journal} {\bibinfo  {journal} {IEEE Transactions on Circuits and Systems}\ }\textbf {\bibinfo {volume} {27}},\ \bibinfo {pages} {1014} (\bibinfo {year} {1980})}\BibitemShut {NoStop}%
\bibitem [{\citenamefont {Ostrogradsky}(1850)}]{Ostrogradsky:1850}%
  \BibitemOpen
  \bibfield  {author} {\bibinfo {author} {\bibfnamefont {M.}~\bibnamefont {Ostrogradsky}},\ }\href@noop {} {\bibfield  {journal} {\bibinfo  {journal} {Mem. Acad. St. Petersbourg}\ }\textbf {\bibinfo {volume} {6}},\ \bibinfo {pages} {385} (\bibinfo {year} {1850})}\BibitemShut {NoStop}%
\bibitem [{\citenamefont {Woodard}(2015)}]{Woodard:2015}%
  \BibitemOpen
  \bibfield  {author} {\bibinfo {author} {\bibfnamefont {R.~P.}\ \bibnamefont {Woodard}},\ }\href {https://doi.org/10.4249/scholarpedia.32243} {\bibfield  {journal} {\bibinfo  {journal} {Scholarpedia}\ }\textbf {\bibinfo {volume} {10}},\ \bibinfo {pages} {32243} (\bibinfo {year} {2015})},\ \bibinfo {note} {revision \#186559}\BibitemShut {NoStop}%
\bibitem [{\citenamefont {Nakahara}(2003)}]{nakahara2003geometry}%
  \BibitemOpen
  \bibfield  {author} {\bibinfo {author} {\bibfnamefont {M.}~\bibnamefont {Nakahara}},\ }\href {https://doi.org/10.1201/9781315275826} {\emph {\bibinfo {title} {Geometry, Topology and Physics, Second Edition}}},\ Graduate student series in physics\ (\bibinfo  {publisher} {Taylor \& Francis},\ \bibinfo {year} {2003})\BibitemShut {NoStop}%
\bibitem [{\citenamefont {Choquet-Bruhat}(1968)}]{Choquet-Bruhat:1968}%
  \BibitemOpen
  \bibfield  {author} {\bibinfo {author} {\bibfnamefont {Y.}~\bibnamefont {Choquet-Bruhat}},\ }\href@noop {} {\emph {\bibinfo {title} {G{\'e}om{\'e}trie differentielle et systemes ext{\'e}rieurs}}},\ Monographies Universitaires de Math{\'e}matiques ; 28\ (\bibinfo  {publisher} {Dunod},\ \bibinfo {address} {Paris},\ \bibinfo {year} {1968})\ Chap.\ \bibinfo {chapter} {XVII}\BibitemShut {NoStop}%
\bibitem [{\citenamefont {Weyl}(1923)}]{Weyl:1923}%
  \BibitemOpen
  \bibfield  {author} {\bibinfo {author} {\bibfnamefont {H.}~\bibnamefont {Weyl}},\ }\href@noop {} {\bibfield  {journal} {\bibinfo  {journal} {Revista Matem{\'a}tica Hispano-Americana}\ }\textbf {\bibinfo {volume} {5}},\ \bibinfo {pages} {153} (\bibinfo {year} {1923})}\BibitemShut {NoStop}%
\bibitem [{\citenamefont {Tellegen}(1948{\natexlab{b}})}]{Tellegen:1948}%
  \BibitemOpen
  \bibfield  {author} {\bibinfo {author} {\bibfnamefont {B.~D.~H.}\ \bibnamefont {Tellegen}},\ }\href@noop {} {\bibfield  {journal} {\bibinfo  {journal} {Philips Research Reports}\ }\textbf {\bibinfo {volume} {3}},\ \bibinfo {pages} {81} (\bibinfo {year} {1948}{\natexlab{b}})}\BibitemShut {NoStop}%
\bibitem [{\citenamefont {De~Gosson}(2006)}]{de2006symplectic}%
  \BibitemOpen
  \bibfield  {author} {\bibinfo {author} {\bibfnamefont {M.~A.}\ \bibnamefont {De~Gosson}},\ }\href {https://doi.org/10.1007/3-7643-7575-2} {\emph {\bibinfo {title} {Symplectic geometry and quantum mechanics}}},\ Vol.\ \bibinfo {volume} {166}\ (\bibinfo  {publisher} {Springer Science \& Business Media},\ \bibinfo {address} {Basel},\ \bibinfo {year} {2006})\BibitemShut {NoStop}%
\bibitem [{\citenamefont {Kostant}(1970)}]{Kostant:1970}%
  \BibitemOpen
  \bibfield  {author} {\bibinfo {author} {\bibfnamefont {B.}~\bibnamefont {Kostant}},\ }in\ \href@noop {} {\emph {\bibinfo {booktitle} {Lectures in Modern Analysis and Applications III}}},\ \bibinfo {editor} {edited by\ \bibinfo {editor} {\bibfnamefont {C.~T.}\ \bibnamefont {Taam}}}\ (\bibinfo  {publisher} {Springer Berlin Heidelberg},\ \bibinfo {address} {Berlin, Heidelberg},\ \bibinfo {year} {1970})\ pp.\ \bibinfo {pages} {87--208}\BibitemShut {NoStop}%
\bibitem [{\citenamefont {Souriau}(1997)}]{Souriau:1997}%
  \BibitemOpen
  \bibfield  {author} {\bibinfo {author} {\bibfnamefont {J.-M.}\ \bibnamefont {Souriau}},\ }\href@noop {} {\emph {\bibinfo {title} {Structure of Dynamical Systems: A Symplectic View of Physics}}},\ \bibinfo {series} {Progress in Mathematics}, Vol.\ \bibinfo {volume} {149}\ (\bibinfo  {publisher} {Birkh\"{a}user},\ \bibinfo {address} {Boston},\ \bibinfo {year} {1997})\BibitemShut {NoStop}%
\bibitem [{\citenamefont {Egusquiza}\ \emph {et~al.}(2022)\citenamefont {Egusquiza}, \citenamefont {I\~niguez}, \citenamefont {Rico},\ and\ \citenamefont {Villarino}}]{Egusquiza0pi:2022}%
  \BibitemOpen
  \bibfield  {author} {\bibinfo {author} {\bibfnamefont {I.~L.}\ \bibnamefont {Egusquiza}}, \bibinfo {author} {\bibfnamefont {A.}~\bibnamefont {I\~niguez}}, \bibinfo {author} {\bibfnamefont {E.}~\bibnamefont {Rico}},\ and\ \bibinfo {author} {\bibfnamefont {A.}~\bibnamefont {Villarino}},\ }\href {https://doi.org/10.1103/PhysRevB.105.L201104} {\bibfield  {journal} {\bibinfo  {journal} {Physical Review B}\ }\textbf {\bibinfo {volume} {105}},\ \bibinfo {pages} {L201104} (\bibinfo {year} {2022})}\BibitemShut {NoStop}%
\bibitem [{\citenamefont {Galindo}\ and\ \citenamefont {Pascual}(2012)}]{galindo:2012}%
  \BibitemOpen
  \bibfield  {author} {\bibinfo {author} {\bibfnamefont {A.}~\bibnamefont {Galindo}}\ and\ \bibinfo {author} {\bibfnamefont {P.}~\bibnamefont {Pascual}},\ }\href@noop {} {\emph {\bibinfo {title} {Quantum Mechanics I}}},\ Theoretical and Mathematical Physics\ (\bibinfo  {publisher} {Springer Berlin Heidelberg},\ \bibinfo {year} {2012})\BibitemShut {NoStop}%
\bibitem [{\citenamefont {Reed}\ and\ \citenamefont {Simon}(1975)}]{Reed:1975}%
  \BibitemOpen
  \bibfield  {author} {\bibinfo {author} {\bibfnamefont {M.}~\bibnamefont {Reed}}\ and\ \bibinfo {author} {\bibfnamefont {B.}~\bibnamefont {Simon}},\ }\href {https://www.elsevier.com/books/ii-fourier-analysis-self-adjointness/reed/978-0-08-092537-0} {\emph {\bibinfo {title} {{Methods of Modern Mathematical Physics II: Fourier Analysis, Self-Adjointness}}}},\ \bibinfo {edition} {1st}\ ed.\ (\bibinfo  {publisher} {Academic Press},\ \bibinfo {address} {New York},\ \bibinfo {year} {1975})\BibitemShut {NoStop}%
\bibitem [{\citenamefont {Apenko}(1989)}]{Apenko:1989}%
  \BibitemOpen
  \bibfield  {author} {\bibinfo {author} {\bibfnamefont {S.~M.}\ \bibnamefont {Apenko}},\ }\href {https://doi.org/10.1016/0375-9601(89)90329-0} {\bibfield  {journal} {\bibinfo  {journal} {Physics Letters A}\ }\textbf {\bibinfo {volume} {142}},\ \bibinfo {pages} {277} (\bibinfo {year} {1989})}\BibitemShut {NoStop}%
\bibitem [{\citenamefont {Zaikin}(1990)}]{Zaikin:1990}%
  \BibitemOpen
  \bibfield  {author} {\bibinfo {author} {\bibfnamefont {A.~D.}\ \bibnamefont {Zaikin}},\ }\href {https://doi.org/10.1007/BF00683632} {\bibfield  {journal} {\bibinfo  {journal} {Journal of Low Temperature Physics}\ }\textbf {\bibinfo {volume} {80}},\ \bibinfo {pages} {223} (\bibinfo {year} {1990})}\BibitemShut {NoStop}%
\bibitem [{\citenamefont {Sch\"on}\ and\ \citenamefont {Zaikin}(1990)}]{Schoen:1990}%
  \BibitemOpen
  \bibfield  {author} {\bibinfo {author} {\bibfnamefont {G.}~\bibnamefont {Sch\"on}}\ and\ \bibinfo {author} {\bibfnamefont {A.~D.}\ \bibnamefont {Zaikin}},\ }\href {https://doi.org/10.1016/0370-1573(90)90156-V} {\bibfield  {journal} {\bibinfo  {journal} {Physics Reports}\ }\textbf {\bibinfo {volume} {198}},\ \bibinfo {pages} {237} (\bibinfo {year} {1990})}\BibitemShut {NoStop}%
\bibitem [{\citenamefont {Likharev}\ and\ \citenamefont {Zorin}(1985)}]{Likharev:1985}%
  \BibitemOpen
  \bibfield  {author} {\bibinfo {author} {\bibfnamefont {K.~K.}\ \bibnamefont {Likharev}}\ and\ \bibinfo {author} {\bibfnamefont {A.~B.}\ \bibnamefont {Zorin}},\ }\href {https://doi.org/10.1007/BF00683782} {\bibfield  {journal} {\bibinfo  {journal} {Journal of Low Temperature Physics}\ }\textbf {\bibinfo {volume} {59}},\ \bibinfo {pages} {347} (\bibinfo {year} {1985})}\BibitemShut {NoStop}%
\bibitem [{\citenamefont {Nakamura}\ \emph {et~al.}(1999)\citenamefont {Nakamura}, \citenamefont {Pashkin},\ and\ \citenamefont {Tsai}}]{Nakamura:1999}%
  \BibitemOpen
  \bibfield  {author} {\bibinfo {author} {\bibfnamefont {Y.}~\bibnamefont {Nakamura}}, \bibinfo {author} {\bibfnamefont {Y.~A.}\ \bibnamefont {Pashkin}},\ and\ \bibinfo {author} {\bibfnamefont {J.~S.}\ \bibnamefont {Tsai}},\ }\href {https://doi.org/10.1038/19718} {\bibfield  {journal} {\bibinfo  {journal} {Nature}\ }\textbf {\bibinfo {volume} {398}},\ \bibinfo {pages} {786} (\bibinfo {year} {1999})}\BibitemShut {NoStop}%
\bibitem [{\citenamefont {Koch}\ \emph {et~al.}(2007)\citenamefont {Koch}, \citenamefont {Yu}, \citenamefont {Gambetta}, \citenamefont {Houck}, \citenamefont {Schuster}, \citenamefont {Majer}, \citenamefont {Blais}, \citenamefont {Devoret}, \citenamefont {Girvin},\ and\ \citenamefont {Schoelkopf}}]{Koch:2007}%
  \BibitemOpen
  \bibfield  {author} {\bibinfo {author} {\bibfnamefont {J.}~\bibnamefont {Koch}}, \bibinfo {author} {\bibfnamefont {T.~M.}\ \bibnamefont {Yu}}, \bibinfo {author} {\bibfnamefont {J.}~\bibnamefont {Gambetta}}, \bibinfo {author} {\bibfnamefont {A.~A.}\ \bibnamefont {Houck}}, \bibinfo {author} {\bibfnamefont {D.~I.}\ \bibnamefont {Schuster}}, \bibinfo {author} {\bibfnamefont {J.}~\bibnamefont {Majer}}, \bibinfo {author} {\bibfnamefont {A.}~\bibnamefont {Blais}}, \bibinfo {author} {\bibfnamefont {M.~H.}\ \bibnamefont {Devoret}}, \bibinfo {author} {\bibfnamefont {S.~M.}\ \bibnamefont {Girvin}},\ and\ \bibinfo {author} {\bibfnamefont {R.~J.}\ \bibnamefont {Schoelkopf}},\ }\href {https://doi.org/10.1103/PhysRevA.76.042319} {\bibfield  {journal} {\bibinfo  {journal} {Physical Review A}\ }\textbf {\bibinfo {volume} {76}},\ \bibinfo {pages} {042319} (\bibinfo {year} {2007})}\BibitemShut {NoStop}%
\bibitem [{\citenamefont {Koch}\ \emph {et~al.}(2009)\citenamefont {Koch}, \citenamefont {Manucharyan}, \citenamefont {Devoret},\ and\ \citenamefont {Glazman}}]{Koch:2009}%
  \BibitemOpen
  \bibfield  {author} {\bibinfo {author} {\bibfnamefont {J.}~\bibnamefont {Koch}}, \bibinfo {author} {\bibfnamefont {V.}~\bibnamefont {Manucharyan}}, \bibinfo {author} {\bibfnamefont {M.~H.}\ \bibnamefont {Devoret}},\ and\ \bibinfo {author} {\bibfnamefont {L.~I.}\ \bibnamefont {Glazman}},\ }\href {https://doi.org/10.1103/PhysRevLett.103.217004} {\bibfield  {journal} {\bibinfo  {journal} {Physical Review Letters}\ }\textbf {\bibinfo {volume} {103}},\ \bibinfo {pages} {217004} (\bibinfo {year} {2009})}\BibitemShut {NoStop}%
\bibitem [{\citenamefont {Devoret}(2021)}]{Devoret:2021}%
  \BibitemOpen
  \bibfield  {author} {\bibinfo {author} {\bibfnamefont {M.~H.}\ \bibnamefont {Devoret}},\ }\href {https://doi.org/10.1007/s10948-020-05784-9} {\bibfield  {journal} {\bibinfo  {journal} {Journal of Superconductivity and Novel Magnetism}\ }\textbf {\bibinfo {volume} {34}},\ \bibinfo {pages} {1633} (\bibinfo {year} {2021})}\BibitemShut {NoStop}%
\bibitem [{\citenamefont {Sonin}(2022)}]{Sonin:2022}%
  \BibitemOpen
  \bibfield  {author} {\bibinfo {author} {\bibfnamefont {E.~B.}\ \bibnamefont {Sonin}},\ }\href {https://doi.org/10.1063/10.0010205} {\bibfield  {journal} {\bibinfo  {journal} {Low Temperature Physics}\ }\textbf {\bibinfo {volume} {48}},\ \bibinfo {pages} {400} (\bibinfo {year} {2022})}\BibitemShut {NoStop}%
\bibitem [{\citenamefont {Mooij}\ and\ \citenamefont {Nazarov}(2006)}]{Mooij:2006}%
  \BibitemOpen
  \bibfield  {author} {\bibinfo {author} {\bibfnamefont {J.}~\bibnamefont {Mooij}}\ and\ \bibinfo {author} {\bibfnamefont {Y.}~\bibnamefont {Nazarov}},\ }\href {https://doi.org/10.1038/nphys234} {\bibfield  {journal} {\bibinfo  {journal} {Nature Physics}\ }\textbf {\bibinfo {volume} {2}},\ \bibinfo {pages} {169} (\bibinfo {year} {2006})}\BibitemShut {NoStop}%
\bibitem [{\citenamefont {Thanh~Le}\ \emph {et~al.}(2020)\citenamefont {Thanh~Le}, \citenamefont {Cole},\ and\ \citenamefont {Stace}}]{ThanhLe:2020}%
  \BibitemOpen
  \bibfield  {author} {\bibinfo {author} {\bibfnamefont {D.}~\bibnamefont {Thanh~Le}}, \bibinfo {author} {\bibfnamefont {J.~H.}\ \bibnamefont {Cole}},\ and\ \bibinfo {author} {\bibfnamefont {T.~M.}\ \bibnamefont {Stace}},\ }\href {https://doi.org/10.1103/PhysRevResearch.2.013245} {\bibfield  {journal} {\bibinfo  {journal} {Physical Review Research}\ }\textbf {\bibinfo {volume} {2}},\ \bibinfo {pages} {013245} (\bibinfo {year} {2020})}\BibitemShut {NoStop}%
\bibitem [{\citenamefont {Koliofoti}\ and\ \citenamefont {Riwar}(2023)}]{Koliofoti:2023}%
  \BibitemOpen
  \bibfield  {author} {\bibinfo {author} {\bibfnamefont {C.}~\bibnamefont {Koliofoti}}\ and\ \bibinfo {author} {\bibfnamefont {R.-P.}\ \bibnamefont {Riwar}},\ }\href {https://doi.org/10.1038/s41534-023-00790-w} {\bibfield  {journal} {\bibinfo  {journal} {npj Quantum Information}\ }\textbf {\bibinfo {volume} {9}},\ \bibinfo {pages} {1} (\bibinfo {year} {2023})},\ \bibinfo {note} {publisher: Nature Publishing Group}\BibitemShut {NoStop}%
\bibitem [{\citenamefont {Josephson}(1962)}]{Josephson:1962}%
  \BibitemOpen
  \bibfield  {author} {\bibinfo {author} {\bibfnamefont {B.~D.}\ \bibnamefont {Josephson}},\ }\href {https://doi.org/https://doi.org/10.1016/0031-9163(62)91369-0} {\bibfield  {journal} {\bibinfo  {journal} {Physics Letters}\ }\textbf {\bibinfo {volume} {1}},\ \bibinfo {pages} {251} (\bibinfo {year} {1962})}\BibitemShut {NoStop}%
\bibitem [{\citenamefont {Astafiev}\ \emph {et~al.}(2012)\citenamefont {Astafiev}, \citenamefont {Ioffe}, \citenamefont {Kafanov}, \citenamefont {Pashkin}, \citenamefont {Arutyunov}, \citenamefont {Shahar}, \citenamefont {Cohen},\ and\ \citenamefont {Tsai}}]{Astafiev:2012}%
  \BibitemOpen
  \bibfield  {author} {\bibinfo {author} {\bibfnamefont {O.}~\bibnamefont {Astafiev}}, \bibinfo {author} {\bibfnamefont {L.}~\bibnamefont {Ioffe}}, \bibinfo {author} {\bibfnamefont {S.}~\bibnamefont {Kafanov}}, \bibinfo {author} {\bibfnamefont {Y.}~\bibnamefont {Pashkin}}, \bibinfo {author} {\bibfnamefont {K.}~\bibnamefont {Arutyunov}}, \bibinfo {author} {\bibfnamefont {D.}~\bibnamefont {Shahar}}, \bibinfo {author} {\bibfnamefont {O.}~\bibnamefont {Cohen}},\ and\ \bibinfo {author} {\bibfnamefont {J.}~\bibnamefont {Tsai}},\ }\href {https://doi.org/10.1038/nature10930} {\bibfield  {journal} {\bibinfo  {journal} {Nature}\ }\textbf {\bibinfo {volume} {484}},\ \bibinfo {pages} {355} (\bibinfo {year} {2012})}\BibitemShut {NoStop}%
\bibitem [{\citenamefont {Pechenezhskiy}\ \emph {et~al.}(2020)\citenamefont {Pechenezhskiy}, \citenamefont {Mencia}, \citenamefont {Nguyen}, \citenamefont {Lin},\ and\ \citenamefont {Manucharyan}}]{Pechenezhskiy:2020}%
  \BibitemOpen
  \bibfield  {author} {\bibinfo {author} {\bibfnamefont {I.~V.}\ \bibnamefont {Pechenezhskiy}}, \bibinfo {author} {\bibfnamefont {R.~A.}\ \bibnamefont {Mencia}}, \bibinfo {author} {\bibfnamefont {L.~B.}\ \bibnamefont {Nguyen}}, \bibinfo {author} {\bibfnamefont {Y.-H.}\ \bibnamefont {Lin}},\ and\ \bibinfo {author} {\bibfnamefont {V.~E.}\ \bibnamefont {Manucharyan}},\ }\href {https://doi.org/10.1038/s41586-020-2687-9} {\bibfield  {journal} {\bibinfo  {journal} {Nature}\ }\textbf {\bibinfo {volume} {585}},\ \bibinfo {pages} {368} (\bibinfo {year} {2020})}\BibitemShut {NoStop}%
\bibitem [{\citenamefont {You}\ \emph {et~al.}(2019)\citenamefont {You}, \citenamefont {Sauls},\ and\ \citenamefont {Koch}}]{You:2019}%
  \BibitemOpen
  \bibfield  {author} {\bibinfo {author} {\bibfnamefont {X.}~\bibnamefont {You}}, \bibinfo {author} {\bibfnamefont {J.~A.}\ \bibnamefont {Sauls}},\ and\ \bibinfo {author} {\bibfnamefont {J.}~\bibnamefont {Koch}},\ }\href {https://doi.org/10.1103/PhysRevB.99.174512} {\bibfield  {journal} {\bibinfo  {journal} {Physical Review B}\ }\textbf {\bibinfo {volume} {99}},\ \bibinfo {pages} {174512} (\bibinfo {year} {2019})}\BibitemShut {NoStop}%
\bibitem [{\citenamefont {Kennelly}(1899)}]{Kennelly:1899}%
  \BibitemOpen
  \bibfield  {author} {\bibinfo {author} {\bibfnamefont {A.~E.}\ \bibnamefont {Kennelly}},\ }\href@noop {} {\bibfield  {journal} {\bibinfo  {journal} {Electrical World and Engineer}\ }\textbf {\bibinfo {volume} {34}},\ \bibinfo {pages} {413} (\bibinfo {year} {1899})}\BibitemShut {NoStop}%
\bibitem [{\citenamefont {Williamson}(1936)}]{Williamson:1936}%
  \BibitemOpen
  \bibfield  {author} {\bibinfo {author} {\bibfnamefont {J.}~\bibnamefont {Williamson}},\ }\href {http://www.jstor.org/stable/2371062} {\bibfield  {journal} {\bibinfo  {journal} {American Journal of Mathematics}\ }\textbf {\bibinfo {volume} {58}},\ \bibinfo {pages} {141} (\bibinfo {year} {1936})}\BibitemShut {NoStop}%
\bibitem [{\citenamefont {Janhsen}\ \emph {et~al.}(1992)\citenamefont {Janhsen}, \citenamefont {Schiek},\ and\ \citenamefont {Hansen}}]{Janhsen:1992}%
  \BibitemOpen
  \bibfield  {author} {\bibinfo {author} {\bibfnamefont {A.}~\bibnamefont {Janhsen}}, \bibinfo {author} {\bibfnamefont {B.}~\bibnamefont {Schiek}},\ and\ \bibinfo {author} {\bibfnamefont {V.}~\bibnamefont {Hansen}},\ }in\ \href {https://doi.org/10.1109/EUMA.1992.335748} {\emph {\bibinfo {booktitle} {1992 22nd European Microwave Conference}}},\ Vol.~\bibinfo {volume} {1}\ (\bibinfo {year} {1992})\ p.\ \bibinfo {pages} {251}\BibitemShut {NoStop}%
\bibitem [{\citenamefont {Hassler}\ \emph {et~al.}(2019)\citenamefont {Hassler}, \citenamefont {Stubenrauch},\ and\ \citenamefont {Ciani}}]{Hassler:2019}%
  \BibitemOpen
  \bibfield  {author} {\bibinfo {author} {\bibfnamefont {F.}~\bibnamefont {Hassler}}, \bibinfo {author} {\bibfnamefont {J.}~\bibnamefont {Stubenrauch}},\ and\ \bibinfo {author} {\bibfnamefont {A.}~\bibnamefont {Ciani}},\ }\href {https://doi.org/10.1103/PhysRevB.99.014515} {\bibfield  {journal} {\bibinfo  {journal} {Physical Review B}\ }\textbf {\bibinfo {volume} {99}},\ \bibinfo {pages} {014515} (\bibinfo {year} {2019})}\BibitemShut {NoStop}%
\bibitem [{\citenamefont {Osborne}\ \emph {et~al.}(2023)\citenamefont {Osborne}, \citenamefont {Larson}, \citenamefont {Jones}, \citenamefont {Simmonds}, \citenamefont {Gyenis},\ and\ \citenamefont {Lucas}}]{Osborne:2023}%
  \BibitemOpen
  \bibfield  {author} {\bibinfo {author} {\bibfnamefont {A.}~\bibnamefont {Osborne}}, \bibinfo {author} {\bibfnamefont {T.}~\bibnamefont {Larson}}, \bibinfo {author} {\bibfnamefont {S.}~\bibnamefont {Jones}}, \bibinfo {author} {\bibfnamefont {R.~W.}\ \bibnamefont {Simmonds}}, \bibinfo {author} {\bibfnamefont {A.}~\bibnamefont {Gyenis}},\ and\ \bibinfo {author} {\bibfnamefont {A.}~\bibnamefont {Lucas}},\ }\Eprint {https://arxiv.org/abs/2304.08531} {arXiv:2304.08531 [quant-ph]}  (\bibinfo {year} {2023})\BibitemShut {NoStop}%
\bibitem [{\citenamefont {Osborne}\ \emph {et~al.}(2024)\citenamefont {Osborne}, \citenamefont {Larson}, \citenamefont {Jones}, \citenamefont {Simmonds}, \citenamefont {Gyenis},\ and\ \citenamefont {Lucas}}]{Osborne_v2:2024}%
  \BibitemOpen
  \bibfield  {author} {\bibinfo {author} {\bibfnamefont {A.}~\bibnamefont {Osborne}}, \bibinfo {author} {\bibfnamefont {T.}~\bibnamefont {Larson}}, \bibinfo {author} {\bibfnamefont {S.~G.}\ \bibnamefont {Jones}}, \bibinfo {author} {\bibfnamefont {R.~W.}\ \bibnamefont {Simmonds}}, \bibinfo {author} {\bibfnamefont {A.}~\bibnamefont {Gyenis}},\ and\ \bibinfo {author} {\bibfnamefont {A.}~\bibnamefont {Lucas}},\ }\href {https://doi.org/10.1103/PRXQuantum.5.020309} {\bibfield  {journal} {\bibinfo  {journal} {PRX Quantum}\ }\textbf {\bibinfo {volume} {5}},\ \bibinfo {pages} {020309} (\bibinfo {year} {2024})}\BibitemShut {NoStop}%
\bibitem [{\citenamefont {Choquet-Bruhat}\ \emph {et~al.}(1982)\citenamefont {Choquet-Bruhat}, \citenamefont {DeWitt-Morette},\ and\ \citenamefont {Bleick}}]{Choquet-Bruhat:1982}%
  \BibitemOpen
  \bibfield  {author} {\bibinfo {author} {\bibfnamefont {Y.}~\bibnamefont {Choquet-Bruhat}}, \bibinfo {author} {\bibfnamefont {C.}~\bibnamefont {DeWitt-Morette}},\ and\ \bibinfo {author} {\bibfnamefont {M.}~\bibnamefont {Bleick}},\ }\href@noop {} {\emph {\bibinfo {title} {Analysis, Manifolds and Physics, Part 1: Basics}}},\ Analysis, Manifolds, and Physics\ (\bibinfo  {publisher} {North-Holland},\ \bibinfo {year} {1982})\BibitemShut {NoStop}%
\bibitem [{\citenamefont {Jos{\'e}}\ and\ \citenamefont {Saletan}(1998)}]{Jose:1998}%
  \BibitemOpen
  \bibfield  {author} {\bibinfo {author} {\bibfnamefont {J.}~\bibnamefont {Jos{\'e}}}\ and\ \bibinfo {author} {\bibfnamefont {E.}~\bibnamefont {Saletan}},\ }\href {https://doi.org/10.1017/CBO9780511803772} {\emph {\bibinfo {title} {Classical Dynamics: A Contemporary Approach}}},\ Classical Dynamics: A Contemporary Approach\ (\bibinfo  {publisher} {Cambridge University Press},\ \bibinfo {year} {1998})\BibitemShut {NoStop}%
\bibitem [{\citenamefont {Abraham}\ and\ \citenamefont {Marsden}(2008)}]{Abraham:2008}%
  \BibitemOpen
  \bibfield  {author} {\bibinfo {author} {\bibfnamefont {R.}~\bibnamefont {Abraham}}\ and\ \bibinfo {author} {\bibfnamefont {J.}~\bibnamefont {Marsden}},\ }\href@noop {} {\emph {\bibinfo {title} {Foundations of Mechanics}}},\ AMS Chelsea publishing\ (\bibinfo  {publisher} {Benjamin/Cummings Publishing Company},\ \bibinfo {year} {2008})\BibitemShut {NoStop}%
\bibitem [{\citenamefont {Carosso}(2022)}]{Carosso:2022}%
  \BibitemOpen
  \bibfield  {author} {\bibinfo {author} {\bibfnamefont {A.}~\bibnamefont {Carosso}},\ }\href@noop {} {\bibfield  {journal} {\bibinfo  {journal} {Studies in History and Philosophy of Science}\ }\textbf {\bibinfo {volume} {96}},\ \bibinfo {pages} {35} (\bibinfo {year} {2022})}\BibitemShut {NoStop}%
\bibitem [{\citenamefont {Belevitch}(1968)}]{Belevitch:1968}%
  \BibitemOpen
  \bibfield  {author} {\bibinfo {author} {\bibfnamefont {V.}~\bibnamefont {Belevitch}},\ }\href@noop {} {\emph {\bibinfo {title} {Classical Network Theory}}},\ Holden-Day series in information systems\ (\bibinfo  {publisher} {Holden-Day},\ \bibinfo {year} {1968})\BibitemShut {NoStop}%
\bibitem [{\citenamefont {Goldstein}\ \emph {et~al.}(2002)\citenamefont {Goldstein}, \citenamefont {Poole},\ and\ \citenamefont {Safko}}]{Goldstein:2002}%
  \BibitemOpen
  \bibfield  {author} {\bibinfo {author} {\bibfnamefont {H.}~\bibnamefont {Goldstein}}, \bibinfo {author} {\bibfnamefont {C.~P.}\ \bibnamefont {Poole}},\ and\ \bibinfo {author} {\bibfnamefont {J.~L.}\ \bibnamefont {Safko}},\ }\href@noop {} {\emph {\bibinfo {title} {Classical Mechanics}}},\ \bibinfo {edition} {3rd}\ ed.\ (\bibinfo  {publisher} {Pearson Education, Inc, publishing as Addison Wesley},\ \bibinfo {year} {2002})\BibitemShut {NoStop}%
\bibitem [{\citenamefont {Thiemann}(2008)}]{Thiemann:2008}%
  \BibitemOpen
  \bibfield  {author} {\bibinfo {author} {\bibfnamefont {T.}~\bibnamefont {Thiemann}},\ }\href@noop {} {\emph {\bibinfo {title} {Modern canonical quantum general relativity}}}\ (\bibinfo  {publisher} {Cambridge University Press},\ \bibinfo {year} {2008})\BibitemShut {NoStop}%
\bibitem [{\citenamefont {Mukunda}(1980)}]{Mukunda:1980}%
  \BibitemOpen
  \bibfield  {author} {\bibinfo {author} {\bibfnamefont {N.}~\bibnamefont {Mukunda}},\ }\href {https://doi.org/10.1088/0031-8949/21/6/004} {\bibfield  {journal} {\bibinfo  {journal} {Physica Scripta}\ }\textbf {\bibinfo {volume} {21}},\ \bibinfo {pages} {801} (\bibinfo {year} {1980})}\BibitemShut {NoStop}%
\bibitem [{\citenamefont {{Evans}}\ and\ \citenamefont {{Tuckey}}(1993)}]{Evans:1993}%
  \BibitemOpen
  \bibfield  {author} {\bibinfo {author} {\bibfnamefont {J.~M.}\ \bibnamefont {{Evans}}}\ and\ \bibinfo {author} {\bibfnamefont {P.~A.}\ \bibnamefont {{Tuckey}}},\ }\href {https://doi.org/10.1142/S0217751X93001661} {\bibfield  {journal} {\bibinfo  {journal} {International Journal of Modern Physics A}\ }\textbf {\bibinfo {volume} {08}},\ \bibinfo {pages} {4055} (\bibinfo {year} {1993})}\BibitemShut {NoStop}%
\bibitem [{\citenamefont {Gitman}\ and\ \citenamefont {Tyutin}(2012)}]{Gitman:2012}%
  \BibitemOpen
  \bibfield  {author} {\bibinfo {author} {\bibfnamefont {D.}~\bibnamefont {Gitman}}\ and\ \bibinfo {author} {\bibfnamefont {I.~V.}\ \bibnamefont {Tyutin}},\ }\href@noop {} {\emph {\bibinfo {title} {Quantization of fields with constraints}}}\ (\bibinfo  {publisher} {Springer Science \& Business Media},\ \bibinfo {year} {2012})\BibitemShut {NoStop}%
\bibitem [{\citenamefont {Belhadi}\ \emph {et~al.}(2016)\citenamefont {Belhadi}, \citenamefont {B{\'e}rard},\ and\ \citenamefont {Mohrbach}}]{Belhadi:2016}%
  \BibitemOpen
  \bibfield  {author} {\bibinfo {author} {\bibfnamefont {Z.}~\bibnamefont {Belhadi}}, \bibinfo {author} {\bibfnamefont {A.}~\bibnamefont {B{\'e}rard}},\ and\ \bibinfo {author} {\bibfnamefont {H.}~\bibnamefont {Mohrbach}},\ }\href {https://doi.org/10.1016/j.physleta.2016.08.018} {\bibfield  {journal} {\bibinfo  {journal} {Physics Letters A}\ }\textbf {\bibinfo {volume} {380}},\ \bibinfo {pages} {3355} (\bibinfo {year} {2016})}\BibitemShut {NoStop}%
\bibitem [{\citenamefont {Riwar}\ and\ \citenamefont {DiVincenzo}(2022)}]{Riwar:2022}%
  \BibitemOpen
  \bibfield  {author} {\bibinfo {author} {\bibfnamefont {R.-P.}\ \bibnamefont {Riwar}}\ and\ \bibinfo {author} {\bibfnamefont {D.~P.}\ \bibnamefont {DiVincenzo}},\ }\href {https://doi.org/10.1038/s41534-022-00539-x} {\bibfield  {journal} {\bibinfo  {journal} {npj Quantum Information}\ }\textbf {\bibinfo {volume} {8}},\ \bibinfo {pages} {36} (\bibinfo {year} {2022})}\BibitemShut {NoStop}%
\bibitem [{\citenamefont {Bryon}\ \emph {et~al.}(2023)\citenamefont {Bryon}, \citenamefont {Weiss}, \citenamefont {You}, \citenamefont {Sussman}, \citenamefont {Croot}, \citenamefont {Huang}, \citenamefont {Koch},\ and\ \citenamefont {Houck}}]{Bryon:2023}%
  \BibitemOpen
  \bibfield  {author} {\bibinfo {author} {\bibfnamefont {J.}~\bibnamefont {Bryon}}, \bibinfo {author} {\bibfnamefont {D.}~\bibnamefont {Weiss}}, \bibinfo {author} {\bibfnamefont {X.}~\bibnamefont {You}}, \bibinfo {author} {\bibfnamefont {S.}~\bibnamefont {Sussman}}, \bibinfo {author} {\bibfnamefont {X.}~\bibnamefont {Croot}}, \bibinfo {author} {\bibfnamefont {Z.}~\bibnamefont {Huang}}, \bibinfo {author} {\bibfnamefont {J.}~\bibnamefont {Koch}},\ and\ \bibinfo {author} {\bibfnamefont {A.~A.}\ \bibnamefont {Houck}},\ }\href {https://doi.org/10.1103/PhysRevApplied.19.034031} {\bibfield  {journal} {\bibinfo  {journal} {Physical Review Applied}\ }\textbf {\bibinfo {volume} {19}},\ \bibinfo {pages} {034031} (\bibinfo {year} {2023})}\BibitemShut {NoStop}%
\bibitem [{\citenamefont {Ince}(1956)}]{ince:1956}%
  \BibitemOpen
  \bibfield  {author} {\bibinfo {author} {\bibfnamefont {E.}~\bibnamefont {Ince}},\ }\href@noop {} {\emph {\bibinfo {title} {Ordinary Differential Equations}}},\ Dover Books on Mathematics\ (\bibinfo  {publisher} {Dover Publications},\ \bibinfo {year} {1956})\BibitemShut {NoStop}%
\bibitem [{\citenamefont {Shapere}\ and\ \citenamefont {Wilczek}(2012)}]{Shapere:2012}%
  \BibitemOpen
  \bibfield  {author} {\bibinfo {author} {\bibfnamefont {A.}~\bibnamefont {Shapere}}\ and\ \bibinfo {author} {\bibfnamefont {F.}~\bibnamefont {Wilczek}},\ }\href {https://doi.org/10.1103/PhysRevLett.109.200402} {\bibfield  {journal} {\bibinfo  {journal} {Physical Review Letters}\ }\textbf {\bibinfo {volume} {109}},\ \bibinfo {pages} {200402} (\bibinfo {year} {2012})}\BibitemShut {NoStop}%
\bibitem [{\citenamefont {Schreier}\ \emph {et~al.}(2008)\citenamefont {Schreier}, \citenamefont {Houck}, \citenamefont {Koch}, \citenamefont {Schuster}, \citenamefont {Johnson}, \citenamefont {Chow}, \citenamefont {Gambetta}, \citenamefont {Majer}, \citenamefont {Frunzio}, \citenamefont {Devoret}, \citenamefont {Girvin},\ and\ \citenamefont {Schoelkopf}}]{Schreier:2008}%
  \BibitemOpen
  \bibfield  {author} {\bibinfo {author} {\bibfnamefont {J.~A.}\ \bibnamefont {Schreier}}, \bibinfo {author} {\bibfnamefont {A.~A.}\ \bibnamefont {Houck}}, \bibinfo {author} {\bibfnamefont {J.}~\bibnamefont {Koch}}, \bibinfo {author} {\bibfnamefont {D.~I.}\ \bibnamefont {Schuster}}, \bibinfo {author} {\bibfnamefont {B.~R.}\ \bibnamefont {Johnson}}, \bibinfo {author} {\bibfnamefont {J.~M.}\ \bibnamefont {Chow}}, \bibinfo {author} {\bibfnamefont {J.~M.}\ \bibnamefont {Gambetta}}, \bibinfo {author} {\bibfnamefont {J.}~\bibnamefont {Majer}}, \bibinfo {author} {\bibfnamefont {L.}~\bibnamefont {Frunzio}}, \bibinfo {author} {\bibfnamefont {M.~H.}\ \bibnamefont {Devoret}}, \bibinfo {author} {\bibfnamefont {S.~M.}\ \bibnamefont {Girvin}},\ and\ \bibinfo {author} {\bibfnamefont {R.~J.}\ \bibnamefont {Schoelkopf}},\ }\href {https://doi.org/10.1103/PhysRevB.77.180502} {\bibfield  {journal} {\bibinfo  {journal} {Physical Review B}\ }\textbf {\bibinfo {volume} {77}},\ \bibinfo {pages} {180502} (\bibinfo {year}
  {2008})}\BibitemShut {NoStop}%
\end{thebibliography}%

\end{document}